\documentclass[nojss]{jss}

\usepackage{enumitem}

\usepackage{clipboard}
\newclipboard{quotes}
\usepackage{xr}
\usepackage[all]{nowidow}
\usepackage{tabularx}
\makeatletter
\newcommand{\multiline}[1]{%
  \begin{tabularx}{\dimexpr\linewidth-\ALG@thistlm}[t]{@{}X@{}}
    #1
  \end{tabularx}
}
\makeatother

\usepackage{algpseudocode,algorithm,algorithmicx}

\algrenewcommand\algorithmicrequire{\textbf{Precondition:}}
\algrenewcommand\algorithmicensure{\textbf{Postcondition:}}

\usepackage{siunitx} % si units

\definecolor{darkgreen}{rgb}{0.0, 0.5, 0.0}
 
\usepackage{layouts} % \printinunitsof to get the size of \textwidth and \linewidth

\usepackage{rotating} % \rotatebox

\usepackage{amsmath}	% align environment.
\usepackage{amsfonts}	% \mathbb{} (used for Real number symbol, etc.)
\usepackage{amsthm}		% mathy stuff (Theorems, Lemmas, etc.)
\usepackage{commath}
\usepackage{bbm} % \mathbb{} doesn't support digits (1, 2, 3, etc.), so use \mathbbm{} in these instances
\usepackage{mathtools} % \vdotswithin command to have vertical dots between equals signs

\usepackage{enumerate}

\usepackage{pbox} % formatting cells with a table (forced line break within a cell)
\usepackage{multirow}

\usepackage{thumbpdf,lmodern}

\usepackage{framed}

\newcommand{\class}[1]{`\code{#1}'}
\newcommand{\fct}[1]{\code{#1()}}

\def\mbf#1{{%         \mbf{X} makes X a math bold letter
\mathchoice%          selects with respect to current style
{\hbox{\boldmath$\displaystyle{#1}$}}%      case 1 is displaystyle
{\hbox{\boldmath$\textstyle{#1}$}}%         case 2 is textstyle
{\hbox{\boldmath$\scriptstyle{#1}$}}%       case 3 is scriptstyle
{\hbox{\boldmath$\scriptscriptstyle{#1}$}}% case 4 is scriptscriptstyle
}}
\def\vec{\mbf}

\def\d{\textrm{d}} % Define the "d" for use in integrals.
\newcommand{\Gau}{{\text{Gau}}}
\def\inddist{\:\stackrel{\text{ind}}{\sim}\:}

\renewcommand{\E}[1]{\mathbb{E}(#1)} % Expectation operator
\newcommand{\sd}[1]{{\rm sd\!}\left(#1\right)} % standard deviation operator
\newcommand{\cov}[2]{{\rm cov\!}\left(#1,\, #2\right)} % covariance operator
\newcommand{\tp}{{\!\scriptscriptstyle \top}}
\usepackage{scalerel,stackengine}
\usepackage{scalerel}
\stackMath
\newcommand\reallywidehat[1]{%
\savestack{\tmpbox}{\stretchto{%
  \scaleto{%
    \scalerel*[\widthof{\ensuremath{#1}}]{\kern-.6pt\bigwedge\kern-.6pt}%
    {\rule[-\textheight/2]{1ex}{\textheight}}%WIDTH-LIMITED BIG WEDGE
  }{\textheight}% 
}{0.5ex}}%
\stackon[1pt]{#1}{\tmpbox}%
}

\stackMath
\usepackage{verbatimbox} % For \addvbuffer
\usepackage{xparse}
\newlength\glyphwidth
\newlength\widthofx
\newsavebox\hatglyphCONTENT
\sbox\hatglyphCONTENT{%
    \addvbuffer[-0.05ex -1.3ex]{$\hat{\phantom{.}}$}%
}
\newcommand\hatglyph{\resizebox{0.6\widthofx}{!}{\usebox{\hatglyphCONTENT}}}
\newcommand\shifthat[2]{%
    \stackengine{0.2\widthofx}{%
        \SavedStyle#2}{%
        \rule{#1}{0ex}\hatglyph}{O}{c}{F}{T}{S}%
}
\ExplSyntaxOn
\newcommand\relativeGlyphOffset[1]{%
    \str_case:nnF{#1}{%
        {A}{0.18}%
        {B}{0.1}%
        {W}{0.02}%
        {J}{0.18}%
        {\phi}{0.17}%
    }{0.05}% Default
}\ExplSyntaxOff
\NewDocumentCommand{\hatt}{mO{#1}}{%
    \ThisStyle{%
        \setlength\glyphwidth{\widthof{$\SavedStyle{}\longleftarrow$}}%
        \setlength\widthofx{\widthof{$\SavedStyle{}x$}}%
        \shifthat{\relativeGlyphOffset{#2}\glyphwidth}{#1}%
  }%
}

\usepackage{tikz}
\usepackage{tikz-qtree}
\usepackage{pgf}
\usetikzlibrary{positioning}
\usetikzlibrary{arrows, automata}
\usetikzlibrary{shapes.geometric}
\usetikzlibrary{arrows.meta}
\usetikzlibrary{matrix} % for the grid

\makeatletter
\def\squarecorner#1{
    \pgf@x=\the\wd\pgfnodeparttextbox%
    \pgfmathsetlength\pgf@xc{\pgfkeysvalueof{/pgf/inner xsep}}%
    \advance\pgf@x by 2\pgf@xc%
    \pgfmathsetlength\pgf@xb{\pgfkeysvalueof{/pgf/minimum width}}%
    \ifdim\pgf@x<\pgf@xb%
        \pgf@x=\pgf@xb%
    \fi%
    \pgf@y=\ht\pgfnodeparttextbox%
    \advance\pgf@y by\dp\pgfnodeparttextbox%
    \pgfmathsetlength\pgf@yc{\pgfkeysvalueof{/pgf/inner ysep}}%
    \advance\pgf@y by 2\pgf@yc%
    \pgfmathsetlength\pgf@yb{\pgfkeysvalueof{/pgf/minimum height}}%
    \ifdim\pgf@y<\pgf@yb%
        \pgf@y=\pgf@yb%
    \fi%
    \ifdim\pgf@x<\pgf@y%
        \pgf@x=\pgf@y%
    \else
        \pgf@y=\pgf@x%
    \fi
    \pgf@x=#1.5\pgf@x%
    \advance\pgf@x by.5\wd\pgfnodeparttextbox%
    \pgfmathsetlength\pgf@xa{\pgfkeysvalueof{/pgf/outer xsep}}%
    \advance\pgf@x by#1\pgf@xa%
    \pgf@y=#1.5\pgf@y%
    \advance\pgf@y by-.5\dp\pgfnodeparttextbox%
    \advance\pgf@y by.5\ht\pgfnodeparttextbox%
    \pgfmathsetlength\pgf@ya{\pgfkeysvalueof{/pgf/outer ysep}}%
    \advance\pgf@y by#1\pgf@ya%
}
\makeatother

\pgfdeclareshape{square}{
    \savedanchor\northeast{\squarecorner{}}
    \savedanchor\southwest{\squarecorner{-}}

    \foreach \x in {east,west} \foreach \y in {north,mid,base,south} {
        \inheritanchor[from=rectangle]{\y\space\x}
    }
    \foreach \x in {east,west,north,mid,base,south,center,text} {
        \inheritanchor[from=rectangle]{\x}
    }
    \inheritanchorborder[from=rectangle]
    \inheritbackgroundpath[from=rectangle]
}

\newcommand{\FRKgeneric}{\pkg{FRK} }

\author{Matthew Sainsbury-Dale\\University of Wollongong
   \And \quad\quad Andrew Zammit-Mangion\\\quad\quad University of Wollongong 
   \And Noel Cressie\\University of Wollongong}
\Plainauthor{Matthew Sainsbury-Dale, Andrew Zammit-Mangion, Noel Cressie}

\title{\Copy{Title}{Modelling Big, Heterogeneous, Non-Gaussian Spatial and Spatio-Temporal Data using \pkg{FRK}}}
\Plaintitle{\Copy{PlainTitle}{Modelling Big, Heterogeneous, Non-Gaussian Spatial and Spatio-Temporal Data using FRK}}
\Shorttitle{Modelling Non-Gaussian Spatial and Spatio-Temporal Data using \pkg{FRK}}

\Abstract{
 Non-Gaussian spatial and spatio-temporal data are becoming increasingly prevalent, and their analysis is needed in a variety of disciplines. \pkg{FRK} is an \proglang{R} package for spatial/spatio-temporal modelling and prediction with very large data sets that, to date, has only supported linear process models and Gaussian data models. In this paper, we describe a major upgrade to \pkg{FRK} that allows for non-Gaussian data to be analysed in a generalised linear mixed model framework. These vastly more general spatial and spatio-temporal models are fitted using the Laplace approximation via the software \pkg{TMB}. The existing functionality of \pkg{FRK} is retained with this advance into non-Gaussian models; in particular, it allows for automatic basis-function construction, it can handle both point-referenced and areal data simultaneously, and it can predict process values at any spatial support from these data. This new version of \pkg{FRK} also allows for the use of a large number of basis functions when modelling the spatial process, and is thus often able to achieve more accurate predictions than previous versions of the package in a Gaussian setting. We demonstrate innovative features in this new version of \pkg{FRK}, highlight its ease of use, and compare it to alternative packages using both simulated and real data sets. 
}

\Plainkeywords{areal data, basis functions, big data, change-of-support, fixed rank kriging, non-Gaussian data, spatial statistics}

\Address{
  Matthew Sainsbury-Dale\\
  National Institute for Applied Statistics Research Australia (NIASRA)\\
  School of Mathematics and Applied Statistics\\
  University of Wollongong\\
  Wollongong, Australia\\
  E-mail: \email{msainsburydale@gmail.com}\\
  URL: \url{https://github.com/msainsburydale}
}

\begin{document}

\sloppy % Make latex less fussy and prevent words going into the margin.

%    \printinunitsof{in}\prntlen{\textwidth}
%    \printinunitsof{in}\prntlen{\linewidth}

\section{Introduction}\label{sec:intro}

 Non-Gaussian spatial and spatio-temporal data arise from a vast array of sources, and % studies in the geophysical sciences and social sciences. 
% contaminated soil \citep[e.g.,][]{Paul_Cressie_2011_lognormal_kriging_block_prediction}, satellite remote sensing  \citep[e.g.,][]{Sengupta_2016_MODIS}, small-area demographics \citep[e.g.,][]{Bradley_2016_Bayesian_spatial_COS_lattice_data}, and earthquake magnitudes \citep[e.g.,][]{Hu_2018_log-gamma_earthquake_magnitudes}.  
 the statistical modelling of these data is pertinent, as accurate predictions, and uncertainty quantification of those predictions, give informed answers to real-world problems. 
 
 There are, by now, several approaches to statistical modelling and spatial/spatio-temporal prediction with non-Gaussian data. 
 One widespread method to deal with non-Gaussian data is \textit{trans-Gaussian kriging} \citep[pg.~137--138]{Cressie_1993_stats_for_spatial_data}, in which standard kriging (i.e., spatial optimal linear prediction) is used after applying a non-linear transformation to the data, and approximately unbiased predictions are made on the original scale using a delta-method approximation. 
Several other approaches hinge on the use of a spatial version of the generalised linear mixed model (GLMM), whereby the response distribution is assumed to be a member of the exponential family of distributions \citep[e.g.,][]{McCullagh_Nelder_1989_GLM}, and the mean is modelled using a transformation of some latent spatial process $Y(\cdot)$ \citep{Diggle_1998_spatial_GLMM}. 
 In their seminal work, \cite{Diggle_1998_spatial_GLMM} employed a stationary model for $Y(\cdot)$ within the spatial GLMM framework, and a Markov chain Monte Carlo (MCMC) algorithm to obtain predictive distributions. 
 Optimal prediction or estimation of unknown quantities from $m$ observations entails the inversion of an $m\times m$ covariance matrix for many statistical models. 
% Since this task is generally $O(m^3)$ in computational complexity, some form of dimension-reduction is often employed in `big data' settings. 
 \Copy{FixedRankMethods}{Since this task is generally $O(m^3)$ in computational complexity, alternative approaches that scale well with sample size, which we collectively refer to as \textit{fixed-rank} approaches, are often employed in `big data' settings.}

 Fixed-rank variants of trans-Gaussian kriging are relatively under-developed 
% (see \cite{Cressie_2021_review_spatial-basis-function_models}, sec.~3.2.1, for a discussion), 
 \citep[see][sec.~3.2.1]{Cressie_2021_review_spatial-basis-function_models}, 
 however many modellers have used fixed-rank variants of the spatial GLMM.  
  A popular fixed-rank model for $Y(\cdot)$ is the so-called spatial random effects (SRE) model, where $Y(\cdot)$ is modelled as a linear combination of a fixed number of spatial basis functions with spatially correlated random coefficients \citep{Cressie_Johannesson_2008_FRK}: For example, \cite{Sengupta_Cressie_2013_spatial_GLMM_FRK} and \citet{Bradley_2016_Bayesian_spatial_COS_lattice_data} use it in the spatial GLMM context. 
%  They obtained parameter estimates using a Laplace approximation in an expectation maximisation (EM) algorithm, and the empirical predictive distribution was generated using an MCMC algorithm.  
% A latent SRE model was also employed by \citet{Bradley_2016_Bayesian_spatial_COS_lattice_data} to facilitate spatial change-of-support for count-valued survey data. % (in particular, up-scaling to larger target supports).  
% \citet{Bradley_2020_Bayesian_Hierarchical_Models_With_Conjugate_Full-Conditional_Distributions_for_Dependent_Data_From_the_Natural_Exponential_Family} modelled data from the natural exponential family using the so-called `conjugate multivariate distribution'. %, which is the multivariate version of the conjugate prior for distributions in the natural exponential family presented by \cite{Diaconis_Ylvisaker_1979_conjugate_prior_natural_exponential_family}. 
\cite{Lindgren_Rue_2011_GF_GMRF_SPDE} modelled $Y(\cdot)$ by linking Gaussian fields (GFs) with Gaussian Markov random fields (GMRFs) via stochastic partial differential equations (SPDEs), with dimension-reduction facilitated by the finite-element method.  
 \cite{Finley_2020_spNNGP} modelled binomial data using a spatial GLMM with $Y(\cdot)$ a nearest neighbour Gaussian process \citep[NNGP; ][]{Datta_2016_NNGP_spatial}.  
% Another approach that aids scalability is the partitioning of the spatial domain into disjoint regions, which can facilitate code parallelism. 
% In the spatial GLMM context, 
 \cite{Lee_2020_partitioned_domain_basis_function_non_Gaussian} %used a clustering algorithm to partition 
 partitioned the spatial domain into disjoint subregions and, for each subregion, 
 a spatial GLMM model 
 %with a thin-plate-splines representation for $Y(\cdot)$ 
 was used independently of the other subregions. 
 Then the global process was constructed as a weighted sum of the mutually independent local processes. 
% Their local models for $Y(\cdot)$ resembled the SRE models used by other authors; however, the basis-function covariance matrices in each subregion were assumed to be diagonal 
% (a reduced set of basis functions was used to lower the potential for overfitting). % using lasso regression  
% and their model did not cater for so-called `fine-scale' process variation.
 The fixed-rank spatial GLMM naturally extends to the spatio-temporal setting; see, for example,  \cite{Lopes_2011_spatial_GLMM_reduced_rank_factor_analytic_model}, 
\citet{Bradley_2018_computationally_efficient_multivariate_ST_models_for_high-dimensional_count-valued_data}, \citet{Bradley_2019_ST_models_for_big_multinomial_data_using_conditional_multivariate_logit_beta_distribution}, and \citet{Zhang_2020_spatio-temporal_Arctic_sea_ice}. 
 
 \Copy{KrigingClause}{Despite the many modelling approaches available, software for spatial and spatio-temporal modelling of non-Gaussian data (for which classical kriging-based approaches are linear and hence may be sub-optimal) are relatively limited.} 
 \Copy{SoftwareReview}{
 Software 
 %that facilitate in a straightforward manner the modelling of non-Gaussian spatial and spatio-temporal data 
 dedicated to this task include the \proglang{R} \citep{Rcoreteam_2021} packages  \pkg{ngspatial} \citep{ngspatial_2014}, \pkg{spBayes} \citep{Finley_2015_spBayes}, \pkg{mgcv} \citep{Wood_2017_GAM:R}, \pkg{spNNGP} \citep{Finley_2020_spNNGP}, \pkg{georob} \citep{georob}, and \pkg{spatialfusion} \citep{spatialfusion}. 
 Each of these packages has a different set of limitations: \pkg{spBayes}, \pkg{mgcv}, and \pkg{spNNGP} are limited to point-referenced data; \Copy{spBayesLimitation}{\pkg{spBayes} uses basis functions that depend on covariance-function parameters, so that computationally it can only handle a small number of predictive-process knots, which in turn yields a high degree of smoothing}; \pkg{georob} is not designed for large data sets; and \pkg{ngspatial}, \pkg{spBayes}, \pkg{spNNGP}, \pkg{georob}, and \pkg{spatialfusion} are restricted to the spatial setting, where they cater for only a small number of non-Gaussian distributions.
 Further, with the exception of \pkg{spatialfusion}, these software packages do not cater for spatial change-of-support.
 }
  Some general-purpose packages  \citep[e.g., \pkg{INLA};][]{Rue_2009_INLA, Lindgren_2015_R-INLA} can, in principle, handle the wide array of modelling challenges posed by non-Gaussian spatial and spatio-temporal data; however, they are not specifically designed for this purpose and can be difficult for an unfamiliar user to implement. 
% A project aimed at facilitating Gaussian and non-Gaussian spatial statistical modelling using \pkg{INLA} is the \pkg{inlabru} package \citep{Bachl_2019_inlabru}; at the time of writing,  spatio-temporal modelling was not implemented in \pkg{inlabru}. 
 The package \pkg{inlabru} \citep{Bachl_2019_inlabru} aims to facilitate Gaussian and non-Gaussian spatial modelling using \pkg{INLA} but, at the time of writing, it does not implement spatio-temporal modelling.

 \pkg{FRK} \citep{FRK_paper} is an \proglang{R} package for spatial/spatio-temporal statistical modelling and prediction. 
 In this article, we present a major upgrade to \pkg{FRK} that allows one to cater for many distributions within the exponential family using the spatial GLMM framework; we henceforth refer to it as \pkg{FRK}~v2 and the original version as \pkg{FRK}~v1. 
 \pkg{FRK}~v2 provides a unifying framework that handles large, spatial and spatio-temporal non-Gaussian (and Gaussian) data, and it can seamlessly ingest point-referenced and area-referenced data to solve spatial change-of-support problems. %facilitates change-of-support over target supports that may be smaller or larger than the source supports (i.e., it caters for statistical down-scaling and up-scaling). 
  User-friendliness is a central focus of the package: Challenging statistical analyses may be tackled with only a few lines of intuitive, readable code. Optimal spatial prediction proceeds through the use of an empirical hierarchical statistical model (where likelihood-based estimates are substituted in place of unknown parameters) and a Monte Carlo (MC) algorithm, where a minimal number of user-level decisions is required.
 \pkg{FRK}~v2 also accommodates the modelling of non-Gaussian spatial and spatio-temporal data on the surface of a sphere, a feature not offered by many other packages. 
 Finally, although the primary motivation for this major upgrade is the modelling of non-Gaussian data, \pkg{FRK}~v2 also allows for the use of substantially more basis functions than \pkg{FRK}~v1, which often results in more accurate predictions when in a Gaussian setting. % when modelling the spatial process. 
% Therefore, in a Gaussian setting, it is often able to achieve more accurate predictions than \pkg{FRK}~v1. 

The remainder of the paper is organised as follows. 
In Section~\ref{SEC:Methodology}, we establish the statistical framework for \pkg{FRK}~v2, and we describe model fitting and prediction. 
In Section~\ref{SEC:IllustrativeExample}, we discuss and illustrate the new functionalities in \pkg{FRK}~v2. 
In Section~\ref{SEC:ApplicationStudy}, we present a comparative study between \pkg{FRK}~v2 and several related packages, as well as real-world applications of \pkg{FRK}~v2. 
 Section~\ref{SEC:Conclusion} gives a discussion and conclusions. 

\section{Methodology}\label{SEC:Methodology}

The model used in \pkg{FRK}~v2 is a spatial or spatio-temporal hierarchical statistical model consisting of two conditional-probability layers.
 In the \textit{process layer}, we model the conditional mean of the data as a transformation of a latent spatial process modelled as a low-rank SRE model; see Section  \ref{subsection:04-01:ProcessLayer}. 
 In the %first layer, which we refer to as the 
\textit{data layer}, we use a conditionally independent exponential-family model for each element of the data vector; see Section~\ref{subsection:DataLayer}. 
In Section~\ref{subsection:02-03:Estimation}, we discuss parameter estimation and, in Section~\ref{subsection:Prediction}, we discuss spatial prediction and uncertainty quantification of the predictions. 
 In Section~\ref{sec:Distributions with size parameters}, we consider two distributions that have an assumed-known `size' parameter, namely the binomial distribution and the negative-binomial distribution.
In Section~\ref{sec:spatio-temporal}, we present the approach of \pkg{FRK}~v2 for spatio-temporal data.

\subsection{The process layer} \label{subsection:04-01:ProcessLayer}

%\Copy{Process}{The process layer, which governs the conditional mean of the data, retains many similarities to that in \pkg{FRK}~v1. 
% Note that here we discuss the spatial case only; the extension to a spatio-temporal setting is outlined in Section~\ref{sec:spatio-temporal}. }

The process layer, which governs the conditional mean of the data, retains many similarities to that in \pkg{FRK}~v1. Note that here we discuss the spatial case only; the extension to a spatio-temporal setting is outlined in Section~\ref{sec:spatio-temporal}.

 We denote the latent spatial process as $Y(\cdot) \equiv \{Y(\vec{s}) \colon \vec{s}\in D\}$, where $\vec{s}$ indexes space in the spatial domain of interest $D$. 
 The model for the latent process is
\begin{equation}\label{eqn:04-01:Y(s)}
    Y(\vec{s}) = \vec{t}(\vec{s})^\tp \vec{\alpha} + v(\vec{s}) + \xi(\vec{s}); \quad \vec{s} \in D,
\end{equation}
where each term in (\ref{eqn:04-01:Y(s)}) models a different type of spatial variability. 
First, spatially referenced covariates $\vec{t}(\cdot)$ and their associated regression parameters $\vec{\alpha}$, capture spatial variation that is linked to known, usually large-scale, explanatory variables that are elements of $\vec{t}(\cdot)$; the model requires that the covariates are known at every location in $D$. 
Second, the spatially correlated random effect $v(\cdot)$ captures medium-to-small-scale spatial variation.
%Accounting for the scales of spatial variation with only $\vec{t}(\cdot)^\tp \vec{\alpha}$ and $v(\cdot)$ can result in an overly smooth spatial model and hence overly optimistic predictions; this problem is alleviated by also including a fine-scale-variation random process, $\xi(\cdot)$. 
 \Copy{MotivationForFineScaleProcess}{If the spatial process being modelled has fine-scale variation, including only $\vec{t}(\cdot)^\tp \vec{\alpha}$ and $v(\cdot)$ can result in an overly smooth spatial model and hence overly optimistic predictions. This problem is alleviated by also including a fine-scale-variation random process, $\xi(\cdot)$, in the model.}
 
In \FRKgeneric, the medium-to-small-scale term $v(\cdot)$ is constructed as a linear combination of $r$ spatial basis functions with random coefficients, where $r$ is fixed and usually smaller than $m$, the number of observations. Specifically,
\[
v(\vec{s}) 
= \sum_{l=1}^r \phi_l(\vec{s})\eta_l
= \vec{\phi}(\vec{s})^\tp \vec{\eta}; \quad \vec{s} \in D,
\]
where $\vec{\eta} \equiv \left(\eta_1, \dots, \eta_r \right)^\tp$ is an $r$-dimensional 
vector of random coefficients for the 
$r$-dimensional 
vector $\vec{\phi}(\cdot) \equiv \left(\phi_1(\cdot), \dots, \phi_r(\cdot) \right)^\tp$ of pre-specified spatial basis functions. 
See Appendix~\ref{Appendix:Basis_fns_and_BAUs} for details on how these basis functions are constructed. 
 The fine-scale term, $\xi(\cdot) \equiv \{\xi(\vec{s}): \vec{s} \in D\}$, is modelled as white noise after discretisation, which we discuss next.

 To cater for different observation supports and facilitate solutions to spatial change-of-support problems, \FRKgeneric assumes a discretised domain of interest, \mbox{$D^G \equiv \{A_i: i = 1, \dots, N\}$}, that is made up of $N$ non-overlapping basic areal units (BAUs) such that $D = \cup_{i = 1}^N A_i$. 
% These BAUs may correspond to physical real-world boundaries, such as regional borders, or simply a fine mesh covering that spatial domain. 
 \Copy{BAUEffectOnPrediction}{The finest resolution at which one can make predictions is at the level of the BAU. Point predictions are not possible; however, when the BAUs are so fine that the chosen basis functions are approximately constant within each BAU, the BAU-level predictions are practically equivalent to point predictions (see \citeauthor{FRK_paper}, \citeyear{FRK_paper}, Sec.~2, for more details). %, and see \citeauthor{FRK_paper}, \citeyear{FRK_paper}, Sec.~4, and   \citeauthor{Cressie_2021_review_spatial-basis-function_models}, \citeyear{Cressie_2021_review_spatial-basis-function_models}, Sec.~5).
 }
 Now, let $Y(A_i)$ denote a representative value of $\{Y(\vec{s}) : \vec{s} \in A_i\}$, where commonly that value is the spatial integral or the spatial average over $A_i$.  
 Define the discretised latent spatial process $Y(\cdot)$ evaluated over the $N$ BAUs as $\vec{Y} \equiv (Y_1, \dots, Y_N)^\tp$, where $Y_i \equiv Y(A_i)$, $i = 1, \dots, N$. Then, a vectorised version of (\ref{eqn:04-01:Y(s)}) is
\begin{equation}\label{Ch4:eqn:vecY}
    \vec{Y} = \vec{T}\vec{\alpha} + \vec{S}\vec{\eta} + \vec{\xi},
\end{equation}
 where $\vec{T}$ and $\vec{S}$ are known design matrices constructed from $\vec{t}(\cdot)$ and $\vec{\phi}(\cdot)$ respectively, $\vec{\alpha}$ is a fixed effect, and $\vec{\xi}$ is a vector associated with the fine-scale process which, like $\vec{\eta}$, is  random. 

As in \pkg{FRK}~v1, \Copy{ParameterisationOfBasisFunctionWeights}{the elements of $\vec{\xi}$ are often modelled as independent and identically distributed (i.i.d.) Gaussian random variables with mean zero and variance $\sigma^2_\xi$, and $\vec{\eta}$ is modelled as a mean-zero multivariate-Gaussian random vector with covariance matrix $\cov{\vec{\eta}}{\vec{\eta}}$. 
 In \pkg{FRK}~v2, $\cov{\vec{\eta}}{\vec{\eta}}$ is modelled either as $\vec{K}$ or as $\vec{Q}^{-1}$, where $\vec{Q}$ is a precision matrix.} 
 Both formulations use block-diagonal matrices, so that basis-function coefficients between basis-function resolutions are independent; see Appendix~\ref{Appendix:CovarianceTapering} for how the intra-resolution dependencies with $\vec{K}$ and $\vec{Q}$ are modelled. 
  Although both $\vec{K}$ and $\vec{Q}$ are generally sparse, use of $\vec{Q}$ instead of $\vec{K}$ is typically computationally advantageous. Irrespective of the parameterisation, we assume that $\cov{\vec{\eta}}{\vec{\eta}}$ depends on an unknown parameter vector $\vec{\vartheta}$.

Following standard generalised-linear-model theory \citep{McCullagh_Nelder_1989_GLM}, 
 we use an invertible link function, $g(\cdot)$, to model $Y(\cdot)$ as a transformation of a mean process, $\mu(\cdot)$: % (that we consider in more detail in Section~\ref{subsection:DataLayer}):
\begin{equation}\label{eqn:mu_linked_to_Y}
g\left(\mu(\vec{s})\right) = Y(\vec{s}); \quad \vec{s} \in D.
\end{equation}
Therefore, the mean process evaluated over the BAUs is $\vec{\mu} \equiv (\mu_i: i = 1, \dots, N)^\tp$, where $\mu_i = g^{-1}(Y_i)$,  $i = 1, \dots, N$, and $g^{-1}(\cdot)$ is the inverse link function. 
 We sometimes write $\vec{\mu} = g^{-1}(\vec{Y})$ and $\vec{Y} = g(\vec{\mu})$, where the functions are applied element-wise.

\subsection{The data layer}\label{subsection:DataLayer}

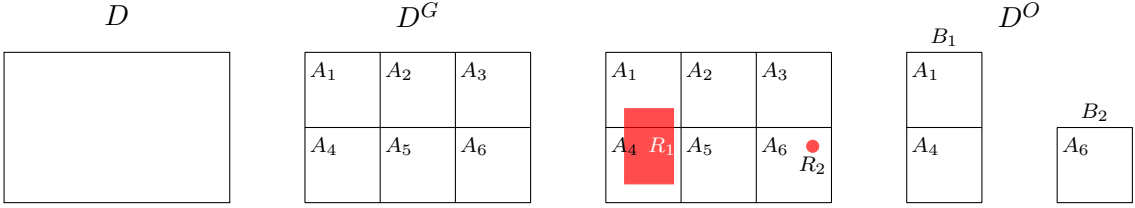
\begin{figure}[t!]
\centering
\begin{tikzpicture}
% Continuous domain D
\begin{scope}[xshift = -4cm]
\node at (1.5,4.5) {$D$};
\draw (0, 2) rectangle (3, 4);% node [below left] {$D$};
\end{scope}
% Discretised domain, D^G
\node at (1.5,4.5) {$D^G$};
\draw[step=1cm] (0,2) grid (3,4);
\node at (0.25,3.75) {\scriptsize $A_1$};
\node at (1.25,3.75) {\scriptsize $A_2$};
\node at (2.25,3.75) {\scriptsize $A_3$};
\node at (0.25,2.75) {\scriptsize $A_4$};
\node at (1.25,2.75) {\scriptsize $A_5$};
\node at (2.25,2.75) {\scriptsize $A_6$};
% Discretised domain, D^G, with observations
\begin{scope}[xshift = 4cm]
\draw[step=1cm] (0,2) grid (3,4);
\filldraw[color=red, fill=red, opacity = 0.7] (0.25, 3.25) rectangle (0.9, 2.25);
\node at (0.75,2.75) {\scriptsize \textcolor{white}{$R_1$}};
\filldraw[red, opacity = 0.7] (2.75,2.75) circle (0.08cm);
\node at (2.75,2.5) {\scriptsize $R_2$};
\node at (0.25,3.75) {\scriptsize $A_1$};
\node at (1.25,3.75) {\scriptsize $A_2$};
\node at (2.25,3.75) {\scriptsize $A_3$};
\node at (0.25,2.75) {\scriptsize $A_4$};
\node at (1.25,2.75) {\scriptsize $A_5$};
\node at (2.25,2.75) {\scriptsize $A_6$};
\end{scope}
% observation supports
\begin{scope}[xshift = 8cm]
\node at (1.5,4.5) {$D^O$};
\draw[step=1cm] (0,4) grid (1,2);
\draw[step=1cm] (2,2) grid (3,3);
\node at (0.5,4.2) {\scriptsize $B_1$};
\node at (0.25,3.75) {\scriptsize $A_1$};
\node at (0.25,2.75) {\scriptsize $A_4$};
\node at (2.5,3.2) {\scriptsize $B_2$};
\node at (2.25,2.75) {\scriptsize $A_6$};
\end{scope}
\end{tikzpicture}
\caption{
An illustration of how the spatial domain, $D$, is discretised into the set $D^G$ of BAUs, and how the observation domain, $D^O$, is derived from the observation supports.
(Left) The continuous spatial domain, $D$.
(Centre left) The spatial domain discretised into $N = 6$ BAUs, $D^G \equiv \{A_i: i = 1, \dots, 6\}$.
(Centre right) $D^G$ superimposed with $m = 2$ observations, one that is areally-referenced ($R_1$), and one that is point-referenced ($R_2$).  
(Right) 
The observation domain, \mbox{$D^O \equiv \{B_j : j = 1, 2\}$}, where $B_1 \equiv A_1 \cup A_4$ and $B_2 \equiv A_6$. 
}\label{fig:BAU_intuition}
\end{figure}

 We denote the vector of $m$ observations (the data vector) as $\vec{Z} \equiv \left(Z_1, \dots, Z_m\right)^\tp$.
   Each datum is originally associated with a spatial support, $R_j \subseteq D$, $j = 1, \dots, m$, which we associate to one or more BAUs. 
 In practice, these spatial supports may not coincide with entire BAUs and, when this is the case, in \pkg{FRK}~v2 we assume that a spatial support contains a BAU if and only if there is a non-empty intersection between the BAU and the spatial support.
 That is, we write the indices of the BAUs associated with spatial support $R_j$ as $c_j \equiv \{i : A_i \cap R_j \neq \emptyset\}$, for $j = 1, \dots, m$. 
%  Attributing point-referenced data to BAUs is straightforward. 
  We then define the set of observation supports in terms of BAUs as 
  $D^O \equiv \{B_j : j = 1, \dots, m\}$, 
  where $B_j \equiv \cup_{i\in c_j} A_i$ is the package's representation of $R_j$ in terms of BAUs. 
% This approach to mapping data supports to BAUs is slightly different from \pkg{FRK}~v1, which assumed that a spatial support contains a BAU if and only if the BAU centroid lies within the spatial support. This modification to \pkg{FRK}~v2 also caters for non-convex BAUs, such as those used in Section~\ref{sec:spatialCOS}, where the centroid of a given BAU may lie outside of the BAU boundary. 
 Figure~\ref{fig:BAU_intuition} shows a pedagogical example with $m = 2$ observations illustrating the relationship between the continuous domain $D$, the BAUs $\{A_1, \dots, A_{6}\}$, the original spatial supports $\{R_1, R_2\}$, and the observation supports $\{B_1, B_2\}$.

Define the conditional mean of the data as $\vec{\mu}_Z \equiv \left(\E{Z_1 \mid \vec{\mu}}, \dots, \E{Z_m \mid \vec{\mu}}\right)^\tp$, where henceforth we treat the supports of $Z_1, \dots, Z_m$ as $B_1, \dots, B_m$, respectively. 
 Since each $B_j \in D^O$ is either a BAU or a union of BAUs, one can construct an $m\times N$ matrix 
\begin{equation}\label{eqn:C_Z}
\vec{C}_Z \equiv \Big(w_{ij}\mathbb{I}(i \in c_j) : i = 1, \dots, N; j = 1, \dots, m\Big),
\end{equation} 
where $\mathbb{I}(\cdot)$ is the indicator function, such that
\begin{equation}\label{eqn:mu_Z}
\vec{\mu}_Z = \vec{C}_Z\vec{\mu}.  
\end{equation} 
 The matrix $\vec{C}_Z$ aggregates the BAU-level process $\vec{\mu}$ over the observation supports and, depending on the weights in (\ref{eqn:C_Z}), it can correspond to a weighted average or a weighted sum over the BAUs. See Appendix~\ref{Appendix:Incidence matrices: Cz and Cp} for details.

Denoting the $j$th element of $\vec{\mu}_Z$ by $\mu_{Z_j}$, we assume that 
\begin{equation}\label{eqn:Z_j|mu(.)}
[Z_j \mid \vec{\mu}, \psi] = \text{EF}(\mu_{Z_j}, \psi); \quad j = 1, \dots, m,
\end{equation}
where EF corresponds to a probability distribution in the exponential family with dispersion parameter $\psi$ and, for generic random quantities $A$ and $B$, $[A \mid B]$ denotes the probability distribution of $A$ given $B$. 
We assume that $\psi$ is spatially invariant (note that $\psi$ is equal to 1 for the binomial, negative-binomial, and Poisson distributions). %; with the parameterisations assumed by \pkg{FRK}~v2, $\psi = 1$ for some distributions in the exponential family (e.g., the binomial, negative-binomial, and Poisson distributions).

Together, (\ref{eqn:mu_Z}) and (\ref{eqn:Z_j|mu(.)}) imply that a given observation depends only on the values of the mean process $\vec{\mu}$ at the corresponding observation support. %, rather than on means elsewhere in the domain. 
Further, we assume that all observations are conditionally independent given the latent spatial process, and that they are all from the same exponential family member. Specifically, 
\[
[\vec{Z} \mid \vec{\mu}_Z, \psi] = \prod_{j=1}^m \text{EF}(\mu_{Zj}, \psi).
\]
As we only consider data models in the exponential family, $\ln{[\vec{Z}  \mid  \vec{\mu}_Z, \psi]}$ may be written as  
\begin{equation}\label{eqn:ln[Z|Y],ExpFam}
\ln{[\vec{Z} \mid \vec{\mu}_Z, \psi]}
=
\sum_{j=1}^m\left\{
\frac{Z_j\lambda(\mu_{Z_j}) - b(\lambda(\mu_{Z_j}))}{a(\psi)} + c(Z_j, \psi)\right\},
\end{equation}
where $a(\cdot)$, $b(\cdot)$, and $c(\cdot, \cdot)$ are deterministic functions specific to the chosen exponential family member, and $\lambda(\cdot)$ is the canonical parameter.

The model employed by \pkg{FRK}~v2 can be summarised as follows.

\begin{gather}
	\Copy{ModelSummary}{ 
    Z_j \mid \vec{\mu}_Z, \psi \inddist \text{EF}(\mu_{Z_j}, \psi); \quad j = 1, \dots, m, \label{eqn:new_model_Z}\\
    \vec{\mu}_Z = \vec{C}_Z \vec{\mu}, \label{eqn:new_model_muZ}\\
    g(\vec{\mu}) = \vec{Y}, \label{eqn:new_model_g(mu)}\\
    \vec{Y} = \vec{T} \vec{\alpha} + \vec{S} \vec{\eta} + \vec{\xi}, \label{eqn:new_model_Y}\\
    \vec{\eta} \mid \vec{\vartheta} \sim \Gau(\vec{0}, \vec{Q}^{-1}), \\
    \vec{\xi} \mid \sigma^2_\xi \sim \Gau(\vec{0}, \sigma^2_\xi \vec{V}), \label{eqn:new_model_priors}
}
\end{gather}
 where $\vec{V}$ is a known, positive-definite diagonal matrix which, in the absence of problem specific fine-scale information, can simply be set to $\vec{I}$, and $\sigma^2_\xi$ is either unknown and estimated, or provided by the user.  
 In a spatio-temporal setting, a more complex model for $\vec{\xi}$ is allowed; see Section~\ref{sec:spatio-temporal}. 
 Note that \pkg{FRK}~v2 is backwards compatible, since an identity link function and a Gaussian data model in (\ref{eqn:new_model_Z}) yields the model used in \pkg{FRK}~v1. %; then $\psi$ is simply the measurement-error variance $\sigma^2_\epsilon$ in $Z_j \mid \mu_{Z_j}, \sigma^2_\epsilon \sim \Gau(\mu_{Z_j}, \sigma^2_\epsilon)$. 

\subsection{Estimation}\label{subsection:02-03:Estimation}

%We now derive the likelihood functions required for model fitting, outline the intractable integrals that arise when non-Gaussian data models are fitted, and describe how \pkg{TMB} \citep{Kristensen_2016_TMB} is used to obtain estimates of the parameters/fixed effects and predictions of the random effects.

Noting that $\vec{\mu}_Z$ is, through (\ref{eqn:new_model_muZ})--(\ref{eqn:new_model_Y}), completely determined by $\vec{\alpha}$, $\vec{\eta}$, and $\vec{\xi}$,  
 the complete-data likelihood function for our model is
\begin{equation}\label{eqn:04:Joint_Likelihood}
    L(\vec{\theta}; \vec{Z}, \vec{\eta}, \vec{\xi})
    \equiv 
    [\vec{Z}, \vec{\eta}, \vec{\xi} \mid \vec{\theta}]
    =
    [\vec{Z} \mid \vec{\mu}_Z, \psi]
    [\vec{\eta} \mid \vec{\vartheta}]
    [\vec{\xi} \mid \sigma^2_\xi], 
\end{equation}
 where $
 \vec{\theta}
 \equiv
 (
 \vec{\alpha}^\tp,
 \vec{\vartheta}^\tp, 
 \sigma^2_\xi, 
 \psi
 )^\tp$, and recall that $\vec{\vartheta}$ denotes the variance-covariance components associated with either $\vec{K}$ or $\vec{Q}$.
The complete-data log-likelihood function, $\ell(\vec{\theta}; \vec{Z}, \vec{\eta}, \vec{\xi})$, is simply the logarithm of (\ref{eqn:04:Joint_Likelihood}). 
Under the modelling assumptions (\ref{eqn:new_model_Z})--(\ref{eqn:new_model_priors}), the conditional density functions  $[\vec{\eta}\mid\vec{\vartheta}]$ and $[\vec{\xi} \mid \sigma^2_\xi]$ are invariant to the specified link function and the assumed distribution of the response variable. 
% Of course, this invariance does not hold for $[\vec{Z} \mid \vec{\mu}_Z, \psi]$. 

%\subsubsection{observed-data likelihood}\label{sec:marginal_likelihood}

The observed-data likelihood, which depends on the observations $\vec{Z}$ and not on the unobserved random effects $\vec{u} \equiv (\vec{\eta}^\tp, \vec{\xi}^\tp)^\tp$, is given by integrating out $\vec{u}$ from (\ref{eqn:04:Joint_Likelihood}):
\begin{equation}\label{eqn:02-04:LikelihoodTheta}
    L^*(\vec{\theta}; \vec{Z}) 
    \equiv
    \int_{\mathbb{R}^{{p}}}
    L(\vec{\theta} ; \vec{Z}, \vec{u}) \d \vec{u}, 
\end{equation}
where ${p}$ is the total number of random effects in the model. 
 The observed-data log-likelihood function is $\ell^*(\vec{\theta}; \vec{Z}) \equiv \log{L^*(\vec{\theta}; \vec{Z})}$. 
When the data are non-Gaussian, the integral in (\ref{eqn:02-04:LikelihoodTheta}) is typically intractable and must be approximated. 
In \pkg{FRK}~v2, a Laplace approximation is used, which we now briefly describe.

Let $\hat{\vec{u}}\equiv\hat{\vec{u}}(\vec{\theta}, \vec{Z})$ be a mode of $\ell(\vec{\theta}; \vec{Z}, \vec{u})$ with respect to $\vec{u}$, 
% \begin{equation*}
% \left.\nabla_{\vec{u}} \ell(\vec{\theta} ; \vec{Z}, \vec{u})\right\rvert_{\vec{u}=\hat{\vec{u}}} = \vec{0},
% \end{equation*}
 and let %$\vec{H}$ be the negative of the inverse Hessian matrix of $\ell(\vec{\theta} ; \vec{Z}, \vec{u})$ with respect to $\vec{u}$, evaluated at $\hat{\vec{u}}$:
\begin{equation*}
    \vec{H} 
    \equiv
    -\left(\left.\nabla_{\vec{u}} \nabla_{\vec{u}} \ell(\vec{\theta} ; \vec{Z}, \vec{u}) \right\rvert_{\vec{u}=\hat{\vec{u}}}\right)^{-1},
\end{equation*}
 where $\nabla_{\vec{u}}$ denotes the gradient with respect to $\vec{u}$. 
 A second-order Taylor-series approximation of $\ell(\vec{\theta} ; \vec{Z}, \vec{u})$ about $\vec{u} =  \hat{\vec{u}}$ results in an approximation of (\ref{eqn:04:Joint_Likelihood}) that has the form of an un-normalised Gaussian density in terms of $\vec{u}$, with mean vector $\hat{\vec{u}}$ and covariance matrix $\vec{H}$. 
 Substitution of this approximation into (\ref{eqn:02-04:LikelihoodTheta}) and evaluation of the integral, yields the Laplace approximation of the observed-data likelihood, $L^*(\vec{\theta}; \vec{Z})
    \approx L(\vec{\theta} ; \vec{Z}, \hat{\vec{u}})
    (2\pi)^{\frac{{p}}{2}}\left|\vec{H}\right|^{\frac{1}{2}}$.
% This allows $\vec{u}$ to be integrated out of $L(\vec{\theta}; \vec{Z}, \vec{u})$, yielding the Laplace approximation of the observed-data log likelihood, $\ell^*(\vec{\theta}; \vec{Z}) \approx \ell(\vec{\theta} ;  \vec{Z}, \hat{\vec{u}}) + \frac{{p}}{2} \ln{2\pi} + \frac{1}{2} \ln{\left|\vec{H}\right|}$.
   \Copy{LaplaceApproximation}{For a more detailed discussion on the Laplace approximation and its properties, see, for example, \cite{Tierney_1986_Laplace_approx}, \cite{Rue_Martino_2007}, and \cite{Rue_2009_INLA}.}

 Note that $[\vec{u} \mid \vec{Z}, \vec{\theta}] \propto [\vec{Z}, \vec{u} \mid \vec{\theta}]$, which is equal to the complete-data likelihood function, $L(\vec{\theta}; \vec{Z}, \vec{u})$. 
 Therefore, since the Laplace approximation replaces $L(\vec{\theta}; \vec{Z}, \vec{u})$ with a term that has the form of an un-normalised Gaussian density in terms of $\vec{u}$, it follows that, approximately, $\vec{u} \mid \vec{Z}, \vec{\theta} \sim \Gau(\hat{\vec{u}}, \vec{H})$. 
  In the software we use (see below), estimates of $\hat{\vec{u}}$ and $\vec{H}^{-1}$ are provided, which makes prediction of $\vec{u}$ and any function of it straightforward via the predictive distribution and its MC simulation (see Section~\ref{subsection:Prediction}).

\subsubsection[Model fitting with TMB]{Model fitting with \pkg{TMB}}

\pkg{FRK}~v2 supplies the \proglang{R} package \pkg{TMB} \citep{Kristensen_2016_TMB} with a \proglang{C++} template function that defines $\ell(\vec{\theta} ; \vec{Z}, \vec{u})$. \pkg{TMB} then computes the Laplace approximation of the observed-data log-likelihood, $\ell^*(\vec{\theta}; \vec{Z})$, and it automatically computes its derivatives; these quantities are then invoked via a user-defined optimising function (\fct{nlminb} is used by default). 
 \pkg{TMB} uses \pkg{CppAD} \citep{CppAD_Package} for automatic differentiation, and it uses the linear-algebra libraries \pkg{Eigen} \citep{Eigen} and \pkg{Matrix} \citep{Matrix_Package} for vector and matrix operations in \proglang{C++} and \proglang{R}, respectively. 
 Use of these packages yields high computational efficiency. 
%\pkg{TMB}'s implementation of automatic differentiation is a key reason why \pkg{FRK}~v2 can easily cater for a large variety of response distributions and link functions, as each response-distribution/link-function combination does not need to be considered on a case-by-case basis.

Note that all unknown quantities are treated as random in \pkg{TMB}. % (with a flat prior assumed if a prior is not provided).
 To retain \pkg{FRK}~v1's mixed-model interpretation, we fix the model parameters and fixed effects to their posterior-mode estimates and then treat them as non-random quantities. 

\subsection{Prediction and uncertainty quantification}\label{subsection:Prediction}

%We now discuss spatial prediction and uncertainty quantification of the predictions. 
There are three principal quantities that could be of interest to the user, namely the latent process 
%$Y(\cdot)$ and mean process $\mu(\cdot)$ in (\ref{eqn:mu_linked_to_Y}), and data at unobserved locations. 
$\vec{Y}$ and mean process $\vec{\mu}$ in (\ref{eqn:new_model_g(mu)}), and data at unobserved locations. 
% To produce predictions and associated uncertainties, we need to determine the posterior distribution of these quantities.
%For each quantity, we use the posterior expectation as our predictor. % ; a decision theoretic justification for this choice is provided by \citet*[ch.~3]{Cressie_1993_stats_for_spatial_data}.
Recall that the Laplace approximation approximates the conditional distribution of $\vec{u} \equiv (\vec{\eta}^\tp, \vec{\xi}^\tp)^\tp$ as \mbox{$\vec{u} \mid \vec{Z}, \vec{\theta} \sim \Gau(\hat{\vec{u}}, \vec{H})$};
since $\vec{Y}$ is a linear function of $\vec{u}$, approximate inference on $\vec{Y}$ can be carried out using well-known formulas. 
%  However, the posterior distribution of a non-linear function of $Y(\cdot)$ (e.g., the mean $\mu(\cdot)$ in (\ref{eqn:mu_linked_to_Y})) is typically not available in closed form, and some approximation is required.
  However, the posterior distribution of a non-linear function of $\vec{Y}$, for example the mean $\vec{\mu}$ in (\ref{eqn:new_model_g(mu)}), is typically not available in closed form, and some approximation is required.
%  In \pkg{FRK}~v2, we use a Monte Carlo (MC) approach to inference on $\vec{u} \equiv (\vec{\eta}^\tp, \vec{\xi}^\tp)^\tp$.
 In \pkg{FRK}~v2 we therefore use a Monte Carlo (MC) approach to inference on non-linear functions of $\vec{Y}$, by first drawing a sample from the approximate conditional distribution of $\vec{u}$ and then transforming the sample accordingly.

 Recall that $\vec{Y} = \vec{T}\vec{\alpha} + \vec{S}\vec{\eta} + \vec{\xi}$, which can be rewritten as $\vec{Y} = \vec{T}\vec{\alpha} + [\vec{S} \; \vec{I}] \,\vec{u}$. 
 We thus define $\vec{Y}_{\!\text{MC}}$, an $N \times n_{\text{MC}}$ matrix whose columns are the $n_{\text{MC}}$ MC samples from $[\vec{Y} \mid \vec{Z}, \vec{\theta}]$, as 
\begin{equation}\label{eqn:Y_MC_simple}
\vec{Y}_{\!\text{MC}}
\equiv \vec{T}\hatt{\vec{A}}[A] + [\vec{S} \; \vec{I}] \,\vec{U},
\end{equation}
where each of the $n_{\text{MC}}$ columns of the matrix $\hatt{\vec{A}}[A]$ is the estimate of $\vec{\alpha}$, and each of the $n_{\text{MC}}$ columns of the matrix $\vec{U}$ is a draw from $\vec{u} \mid \vec{Z}, \vec{\theta} \sim \Gau(\hat{\vec{u}}, \vec{H})$.  
 We obtain MC samples of the $N$-dimensional vector $\vec{\mu}$ from $[\vec{\mu} \mid \vec{Z}, \vec{\theta}]$ via the $N\times n_{MC}$ matrix $\vec{M} \equiv g^{-1}(\vec{Y}_{\!\text{MC}})$, where $g^{-1}(\cdot)$ is applied element-wise.

  \pkg{FRK}~v2 also allows prediction of data over all $N$ BAUs, which we write as $\vec{Z}^* \equiv (Z^*_1, \dots, Z^*_N)^\tp$. 
  We assume that these data are from the same exponential family model as that of the original data, $\vec{Z}$. 
 MC samples of $\vec{Z}^*$ can then be constructed straightforwardly using $\vec{M}$. 
 Note that \pkg{FRK}~v2 provides the user with $\vec{M}$ which, if needed, could be used to predict data with link function $g(\cdot)$ but from a distribution that is different from that of the original data.

For each quantity, we use the posterior expectation as the predictor, which can be estimated by simply taking row-wise averages of the matrices of samples defined above. 
In a Gaussian setting, a commonly used metric for uncertainty quantification is the root-mean-squared prediction error (RMSPE). 
In a non-Gaussian setting, it can be difficult to interpret the RMSPE, and it is often more intuitive to quantify uncertainty through the width of the prediction intervals. Hence, in \pkg{FRK}~v2, we also use the MC sampling approach described above to compute user-specified percentiles of the predictive distribution.

\subsubsection{Arbitrary prediction regions}

 Often, one does not wish to predict over single BAUs but over regions spanning multiple BAUs, $\{\tilde{R}_l$, $l = 1, \dots, N_P\}$, where $N_P$ is the number of prediction regions. 
 These regions may overlap and may not coincide with entire BAUs: Our criterion for determining whether a prediction region contains a particular BAU is the same as that used for the spatial supports originally associated with the observations (see Section~\ref{subsection:DataLayer}). 
 That is, we write the indices of the BAUs associated with $\tilde{R}_l$ as $\tilde{c}_l \equiv \{i : A_i \cap \tilde{R}_l \neq \emptyset\}$, for $l = 1, \dots, N_P$.
 We then define the set of prediction regions in terms of BAUs as $D^P \equiv \{\tilde{B}_l : l = 1, \dots, N_P\}$, where $\tilde{B}_l \equiv \cup_{i\in \tilde{c}_l} A_i$ is the package's representation of $\tilde{R}_l$ in terms of BAUs. %, and where $\tilde{c}_l$ is a non-empty subset of $\{1, \dots, N\}$ that gives the indices of the BAUs associated with the $k$th user-specified region, $\tilde{R}_l$.
%Here, $N_P$ is the number of areas at which spatial prediction takes place, and is equal to $|D^P|$.

%Prediction over $D^P$ requires some form of aggregation across the associated BAUs. 
%Since aggregation must be done on the response scale, we restrict prediction over arbitrary regions to the mean process and the data.  
% Let $\vec{\mu}_P \equiv \{\mu(\tilde{B}_l) : l = 1, \dots, N_P\}$ be the mean process evaluated over the prediction regions. Just as $\vec{\mu}_Z$ was constructed from the BAU-level mean process $\vec{\mu}$ via the matrix $\vec{C}_Z$ given by (\ref{eqn:C_Z}), since each $\tilde{B}_l$ is a BAU or a union of BAUs, one can construct an $N_P \times N$ matrix 
%\[
%\vec{C}_P \equiv \left(\tilde{w}_{ik}\mathbb{I}(i \in \tilde{c}_l) : i = 1, \dots, N; l = 1, \dots, N_P\right),
%\]
%such that
%\[
%\vec{\mu}_P = \vec{C}_P \vec{\mu}.
%\]
%%Specifically,
%%\[
%%    \mu_{P, k} \equiv \mu_P(\tilde{B}_l) = 
%%   \sum_{i = 1}^N \tilde{w}_{ik} \mathbb{I}(A_i \subset \tilde{B}_l)\mu_i; \quad i = 1, \dots, N;\;  l = 1, \dots, N_P;\; \tilde{B}_l \in D^P.
%%\]
%As in Section~\ref{subsection:DataLayer}, the relative weights, $\{v_i: i = 1, \dots, N\}$, for the weights, $\{\tilde{w}_{ik}\}$, are controlled by the \code{wts} field of the BAU object and the argument \mbox{\code{normalise\_wts}}. 
% For consistency between the model fitting and prediction stages, in \pkg{FRK}~v2 we require that the same relative weights, $\{v_i: i = 1, \dots, N\}$, and the same setting of \code{normalise\_wts} are used in construction of both $\vec{C}_Z$ and $\vec{C}_P$.

 Prediction of $\vec{\mu}_P \equiv (\mu_{P, 1}, \dots, \mu_{P, N_P})^\tp$ over $D^P$ requires some form of aggregation across the associated BAUs. 
 In \pkg{FRK}~v2, we aggregate the mean process $\vec{\mu}$ over the associated BAUs. We stress that this is different from aggregation of data (which would lead to a different model for dealing with change-of-support).
 Just as $\vec{\mu}_Z$ was constructed from the BAU-level mean process $\vec{\mu}$ via the matrix $\vec{C}_Z$ given by (\ref{eqn:C_Z}), since each $\tilde{B}_l$ is a BAU or a union of BAUs, one can construct an $N_P \times N$ matrix 
\begin{equation}\label{eqn:C_P}
\vec{C}_P \equiv \left(\tilde{w}_{il}\mathbb{I}(i \in \tilde{c}_l) : i = 1, \dots, N; l = 1, \dots, N_P\right),
\end{equation}
such that
\begin{equation}\label{eqn:mu_P}
\vec{\mu}_P = \vec{C}_P \vec{\mu}.
\end{equation}
%%%%%% C_ZC_pAppendix
%As in Section~\ref{subsection:DataLayer}, the relative weights, $\{\tilde{v}_i: i = 1, \dots, N\}$, such that $\tilde{w}_{ik} \propto \tilde{v}_i$, are controlled by the \code{wts} field of the BAU object, and the argument \mbox{\code{normalise\_wts}} is used to control whether $\vec{C}_P$ represents a weighted sum or a weighted average. 
% For consistency between the model fitting and prediction stages, 
% \pkg{FRK}~v2 enforces the use of the same relative weights, $\tilde{v}_i = v_i$ for $i = 1, \dots, N$, and the same setting of \code{normalise\_wts}, in construction of both $\vec{C}_Z$ and $\vec{C}_P$. 
%%%%%% C_ZC_pAppendix 
 For consistency between the model fitting and prediction stages, \pkg{FRK}~v2 enforces $\vec{C}_P$ to have the same qualitative behaviour as $\vec{C}_Z$ (i.e., if $\vec{C}_Z$ corresponds to a weighted average, then so too does $\vec{C}_P$). See Appendix~\ref{Appendix:Incidence matrices: Cz and Cp} for details.  

 MC samples of $\vec{\mu}_P \mid \vec{Z}, \vec{\theta}$ are constructed via $\vec{M}_P \equiv \vec{C}_P \vec{M}$, where recall that the columns of $\vec{M}$ consist of MC samples from $[\vec{\mu} \mid \vec{Z}, \vec{\theta}]$. 
 Predictions and uncertainty quantification of the predictions can then be computed straightforwardly from $\vec{M}_P$. 
   \pkg{FRK}~v2 also allows inference on data $\{Z^*_{P, 1}, \dots, Z^*_{P, N_P}\}$ over aggregations of BAUs, $\{\tilde{B}_1, \dots, \tilde{B}_{N_P}\}$. 
   We assume that these data are from the same exponential family model as that of the original data, $\vec{Z}$. %: That is, $Z^*_{Pk} \sim \text{EF}(\mu(\tilde{B}_l), \psi)$, $l = 1, \dots, N_P$.  
 MC samples of $\vec{Z}^*_P \equiv (Z^*_{P, 1}, \dots, Z^*_{P, N_P})^\tp$ can then be constructed straightforwardly using $\vec{M}_P$. 
 Note that \pkg{FRK}~v2 provides the user with $\vec{M}_P$ which, if needed, can be used to predict data with link function $g(\cdot)$ but from a distribution that is different from that of the original data.

% For intuition on the preceding discussion, write the matrices $\vec{M}$ and $\vec{C}_P$ as
% \[
% \vec{M} =
% \begin{bmatrix}
% \vec{\mu}_1 & \dots & \vec{\mu}_{n_{\text{MC}}}
% \end{bmatrix},
% \]
% and 
% \[
% \vec{C}_P =
% \begin{bmatrix}
% \vec{c}_{1}^\tp \\ 
% \vdots \\
% \vec{c}_{N_P}^\tp
% \end{bmatrix},
% \]
% respectively, where $\vec{\mu}_j$, for $j = 1, \dots, n_{\text{MC}}$, denotes the $j$th MC sample of $\vec{\mu} \mid \vec{Z}, \vec{\theta}$, and $\vec{c}_{k}^\tp$, for $l = 1, \dots, N_P$, is the vector of weights associated with prediction region $\tilde{B}_l$.
% Then we have that
% \[
% \vec{C}_P\vec{M}
% =
% \begin{bmatrix}
% \vec{c}_{1}^\tp \\ 
% \vdots \\
% \vec{c}_{N_P}^\tp
% \end{bmatrix}
% \begin{bmatrix}
% \vec{\mu}_1 & \dots & \vec{\mu}_{n_{\text{MC}}}
% \end{bmatrix}
% =
% \begin{bmatrix}
% \vec{c}_{1}^\tp\vec{\mu}_1 & \dots & \vec{c}_{1}^\tp\vec{\mu}_{n_{\text{MC}}}\\
% \vec{c}_{2}^\tp\vec{\mu}_1 & \dots & \vec{c}_{2}^\tp\vec{\mu}_{n_{\text{MC}}}\\
% \vdots &  & \vdots\\
% \vec{c}_{N_P}^\tp\vec{\mu}_1 & \dots & \vec{c}_{N_P}^\tp\vec{\mu}_{n_{\text{MC}}}
% \end{bmatrix},
% \]
% where $\vec{c}_{k}^\tp \vec{\mu}_j$ is a scalar quantity, and a MC sample of the $k$th element of $\vec{\mu}_P \mid \vec{Z}, \vec{\theta}$. It is not difficult to see that each column of $\vec{C}_P\vec{M}$ contains a MC of $\vec{\mu}_P \mid \vec{Z}, \vec{\theta}$.

\subsection{Distributions with size parameters}\label{sec:Distributions with size parameters}

 Two distributions considered in this framework, namely the binomial distribution and the negative-binomial distribution, have an assumed-known `size' parameter and a `probability of success' parameter. 
 Given the vector of size parameters associated with the data, \mbox{$\vec{k}_Z \equiv (k_{Z_1}, \dots, k_{Zm})^\tp$}, the parameterisation used in \pkg{FRK}~v2 assumes that $Z_j$ represents either the number of `successes' from $k_{Z_j}$ trials (binomial data model) or that it represents the number of failures before $k_{Z_j}$ successes (negative-binomial data model). 
 Some sources use different parameterisations for the negative-binomial distribution: The parameterisation used in \pkg{FRK}~v2 is the same as that used in the \proglang{R} package \pkg{stats} \citep{Rcoreteam_2021}. 

 Software that cater for these distributions typically allow `link' functions such as the logit, probit, and complementary log-log functions.  
 In \pkg{FRK}~v2, these functions are available to link the latent spatial process, $Y(\cdot)$, to a probability process, $\pi(\cdot)$: 
 \begin{equation*}%\label{eqn:prob_process}
    f(\pi(\vec{s})) = Y(\vec{s}), \quad \vec{s} \in D,
\end{equation*}
where $f(\cdot)$ is one of the aforementioned functions whose inverse has a range of $(0, 1)$. 
 Therefore, the probability process evaluated over the BAUs is $\vec{\pi} \equiv (\pi_i: i = 1, \dots, N)^\tp$, where 
  \begin{equation}\label{eqn:pi_i_equals_f_Y_i}
 \pi_i = f^{-1}(Y_i), \quad i = 1, \dots, N. 
 \end{equation}
 Next, we link the BAU-level mean process to the BAU-level probability process,  
  \begin{equation}\label{eqn:h_mu_equals_pi}
 h(\mu_i; k_i) = \pi_i, \quad i = 1, \dots, N, 
 \end{equation}
 where $h(\cdot\,; \cdot)$ is derived from the expectation of the response distribution (see Appendix~\ref{Appendix:Distributions with size parameters} for details), and 
\mbox{$\vec{k} \equiv (k_1, \dots, k_N)^\tp$} 
 is the vector of BAU-level size parameters. 
 For example, a binomial data model results in $\mu_i / k_i = \pi_i$, for $i = 1, \dots, N$. 
 When modelling negative-binomial data, other popular link functions include the log and square-root functions. 
 In these cases, we use the model \mbox{$\mu_i = k_ig^{-1}(Y_i)$}, $i = 1, \dots, N$, and the elements of $\vec{\pi}$ are obtained using (\ref{eqn:h_mu_equals_pi}). 
%  To facilitate the use of conventional notation when describing these response distributions, it is convenient to 
 Irrespective of the chosen link function, we 
  define the probability process over the observation supports in $D^O$ as \mbox{$\vec{\pi}_Z \equiv (\pi_{Z_j} : j = 1, \dots, m)^\tp$}, where
\begin{equation}
	\pi_{Z_j} = h(\mu_{Z_j}; k_{Z_j}), \quad j = 1, \dots, m.
\end{equation}

 When model fitting, the BAU-level size parameters $\vec{k}$ are needed to compute the BAU-level mean process in (\ref{eqn:h_mu_equals_pi}). 
 %%%% Too verbose; refer the readers to the package manual:
% The simplest case is when each observation is associated with a single BAU only \textit{and} each BAU is associated with at most one observation support; then, it is straightforward to assign elements from $\vec{k}_Z$ to elements of $\vec{k}$ and vice-versa, and so the user may provide either $\vec{k}$ or $\vec{k}_Z$. 
% If each observation is associated with exactly one BAU, but some BAUs are associated with multiple observations, the user must provide $\vec{k}_Z$, which is used to infer $\vec{k}$; in particular, $k_i = \sum_{j \in a_i} k_{Z_j}$, $i = 1, \dots, N$, where $a_i$ denotes the indices of the observations associated with BAU $A_i \in D^G$. 
% If one or more observations encompass multiple BAUs, $\vec{k}$ must be provided with the BAUs, as we cannot meaningfully distribute $k_{Z_j}$ over $\{k_i : i \in c_j\}$ when $|c_j| > 1$.  
% In this case, we infer $\vec{k}_Z$ using $k_{Z_j} = \sum_{i \in c_j} k_i$, $j = 1, \dots, m$. 
 %%%%
 The user must supply these size parameters either through the data or though the BAUs. How this is done depends on whether the data are areal or point-referenced, and whether they overlap common BAUs or not; details are provided in the package manual.

  %% Element-wise version: 
% Similarly, by defining the set of size parameters associated with the prediction regions, $\tilde{B}_l$, $l = 1, \dots, N_P$, 
% as \mbox{$\{k_{P,l} = \sum_{i \in \tilde{c}_l} k_i : l = 1, \dots, N_P\}$}, we define the probability process over the prediction regions as 
%$\vec{\pi}_P \equiv (\pi_{P,l} : l = 1, \dots, N_P)^\tp$, where
%\begin{equation}
%	\pi_{P_l} = h(\mu_{P,l}; k_{P,l}), \quad l = 1, \dots, N_P.
%\end{equation}
 %% Vector version (avoids problems associated with subscripting)
%  Now define the vector of size parameters associated with the prediction regions, \mbox{\red{$D^P \equiv \{\tilde{B}_l : l = 1, \dots, N_P\}$}}, as
  Now, define the prediction-region size parameters as
  \mbox{$\vec{k}_P \equiv (\sum_{i \in \tilde{c}_l} k_i : l = 1, \dots, N_P)^\tp$}, where recall that $\tilde{c}_l$ denotes the indices of the BAUs associated with the prediction region $\tilde{B}_l \in D^P$. 
  Then the probability process evaluated over $D^P$ is 
  \begin{equation}
  \vec{\pi}_P \equiv h(\vec{\mu}_P; \vec{k}_P),
  \end{equation} 
  where $h(\cdot\,; \cdot)$ is applied element-wise. 
 When predicting, BAU-level size parameters are needed to compute the predictive distribution of $\vec{\mu}$, $\vec{\pi}_P$, and $\vec{\mu}_P$. 
 If these size parameters are not available at unobserved BAUs, the user can still obtain predictions of $\vec{\pi}$ if the user chooses to use a logit, probit, or complementary log-log link function, since then $\vec{\pi}$ is linked directly to $\vec{Y}$ in (\ref{eqn:pi_i_equals_f_Y_i}).

 In most applications that consider binomial or negative-binomial data models, 
 the conditional mean of an observation is treated as a simple aggregate of the underlying mean process. 
 Therefore, with these distributions, \pkg{FRK}~v2 enforces the matrices $\vec{C}_Z$ and $\vec{C}_P$ in (\ref{eqn:C_Z}) and (\ref{eqn:C_P}), respectively, to correspond to a simple, unweighted summation over the BAUs. See Appendix~\ref{Appendix:Incidence matrices: Cz and Cp} for details.

\subsection{Spatio-temporal framework}\label{sec:spatio-temporal}

% As with \pkg{FRK}~v1, 
 \pkg{FRK}~v2 accommodates spatio-temporal data by using spatio-temporal basis functions constructed via a tensor product of spatial and temporal basis functions. 
 Since one often requires several thousand basis functions in a spatio-temporal setting, we focus here on the case where $\vec{\eta}$ is modelled using the sparse precision matrix $\vec{Q}$. 
 
 Let $r_t$ and $r_s$ denote the number of temporal and spatial basis functions, respectively. 
 Denote $\vec{Q}_t$ and $\vec{Q}_s$ as the precision matrices of the random coefficients associated with the temporal basis functions and the spatial basis functions, respectively. 
 We model the $r_tr_s \times r_tr_s$ precision matrix of the random coefficients associated with the $r_tr_s$ spatio-temporal basis functions as $\vec{Q} = \vec{Q}_t \otimes \vec{Q}_s$, where $\otimes$ is the Kronecker product. 
 This form for $\vec{Q}$ leads to significant computational savings. 
 For the random coefficients associated with the temporal basis functions, \pkg{FRK}~v2 uses a first-order autoregressive model.

 Recall from the spatial-only case that \pkg{FRK}~v2 assumes \mbox{$\vec{\xi} \sim \Gau(\vec{0}, \sigma^2_\xi \vec{V})$}, where $\vec{V}$ is a known, positive-definite diagonal matrix that can be set to $\vec{I}$ in the absence of problem specific fine-scale information. In a spatio-temporal setting, it is possible that each spatial BAU is observed multiple times. In these situations, \pkg{FRK}~v2 also allows each spatial BAU to be associated with its own fine-scale variance parameter. Specifically, let $N_s$ and $N_t$ denote the number of spatial and temporal BAUs, respectively, so that $N = N_sN_t$. Then, \pkg{FRK}~v2 also allows the model $\vec{\xi} \sim \Gau(\vec{0}, \vec{\Sigma}_\xi)$ where, assuming that the BAUs are ordered such that space runs faster than time,    
\begin{equation}\label{eqn:fine-scale_cov_mat_fs_by_spatial_BAU}
\vec{\Sigma}_\xi \equiv \text{diag}(\text{vec}(\underbrace{\vec{\sigma}^2_\xi, \dots, \vec{\sigma}^2_\xi}_{N_t \; \text{times}})) \odot \vec{V},
\end{equation}
$\vec{\sigma}^2_\xi \equiv (\sigma^2_{\xi, 1}, \dots, \sigma^2_{\xi, N_s})^\tp$, 
$\text{vec}(\cdot)$ `stacks' its arguments into a single vector, $\text{diag}(\cdot)$ returns a diagonal matrix from a vector argument, and $\odot$ denotes element-wise multiplication. This model may be advantageous when the variance of the fine-scale component is spatially varying, when the number of spatial BAUs (and hence the number of fine-scale variance parameters to estimate) is relatively low, and when we have observations from each spatial BAU at many time-points; see, for instance, the example presented in Section~\ref{sec:ST_example}.

\section{New features and their usage}\label{SEC:IllustrativeExample}

 We now focus on the new features in \pkg{FRK}~v2, an overview of which is presented in Table~\ref{tab:new_arguments_in_functions}. 
% The primary new feature in \pkg{FRK}~v2 is the package's ability to cater for non-Gaussian data models: 
% A full list of available data models and link functions is shown in Table~\ref{table:response_and_links}.  
%  The available exponential-family-members include the Gaussian, Poisson, gamma, inverse-Gaussian, negative-binomial, and binomial distributions, and these distributions can be used in combination with the identity, inverse, log, square-root, logit, probit, and complementary-log-log link functions. 
% In Sections \ref{sec:03-01:Poisson} and \ref{sec:03-03:negative-binomial}, we illustrate the use of \pkg{FRK}~v2 with non-Gaussian spatial point-referenced and area-referenced data, respectively. 
 In Section \ref{sec:03-01:Poisson}, we illustrate the use of \pkg{FRK}~v2 with non-Gaussian spatial data. 
 In Section \ref{sec:3.2:modelselection}, we demonstrate how the important tasks of model selection and model validation can be performed with \pkg{FRK}~v2. 
 In Section~\ref{sec:3:increased_resolution}, we show the potential improvement in predictive performance of \pkg{FRK}~v2 over \pkg{FRK}~v1 when the data are Gaussian, owing to an increase in the maximum number of basis functions allowed in \pkg{FRK}~v2.  
 All results presented in the remainder of this paper can be generated using the reproducible code at \mbox{\url{https://github.com/msainsburydale/FRKv2_src}}. 

\begin{table}
    \centering
    \setlength{\tabcolsep}{4pt}
    \begin{tabular}{ccp{9cm}}
    \hline
    Function & Argument & Use\\
    \hline
     \fct{SRE}/\fct{FRK}   & \code{response}  & String indicating the response distribution. \\
        & \code{link}      & String indicating the link function.\\
        & \code{K\_type}   & String indicating the parameterisation of $\cov{\vec{\eta}}{\vec{\eta}}$; the newly permissible value, \code{"precision"}, indicates that a sparse precision matrix should be used. 
        \\
        & \code{normalise\_wts} & Flag controlling whether the weights in $\vec{C}_Z$ and $\vec{C}_P$ should be normalised or not.\\
        & \code{fs\_by\_spatial\_BAU} & Flag controlling whether each spatial BAU is given its own fine-scale variance parameter; only applicable in a spatio-temporal setting.\\
        & & \\
       \fct{SRE.fit}/\fct{FRK} & \code{method} & String indicating the method of model fitting: \code{"TMB"} is required whenever a non-Gaussian data model or non-identity link function is used.\\
%        & \code{optimiser} & Optimising function if \code{method = "TMB"}; default \fct{nlminb}. \\
%        & \code{taper}     & A positive scalar indicating the strength of the covariance tapering (see Appendix~\ref{Appendix:CovarianceTapering}).\\
        & \code{known\_sigma2fs} & Positive number at which to fix the fine-scale variance. 
%        If \code{fs\_by\_spatial\_BAU = TRUE}, the argument \code{known\_sigma2fs} should be a vector of length equal to the number of spatial BAUs.
\\
& & \\
        \fct{predict} 
%        & \code{newdata} & The prediction regions; in addition to \code{Spatial*DataFrame}and \code{STFDF}, \code{newdata} can now be a \code{SpatialPoints} or \code{STI} object, facilitating prediction over spatial or spatio-temporal point-referenced locations.\\
        & 
%        \code{type} & A vector of strings indicating the quantities of interest for which inference is made. 
%        The inclusion of \code{"link"}, \code{"mean"}, and \code{"response"} respectively indicates that inference on the latent process ($\vec{Y}$), the mean process ($\vec{\mu}$ or $\vec{\mu}_P$) and, if applicable, the probability process ($\vec{\pi}$), or the data ($\vec{Z}^*$ or $\vec{Z}^*_P$), is made. 
        \code{type} & Vector of strings indicating the quantities of interest for which inference is made. 
        The inclusion of \code{"link"} indicates that inference on the latent process ($\vec{Y}$) is made; the inclusion of \code{"mean"} indicates that inference on the mean process ($\vec{\mu}$ or $\vec{\mu}_P$) and, if applicable, the probability process ($\vec{\pi}$ or $\vec{\pi}_P$) is made; and the inclusion of \code{"response"} indicates that inference on the data ($\vec{Z}^*$ or $\vec{Z}^*_P$) is made. 
        \\ 
        & \code{percentiles} & Numeric vector %(each element between 0 and 100) 
        indicating the percentiles of the predictive distribution(s) to be returned.\\
%        & \code{kriging} & A character string indicating whether \code{"simple"} or \code{"universal"} kriging is to be performed. \\
%        & \code{k} & Vector of size parameters at each prediction location (applicable only for binomial and negative-binomial data).\\
        & \code{nsim} & The number of MC samples at each BAU.\\
        & & \\
        \fct{auto\_BAUs} & \code{spatial\_BAUs} & 
        The spatial BAUs in a spatio-temporal setting (constructed automatically from the data by default).\\ %, where spatio-temporal BAUs are constructed by taking the Kronecker product of the temporal BAUs (box functions) and spatial BAUs. 
%        If \code{NULL}, the spatial BAUs are constructed automatically from the data.\\
%        & \code{buffer} & Numeric (between 0 and 0.5) indicating the size of the buffer of basis functions along the boundary. 
%        The buffer is added by computing the number of basis functions in each dimension, and increasing this number by a factor of \code{buffer}.
%        As precision matrices are more sensitive to boundary effects, a buffer can be useful when \code{K\_type = "precision"}.\\
& & \\
\fct{plot} & - & Visualise the data, predictions, and uncertainty quantification of the predictions given an \class{SRE} object and the object resulting from a call to \fct{predict}.\\
        \hline
    \end{tabular}
     \caption{Important new or augmented function arguments in \pkg{FRK}~v2. The package now also provides methods for \fct{AIC}, \fct{BIC}, \fct{fitted}, \fct{residuals}, and \fct{simulate}.}
 \label{tab:new_arguments_in_functions} 
\end{table}

\subsection{Non-Gaussian data}\label{sec:03-01:Poisson}
  
   The primary new feature in \pkg{FRK}~v2 is the package's ability to cater for non-Gaussian data models. 
   The available exponential-family-members include the Gaussian, Poisson, gamma, inverse-Gaussian, negative-binomial, and binomial distributions, and these distributions can be used in combination with the identity, inverse, log, square-root, logit, probit, and complementary-log-log link functions. 
 
\begin{figure}[t!]
    \centering
    \includegraphics[width = 0.8\linewidth]{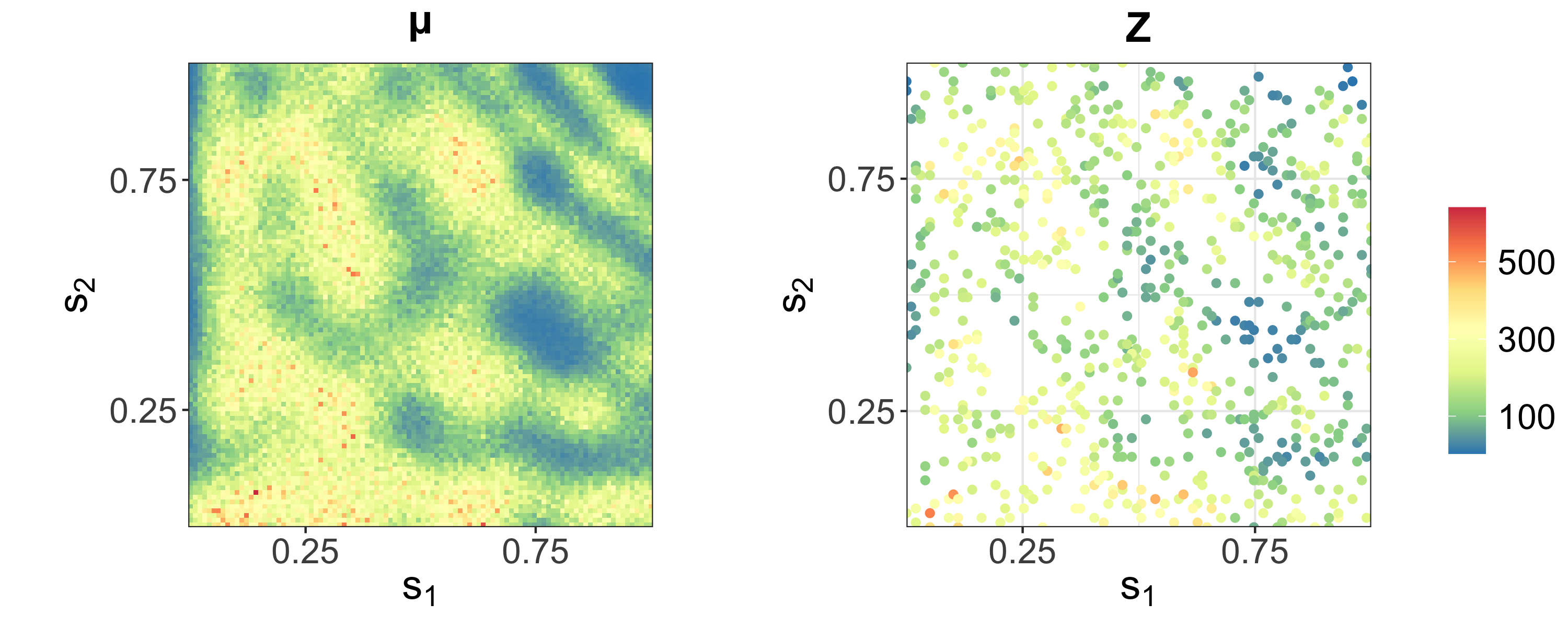}
    \vspace{-10pt}    
    \caption{Simulated, point-referenced, Poisson data set used in the illustrative example of Section~\ref{sec:03-01:Poisson}. (Left) True mean process evaluated over the BAUs. (Right) Simulated data.}
  \label{fig:Poisson_true_and_Z}
\end{figure}

 For illustration, and so that readers can familiarise themselves with the workflow of \pkg{FRK}~v2, we now analyse a simulated Poisson data set containing 750 observations at spatial locations shown in Figure~\ref{fig:Poisson_true_and_Z}. 
 The true mean process evaluated over the BAUs, $\vec{\mu}$, is also shown in Figure~\ref{fig:Poisson_true_and_Z}:  
 It was constructed by passing a sum of trigonometric functions through the exponential function. 
% It was constructed by first simulating $\vec{Y}$ from a Gaussian process with an exponential covariance function (with variance parameter $\sigma^2 = 0.1$ and length-scale parameter $\tau = 0.2$), and then passing $\vec{Y}$ through the exponential function, $\exp(\cdot)$.
 In what follows, $\text{Poi}(\mu)$ refers to a probability distribution on the non-negative integers with probability mass function (PMF) $\{e^{-\mu} \mu^z/z!: z = 0, 1, \dots\}$ for $\mu > 0$. 

The first step when using \FRKgeneric is to create basis functions and BAUs, which can be done automatically using the helper functions, \fct{auto\_BAUs} and \fct{auto\_basis}; see Appendix~\ref{Appendix:Basis_fns_and_BAUs} for details. 
Next, an \class{SRE} object is constructed using \fct{SRE}, within which we specify the data model, the link function, and the parameterisation of $\cov{\vec{\eta}}{\vec{\eta}}$.  
 In this example, we model counts using a Poisson data model, $Z_j \mid \vec{\mu} \sim \text{Poi}(\mu_{Z_j})$, for $j = 1, \dots, m$, and we use the log link function, $g(\cdot) = \log(\cdot)$. 
 These choices are made by setting \code{response = "poisson"} and \code{link = "log"}. 
 We then fit the model using \fct{SRE.fit}.
 These steps may also be performed with the convenient wrapper function \fct{FRK}. 
 Note that when the data are non-Gaussian or when a non-identity link function is chosen, \fct{FRK} automatically enforces \code{method = "TMB"} and selects \code{K\_type = "precision"}, which means that \pkg{TMB} is used for model fitting and that the basis-function coefficients are modelled via a sparse precision matrix, $\vec{Q}$: 
\begin{Code}
R> S <- FRK(f = Z ~ 1, data = Poisson_data, response = "poisson", link = "log")
\end{Code}
Prediction is done using \fct{predict}. 
The argument \code{type} specifies the quantities of interest for which predictions and uncertainty quantification of the predictions are desired. 
In this example, we set \code{type = c("link", "mean")} to obtain predictions for the latent spatial process, $\vec{Y}$, and the mean process, $\vec{\mu}$. 
The \code{percentiles} argument allows the computation of percentiles of the predictive distributions and hence computation of prediction intervals; if left unspecified, the 5th and 95th percentiles are returned: 
\begin{Code}
R> pred <- predict(S, type = c("link", "mean"))
\end{Code}
When \code{method = "TMB"}, the returned object from \fct{predict} is a \class{list} containing two elements. 
The first element is an object of the same class as \code{newdata} (if \code{newdata} is unspecified, prediction is done over the BAUs) and contains the predictions and uncertainty quantification of the predictions for each term in \code{type}. 
The second element is a \class{list} of matrices containing MC samples for each term in \code{type} at each prediction location. 

\begin{figure}[t!]
    \centering
    \includegraphics[width = \linewidth]{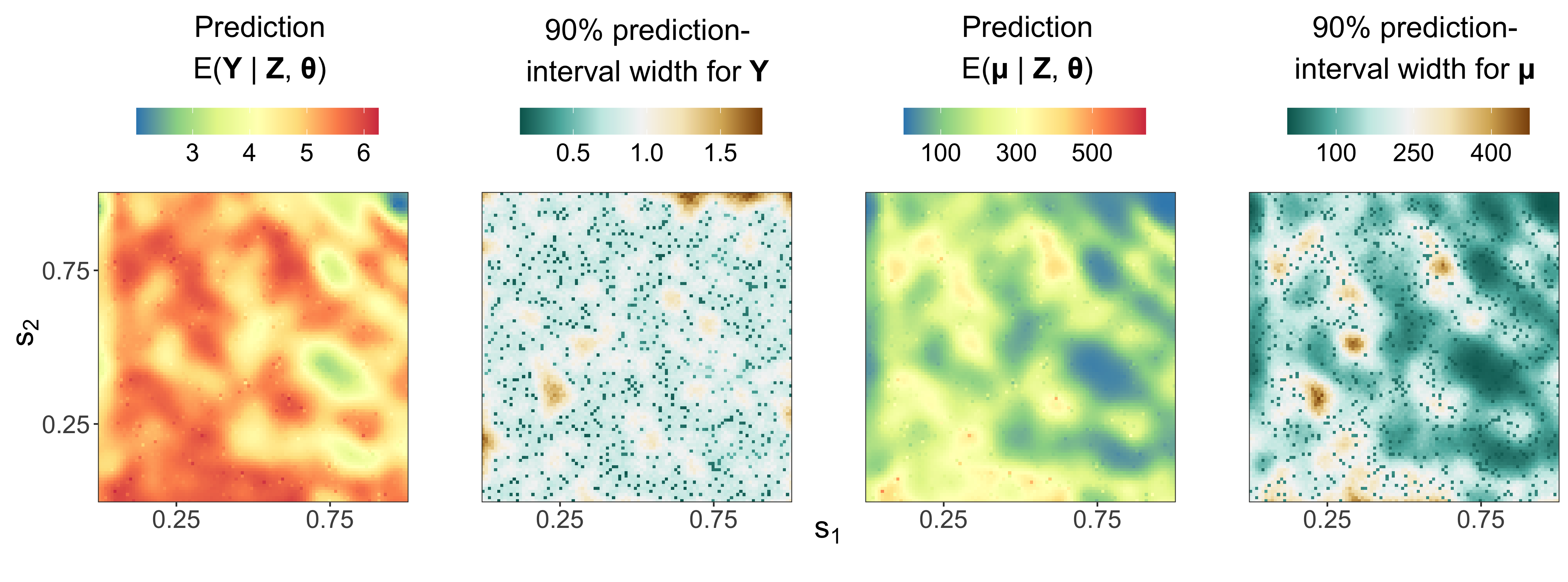}
    \vspace{-20pt}  
    \caption{Predictions and prediction-interval widths returned by \pkg{FRK}~v2 when applied to the simulated Poisson data shown in the right panel of Figure~\ref{fig:Poisson_true_and_Z}. (Left) Prediction of $\vec{Y}$, the latent process over the BAUs. (Centre left) Width of the 90\% prediction interval for each element of $\vec{Y}$. (Centre right) Prediction of $\vec{\mu} = g^{-1}(\vec{Y})$, the mean process over the BAUs. (Right) Width of the 90\% prediction interval for each element of $\vec{\mu}$.}
  \label{fig:Poisson_nres3}
\end{figure}

Finally, a \class{list} of \class{ggplot} \citep{Wickham_2016_ggplot2} objects of the predictions and their associated uncertainty can be generated using \fct{plot}: 
\begin{Code}
R> plots <- plot(S, pred)
\end{Code}
The \class{ggplot} objects can be arranged easily on a grid using various dedicated packages. 
 Figure~\ref{fig:Poisson_nres3} shows predictions and prediction-interval widths for the latent process evaluated over the BAUs, $\vec{Y}$, and the mean process evaluated over the BAUs, $\vec{\mu}$. 
The predictions of $\vec{\mu}$ are reasonable given the data and the true values shown in Figure~\ref{fig:Poisson_true_and_Z}. 
The prediction-interval widths for $\vec{Y}$ overall, do not vary much, but they are larger in regions of data paucity and along the boundary of the spatial domain; on the other hand, the prediction-interval width for $\vec{\mu}$ is large when $\vec{\mu}$ is large, as can be expected when the response is Poisson distributed. 
The  `bullseye' points of low uncertainty, visible for both processes, correspond to the data locations. 
The point-like nature of this reduction in uncertainty arises from the fine-scale random process, $\xi(\cdot)$, being modelled as mutually independent at the BAU level: The fine-scale random effects at unobserved BAUs do not ``borrow strength'' from the inferred fine-scale random effect at neighbouring observed BAUs.

\subsection{Model validation and selection}\label{sec:3.2:modelselection}

% Model validation (i.e., judging the fit of a single model) and model selection (i.e., selecting a preferred model from a set of plausible models) are important stages of any statistical analysis. 
% We now describe how these tasks can be performed with \pkg{FRK}~v2.  
 
\begin{figure}[t!]
    \centering
    \includegraphics[width = \linewidth]{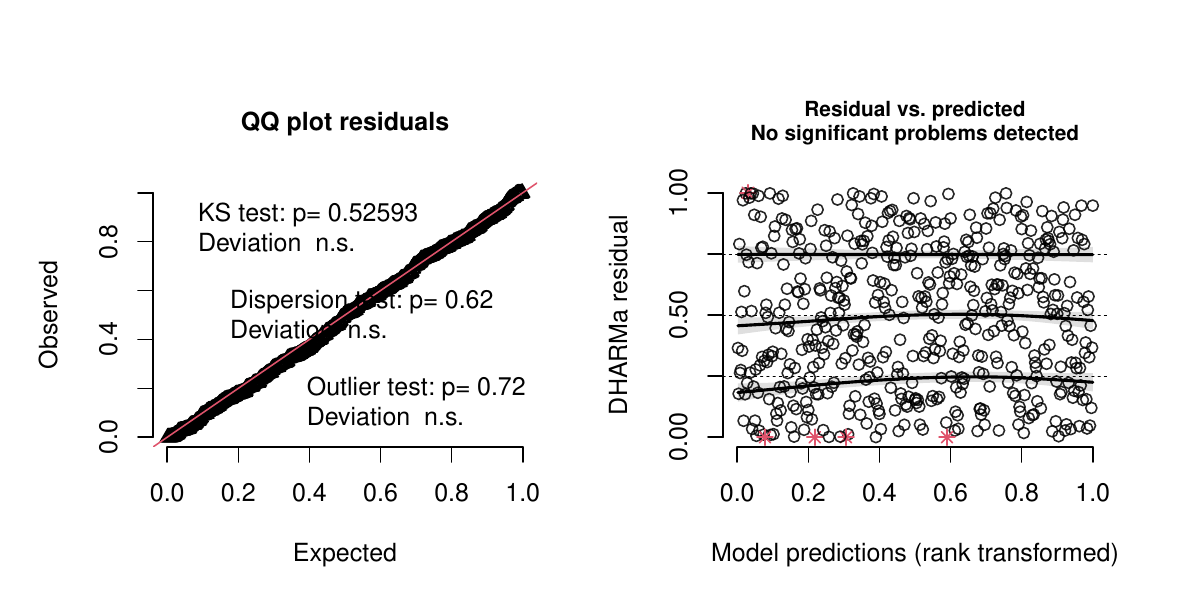}
    \vspace{-20pt}  
    \caption{QQ plot (left) and the residuals against the predicted values (right) for the Poisson example of Section~\ref{sec:03-01:Poisson}.} 
  \label{fig:Poisson_residuals}
\end{figure}

 When working with (spatial) GLMMs, standard residuals are difficult to interpret because their expected properties (e.g., the expected dispersion) change with the fitted value. %Pearson residuals account for this change in expected dispersion, but still do not possess a canonical appearance when plotted, which can spuriously suggest problems even for correctly specified models. 
 Due to these challenges, it is often easier to validate the model via simulation from the fitted model \citep[Ch.~24]{Box_1980_simulation-based_model_checking, Rubin_1984_model_selection, Gelman_1996_posterior_predictive_assessment, Gelman_Hill_2007}. 
 %Specifically, this approach involves simulating new data sets from the fitted model and, using graphical summaries or formal hypothesis tests, checking that these simulations are consistent with observed data.  
 This simulation-based approach is facilitated in \pkg{FRK}~v2 with the function \fct{simulate}. 
 In particular, simulations generated with \fct{simulate} may be used with the \proglang{R} package \pkg{DHARMa} \citep{DHARMa} which, given observed data and simulations from a fitted model, computes interpretable, simulation-based quantile residuals \citep{Cox_Snell_1968_residuals, Dunn_Smyth_1996_randomised_qunatile_residuals}. Under the true model, these residuals always follow a standard uniform distribution, which greatly facilitates their interpretation. 
 Using the Poisson example of Section~\ref{sec:03-01:Poisson}, % with data reserved for model testing, 
 one may create a \class{DHARMa} object as follows: 
%\begin{Code}
%R> DHARMa_object <- createDHARMa(
%+    simulatedResponse = simulate(S), 
%+    observedResponse  = Poisson_data$Z, 
%+    integerResponse = TRUE)
%\end{Code}
%%$
\begin{Code}
R> DHARMa_object <- createDHARMa(
+    simulatedResponse = simulate(S), 
+    observedResponse  = Poisson_data$Z, 
+    integerResponse = TRUE)
\end{Code}
%$
The \class{DHARMa} object can then be used to construct residual plots and to perform a range of hypothesis tests, including tests for residual spatial dependence and zero-inflation, as well as arbitrary, user-defined tests. 
 Figure~\ref{fig:Poisson_residuals} shows the result of calling \fct{plot} on the above \class{DHARMa} object, which generates a QQ plot and a plot of the residuals against the predicted values. Although simulation-based residuals are a natural choice for the spatial GLMM employed by \pkg{FRK}~v2, one may also compute the response or Pearson residuals using \fct{residuals}. 
 %\pkg{FRK}~v2 also provides a function, \fct{diagnostics}, which, given a sample from \fct{posteriorpredictive} and the corresponding observed data, computes several default (e.g., coverage) and user-defined diagnostics.  

 The tools used for model validation may also be employed in model selection. However, model selection also often involves an information criterion, which can typically be decomposed into a term that assesses goodness-of-fit and a term that penalises model complexity.
 Standard likelihood-based criteria are the Akaike information criterion (AIC) and the Bayesian information criterion (BIC). % citations if needed: Akaike 1974 and Schwarz 1978
 The presence of dependent random effects (e.g., basis-function coefficients) complicates the interpretation of these criteria, since the effective number of parameters is more than the number of fixed effects and covariance parameters, and less than this number plus the number of dependent random effects. % (e.g., Hodges and Sargent, 2001; Overholser and Xu, 2014). % Other criteria, such as the conditional AIC, the DIC, and the WAIC, are more suitable for such problems.  
However, they are still often used in such settings and, hence, \pkg{FRK}~v2 also provides methods for the functions \fct{AIC} and \fct{BIC}.

 To demonstrate these approaches to model validation and selection, and to demonstrate that the model employed by \pkg{FRK} has appealing robustness properties with respect to the number of basis functions, we now repeat the analysis of Section~\ref{sec:03-01:Poisson} with models that employ one, two, three, and four resolutions of basis functions, respectively. 
 Table~\ref{tab:03-02:PoissonScoringRules} shows the results for each model. 
   Clearly, predictive performance improves as the number of resolutions, and hence the number of basis functions, increases. However, the coverage remains accurate in all runs, implying that the model is able to accurately quantify uncertainty irrespective of the number of basis functions employed. 
   This important property is due largely to the presence of the fine-scale random variation term, $\xi(\cdot)$, in (\ref{eqn:04-01:Y(s)}). 
  Further, \Copy{PoissonVaryingBasisFunctions}{the relative smoothness in the data (recall the right panel of Figure~\ref{fig:Poisson_true_and_Z}) suggests that only a moderate number of basis functions are needed for this application. 
  Indeed, Table~\ref{tab:03-02:PoissonScoringRules} shows that adding a fourth resolution of basis functions does not substantially improve the predictive performance, which is reflected by an increase in the AIC. %, despite the fact that this criterion does not account for the number of dependent random effects: 
  Importantly, the use of more basis functions than is strictly necessary does not reduce predictive performance. 
  }
  
\begin{table}[t!]
    \centering
    \begin{tabular}{cccccc}
    \hline
    Resolutions (basis functions) & RMSPE  & CRPS & Cvg90  & AIC \\ %& Run time (min.) \\
    \hline
    1 (9)    & 84.71    & 47.58  & 0.901   & 8894.9  \\  % &  0.07 \\
    2 (90)   & 58.60    & 30.97  & 0.878   & 8340.3  \\  % &  0.09 \\
    3 (819)  & 50.12    & 25.77  & 0.885   & 8117.3  \\  % &  0.24 \\
    4 (7380) & 50.10    & 25.78  & 0.886   & 8121.4  \\  % &  3.31 \\
    \hline
    \end{tabular}
        \caption{
    Diagnostics comparing the predictive performance when using a range of basis-function resolutions with point-referenced count data. The diagnostics are the root-mean-squared prediction error (RMPSE), the continuous ranked probability score (CRPS), and the empirical coverage (Cvg90) resulting from a prediction interval with a nominal coverage of 90\% (see Appendix~\ref{app:ScoringRules} for further details). 
    The diagnostics are with regard to prediction of $\vec{\mu}$ and they are averaged over all unobserved BAUs. The Akaike information criterion (AIC) for each model is also shown, and this is computed with respect to the observed data. 
    }
    \label{tab:03-02:PoissonScoringRules}
\end{table}

\subsection[Increased numbers of basis functions]{Increased numbers of basis functions}\label{sec:3:increased_resolution}

 The efficiency of \pkg{TMB} and our use of precision matrix $\vec{Q}$ instead of covariance matrix $\vec{K}$ means that \pkg{FRK}~v2 is well equipped to use a large number of basis functions. 
 This is important, as the predictive performance of fixed rank kriging is often determined by the number of basis functions, as show in the following experiment.
 
 We re-ran the analysis for the comparative study of spatial-prediction methods  published in \citet{Heaton_2019_comparative_study}, which included spatial predictions from \pkg{FRK}~v1.
%The data in this study consists of daytime land surface temperatures as measured by the Terra instrument onboard the MODIS satellite \citep{MODIS_satelitte}, with observations on a $500 \times 300$ grid. 
%The training set consists of 105,569 observations, while the test set consists of 42,740 observations.
 The data used in that study comprised a training data set and a test data set consisting of 105,569 observations and 42,740 observations, respectively. 
%  Table~\ref{tab:Heaton_comparison:full} replicates Table~3 of \citet{Heaton_2019_comparative_study},  
%with an additional entry corresponding to \pkg{FRK}~v2, wherein many more basis functions were used than was practical with \pkg{FRK}~v1. 
% Specifically, \pkg{FRK}~v1 used 485 basis functions, whilst \pkg{FRK}~v2 used 12,114.
% The results show that the increased number of basis functions from \pkg{FRK}~v2 significantly improved the diagnostic scores over \pkg{FRK}~v1. 
 Table~\ref{tab:Heaton_comparison:full} replicates Table~3 of \citet{Heaton_2019_comparative_study},  
with an additional entry corresponding to \pkg{FRK}~v2, wherein many more basis functions (12,144) were used than was practical with \pkg{FRK}~v1 (485). 
%Specifically, \pkg{FRK}~v1 used 485 basis functions, whilst \pkg{FRK}~v2 used 12,114.
The results show that the increased number of basis functions significantly improved the diagnostic scores. 
 To achieve these improvements, we only had to specify \code{nres = 4} rather than \mbox{\code{nres = 3}}; the rest of the code that was used in the comparative study was left unchanged.

\begin{table}[t!]
    \begin{center}
    \setlength{\tabcolsep}{5pt}
    \begin{tabular}{lcccccrr}
    \hline
    Method       & MAE   & RMSPE & CRPS  & IS95  & Cvg95 & Run time (min.) & Cores \\[0pt]
    \hline
    \pkg{FRK}~v1 & 1.96  & 2.44  & 1.44  & 14.08 & 0.79  & 2.32            & 1\\[0pt]
    \pkg{FRK}~v2 & 1.34  & 1.74  & 0.95  & 8.47  & 0.92  & 22.41           & 8    \\[0pt]
    Gapfill & 1.33   & 1.86 & 1.17  & 34.78 & 0.36 & 1.39 & 40\\[0pt]
    Lattice Krig & 1.22   & 1.68 & 0.87  & 7.55 & 0.96 & 27.92 & 1\\[0pt]
    LAGP & 1.65   & 2.08 & 1.17  & 10.81 & 0.83 & 2.27 & 40\\[0pt]
    Metakriging  & 2.08 & 2.50  & 1.44 & 10.77 & 0.89 & 2888.52 & 30\\[0pt]
     MRA & 1.33 & 1.85 & 0.94 & 8.00 & 0.92 & 15.61 & 1\\[0pt]
    NNGP Conjugate & 1.21 & 1.64 & 0.85 & 7.57 & 0.95 & 2.06 & 10\\[0pt]
    NNGP Response & 1.24 & 1.68 & 0.87 & 7.50 & 0.94 & 42.85 & 10\\[0pt]
    Partition & 1.41 & 1.80 & 1.02 & 10.49 & 0.86 & 79.98 & 55 \\[0pt]
    Pred. Proc. & 2.05 & 2.52 & 1.85 & 26.24 & 0.75 & 640.48 & 1\\[0pt]
    SPDE & 1.10 & 1.53 & 0.83 & 8.85 & 0.97 & 120.33 & 2\\[0pt]
    Tapering & 1.87 & 2.45 & 1.32 & 10.31 & 0.93 & 133.26 & 1\\[0pt]
    Periodic Embedding & 1.29 & 1.79 & 0.91 & 7.44 & 0.93 & 9.81 & 1\\\hline
    \end{tabular}
    \end{center}
\vspace{-1em}
    \caption{Scores for each method in the comparative study presented in \cite{Heaton_2019_comparative_study}. The scores are the mean absolute error (MAE), the root-mean-squared prediction error (RMSPE), and the continuous ranked probability score (CRPS), and the empirical coverage (Cvg95) and the interval score (IS95) resulting from a prediction interval with a nominal coverage of 95\% (see Appendix~\ref{app:ScoringRules} for further details). Note that \pkg{FRK}~v2 was implemented in a different computing environment than the other models, and so its run time is not directly comparable. \pkg{FRK}~v2 was implemented using a machine with 16 GB of RAM and an Intel i7-9700 3.00GHz CPU with 8 cores. The other models were implemented using the powerful Becker computing environment with 256 GB of RAM and 2 Intel Xeon E5-2680 v4 2.40GHz CPUs with 14 cores each and 2 threads per core \citep{Heaton_2019_comparative_study}.}
    \label{tab:Heaton_comparison:full}
\end{table}

\section{Application and comparison studies}\label{SEC:ApplicationStudy}

We now provide several application and comparison studies using \pkg{FRK}~v2.
In Section~\ref{sec:04-01:MODIS}, we present a comparison study between \pkg{FRK}~v2 and other packages that cater for non-Gaussian data models. 
In Section~\ref{sec:block_prediction}, we demonstrate block prediction using contaminated soil data.  
In Section~\ref{sec:spatialCOS}, we use data on poverty figures in Sydney, Australia, to demonstrate the spatial change-of-support functionality of \pkg{FRK}~v2 in a non-Gaussian setting; this is a useful example for readers wishing to `try out' \pkg{FRK}~v2 on real-world data.
In Section~\ref{sec:ST_example}, we provide a non-Gaussian spatio-temporal example through modelling crime counts in Chicago during the first two decades of the 21st century.  
 \Copy{sphere}{For the use of \pkg{FRK} on the sphere, see \citet[][Sec.~4.2]{FRK_paper} and the examples given in the package vignette.}

\subsection{Comparative study: MODIS cloud data}\label{sec:04-01:MODIS}

In this section, we compare out-of-sample predictions from \pkg{FRK}~version 2.0.1 to those from the \proglang{R} packages \pkg{INLA} version 20.03.17 \citep{Lindgren_2015_R-INLA}, \pkg{spNNGP} version 0.1.4 \citep{Finley_2020_spNNGP}, \pkg{spBayes} version 0.4.3 \citep{Finley_2015_spBayes}, and \pkg{mgcv} version 1.8.33 \citep{Wood_2017_GAM:R} from a spatial binary data set. 
The data form an image of a cloud taken by the Moderate Resolution Imaging Spectroradiometer (MODIS) instrument aboard the Aqua satellite \citep{MODIS_satelitte} on 05 December 2018 01:00 UTC, over the South Pacific, just east of New Zealand. 
%Data collected from the MODIS instrument have been used in several related works; see, for instance, \cite{Sengupta_2016_MODIS} and \cite{ZammitMangion_2021_Deep_compositional_spatial_model}.
For this comparative study, we pre-processed the data as follows. First, we coarsened the image from over 10 million pixels to a more manageable 33,750 pixels, by creating a 150 $\times$ 225 grid and computing the average value of the response within each grid cell. 
 \Copy{MODISthreshold}{Second, since the data provided by the MODIS instrument is continuous (namely spectral radiances in units of $\text{W/m}^2/\si{\um}/\text{st}$), we applied a threshold to obtain a binary version of the data. Specifically, pixels with spectral radiance greater than 7,000 $\text{W/m}^2/\si{\um}/\text{st}$ were labelled `Cloud', and the remaining pixels were labelled `No-cloud'. This threshold was chosen to illustrate our methodology such that the resulting data have roughly equal proportions of `Cloud' and `No-cloud'.}  

\begin{figure}[t!]
    \centering
    \includegraphics[width = \linewidth]{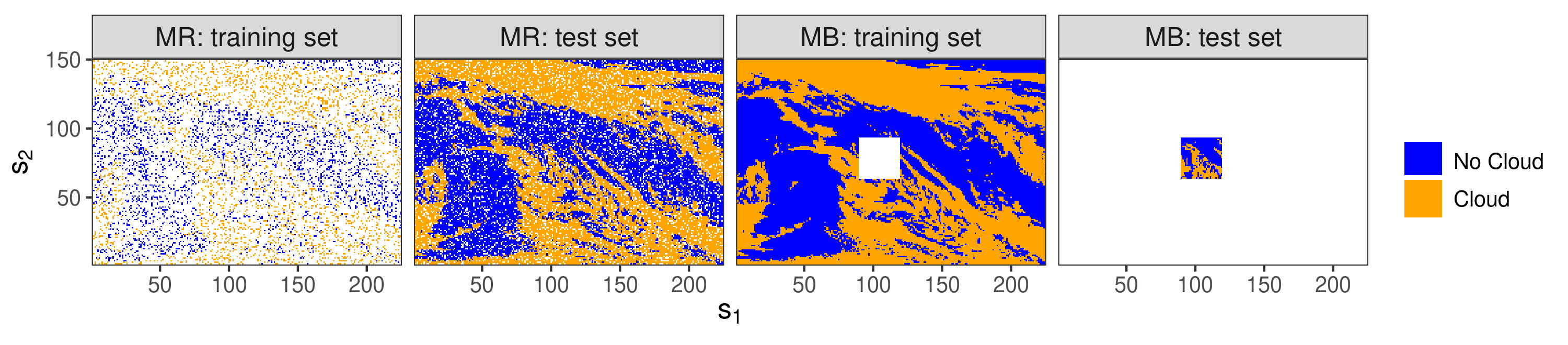}
    \vspace{-30pt}
     \caption{
     MODIS data used in the comparative study of Section~\ref{sec:04-01:MODIS}; a white background is used to denote removed data. (Left) The missing-at-random (MR) sample used for training.  (Centre left) The MR test data. (Centre right) The `missing-in-a-block' (MB) sample used for training. (Right) The MB test data. 
     }\label{fig:03-04-Modis1}
\end{figure}

We considered two types of sampling schemes for model comparison.
 The first was missing-at-random (MR), where we randomly selected a sub-sample of pixels to act as training data.
 Under the MR sampling scheme, we randomly sampled 6,000 pixels for training, leaving 27,750 pixels for testing. 
 The second sampling scheme, which we refer to as `missing-in-a-block' (MB), involved using all pixels outside a central block for training, and using pixels inside the block for testing. 
 The chosen testing block was a 30 $\times$ 30 square (900 pixels) in the middle of the spatial domain of interest, which meant that 32,850 pixels were used for training. 
The training and test sets under the two sampling schemes are shown in Figure~\ref{fig:03-04-Modis1}. 
 In what follows, $\text{Bin}(k, \pi)$ refers to a probability distribution on the non-negative integers with PMF \mbox{$\{{k \choose z}\pi^z(1-\pi)^{k-z} : z = 0, \dots, k\}$} for $\pi \in (0, 1)$ and $k \in \{0, 1, \dots\}$. 
 For the special case of $k = 1$, $\text{Bin}(1, \pi)$ is the Bernoulli distribution.

 In this case study, the data model is $Z_j \mid \vec{\mu} \sim \text{Bin}(1, \pi_{Z_j})$, $j = 1, \dots, m$, where $\pi_{Z_j}$ represents the probability of `Cloud' over the pixel $R_j$. 
 The software packages used in this study that allow use of this Bernoulli data model each required several modelling decisions, which had to be made in a way that balanced predictive performance and run time. 
 We took a systematic approach to model-selection by splitting the training data set equally in two, and then using one half for model fitting and the other half for model evaluation. 
 In this way, we were able to evaluate a large number of arguments for each package and choose the best combination in terms of predictive performance and run time.  
 For the methods requiring specification of a link function, we used the standard logit link function, $f(\pi) = \log(\frac{\pi}{1 - \pi})$.
 
 For \pkg{FRK}~v2, we used four resolutions of basis functions, giving a total of 11,130 basis functions. 
 For \pkg{INLA}, we discretised the domain into 13,494 elements.
 For \pkg{mgcv}, we used the \fct{bam} function, which is similar to the generalised-additive-model function \fct{gam} but optimised for large data sets, with 2,250 knots. 
\Copy{spBayesLimitation2}{For \pkg{spBayes}, we used 400 knots; increasing the number of knots further was computationally prohibitive (\pkg{spBayes} uses basis functions that depend on covariance-function parameters, so that computationally it can only handle a small number of knots).} 
 When using \pkg{spNNGP}, we found that the default option of considering 15 neighbours at a time was appropriate. 
 For \pkg{spNNGP} and \pkg{spBayes}, we used 10,000 MCMC samples in total, with a burn-in of 6,000 samples and a thinning factor of 10. 
The number of cores used for \pkg{spNNGP} can be controlled through the argument \code{n.omp.threads}; choosing a value greater than 1 returned an error (a known issue documented in the \pkg{spNNGP} package manual) and, hence, our reported run-times for \pkg{spNNGP} are for a single core.

\begin{figure}
    \centering
    \includegraphics[width = \linewidth]{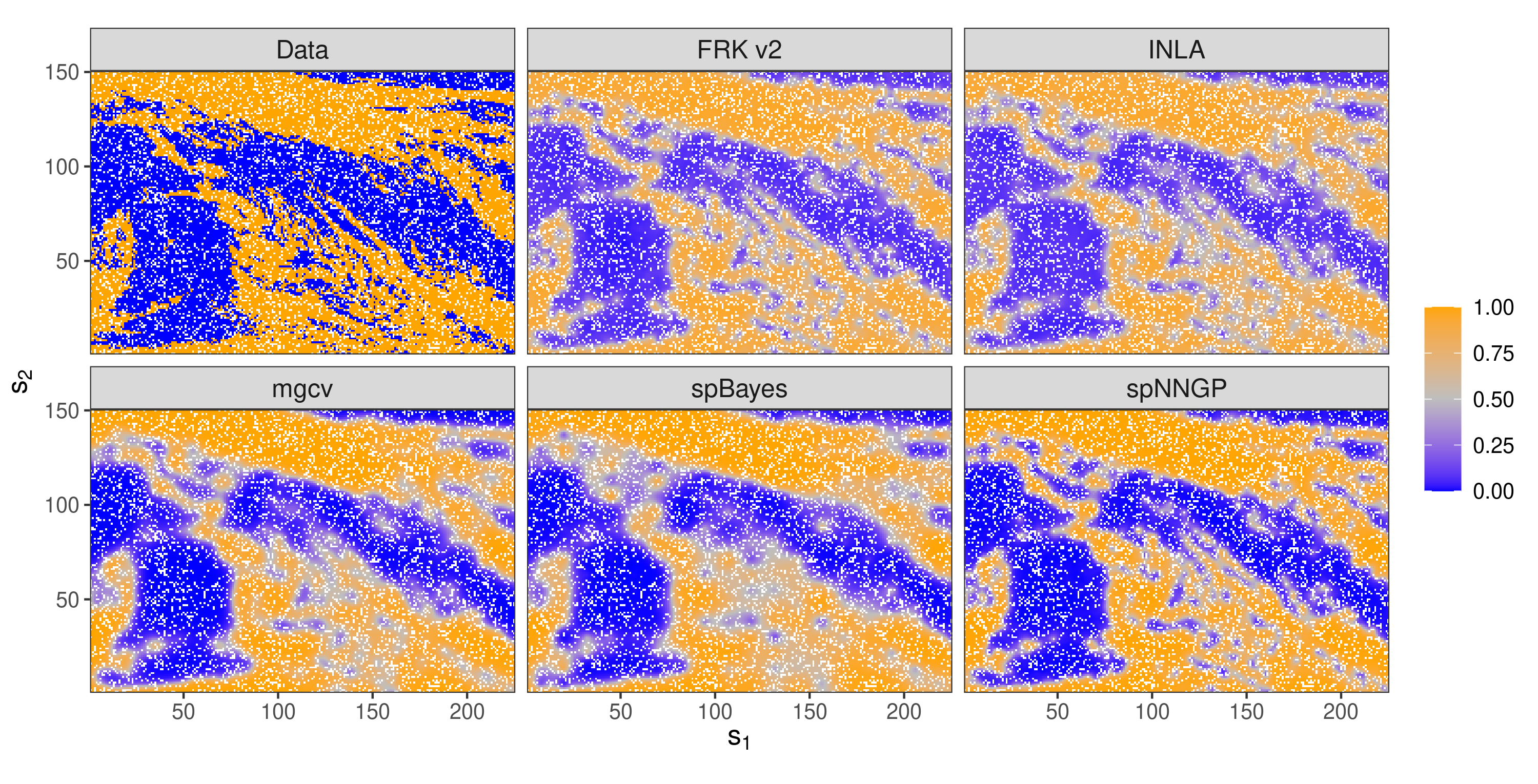}
    \vspace{-30pt}
     \caption{Predictions of the probability of `Cloud' on the test set of the missing-at-random (MR) experiment shown in Figure~\ref{fig:03-04-Modis1}.  The corresponding test data is shown in the top-left panel. Note that the training locations are indicated by white pixels.}   
  \label{fig:MODIS:pred_MR}
\end{figure}

\begin{figure}[t!]
    \centering
    \includegraphics[width = \linewidth]{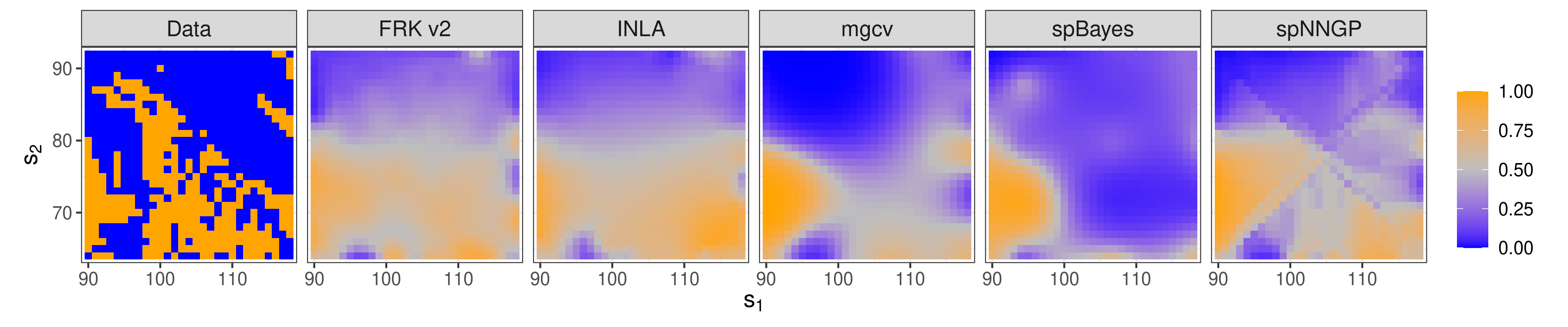}
    \vspace{-30pt}
     \caption{Predictions of the probability of `Cloud' on the test set of the `missing-in-a-block' (MB) experiment shown in Figure~\ref{fig:03-04-Modis1}. Here, we only show the test locations, which correspond to the 30 $\times$ 30 block near the centre of the spatial domain; the test data in this block is shown in the left-most panel of this figure.}   
  \label{fig:MODIS:pred_block}
\end{figure}

 For each method and each sampling scheme, we predicted the probability of `Cloud' at each pixel. 
 Figure~\ref{fig:MODIS:pred_MR} shows the predictions resulting from the MR sampling scheme shown in Figure~\ref{fig:03-04-Modis1}. 
The predictions from \pkg{FRK}~v2, \pkg{INLA}, and \pkg{spNNGP} are similar, 
 while the predictions from \pkg{mgcv} are slightly smoother than those from the aforementioned packages.
 The predictions of \pkg{spBayes} are even smoother, and this is due to the small number of knots it employs. 
 Figure~\ref{fig:MODIS:pred_block} shows the predictions resulting from the MB sampling scheme shown in Figure~\ref{fig:03-04-Modis1}. 
 \Copy{SmallBand}{Close inspection of the corresponding training data, shown in the centre right panel of Figure~\ref{fig:03-04-Modis1}, indicates that predictions within the missing block are largely driven by observations immediately surrounding the block, as expected.} 
 The packages \pkg{FRK}~v2 and \pkg{INLA} return predictive probabilities close to 0.5, while \pkg{mgcv} and \pkg{spBayes} are more confident in their predictions. 
 There is an interesting pattern in the \pkg{spNNGP} predictions; this is an expected artefact of the nearest-neighbour approach. 

%The packages used in this study can provide uncertainty quantification of the predictions for $\vec{\pi}$. 
%% However, when the data is binary, the task of assessing this uncertainty quantification is difficult, as prediction intervals (based on the probability samples) are in the range $(0, 1)$, and so they will \textit{never} contain the validation data, which take a value in the set $\{0, 1\}$.
%However, the underlying distribution of $\vec{\pi}$ is unidentifiable, as the predictive distribution, ${Z^*} \mid \vec{Z}$, for some validation datum ${Z^*}$, depends only on the posterior expectation of the probability parameter at the corresponding location, $\ENoLR{\pi^* \mid \vec{Z}}$.
%For this reason, we do not attempt to validate the prediction intervals, and instead focus our efforts on predictive accuracy.

\begin{figure}[t!]
    \centering
    \includegraphics[width = \linewidth]{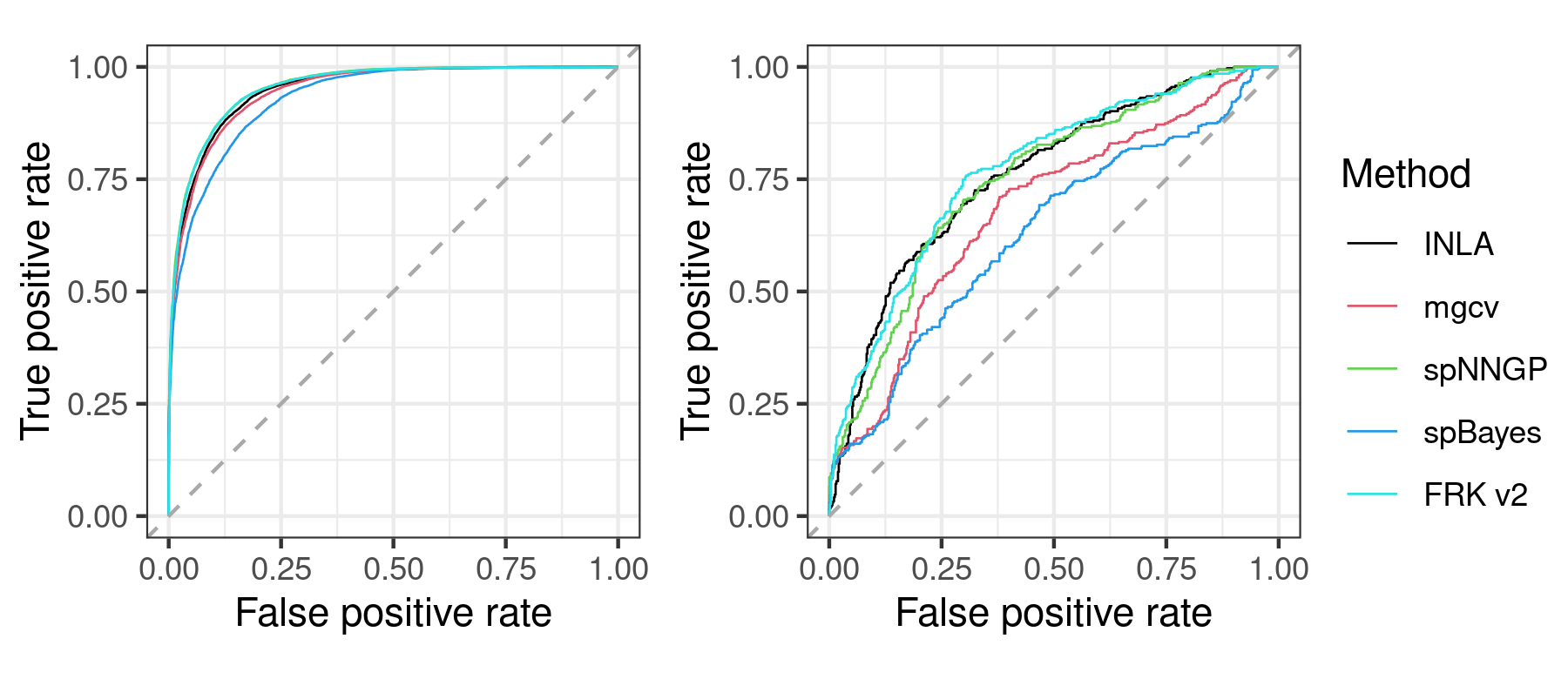}
        \vspace{-25pt}
    \caption{
%    	ROC curves for the training/test sets displayed in Figure~\ref{fig:03-04-Modis1}. 
%    	(Left) ROC curves generated in the `missing-at-random' experiment. 
%    	Note that there is a large degree of overlap between \pkg{FRK}~v2, \pkg{INLA}, \pkg{mgcv}, and \pkg{spNNGP}. 
%    	(Right) ROC curves generated in the `missing-in-a-block' experiment. 
    	ROC curves for the training/test sets displayed in Figure~\ref{fig:03-04-Modis1}. 
ROC curves generated in the `missing-at-random' experiment (left) and the `missing-in-a-block' experiment (right).  
    } 
  \label{fig:MODIS:ROC}
\end{figure}

\begin{table}
    \centering
    \begin{tabular}{lcccc}
    \hline
   Scheme &  Method  & Brier score & AUC &  Run time (min.) \\
   \hline
  MR & FRK v2 & 0.09 & \textbf{0.96} & 9.78 \\ 
   & INLA & 0.09 & 0.95 & \textbf{6.48} \\ 
   & mgcv & 0.09 & 0.95 & 26.53 \\ 
   & spBayes & 0.11 & 0.93 & 73.01 \\ 
   & spNNGP & \textbf{0.08} & \textbf{0.96} & 12.35 \\  
   \hline 
MB & FRK v2 & \textbf{0.19} & \textbf{0.77} & 31.74 \\ 
   & INLA & 0.20 & 0.76 & \textbf{12.19} \\ 
   & mgcv & 0.23 & 0.69 & 125.67 \\ 
   & spBayes & 0.25 & 0.63 & 504.41 \\ 
   & spNNGP & 0.20 & 0.75 & 65.30 \\     
   \hline    
  \end{tabular}
      \caption{Diagnostic results for the MODIS comparison study. %Best performers for a given diagnostic are boldfaced.
      }
    \label{tab:summary_of_analyses}
\end{table}

 To assess predictive accuracy, we compared the predictions from all models using the Brier score (see Appendix~\ref{app:ScoringRules}), and the area under the receiver operating characteristic (ROC) curve (AUC) \citep[see, e.g.,][pg. 317]{Hastie_2009_Elements_Statistical_Learning}.
 The Brier score assesses how close the predicted probability of `Cloud' is to the truth; 
% it is a negatively oriented rule, where 
 lower scores indicate more accurate predictions of the probability of `Cloud'.
 In contrast, higher AUC scores are preferred.
 The results for each method and each sampling scheme are reported in Table~\ref{tab:summary_of_analyses}, and the ROC curves are shown in Figure~\ref{fig:MODIS:ROC}. 
 For the MR sampling scheme, there is little discernible difference between \pkg{FRK}~v2, \pkg{INLA}, \pkg{mgcv}, and \pkg{spNNGP}. 
 However, as one may expect upon viewing the predictions in Figure~\ref{fig:MODIS:pred_MR}, \pkg{spBayes} performs poorly in comparison to the other packages due to the small number of knots it is able to employ.

 The task of prediction over a completely unobserved region is challenging, and so it is no surprise that the diagnostics for the MB sampling scheme are worse than for the MR sampling scheme. 
 In this case, we see \pkg{FRK}~v2, \pkg{INLA}, and \pkg{spNNGP} performing slightly better than \pkg{mgcv}, which in turn performs better than \pkg{spBayes}. 
 \Copy{DataPaucityExplanation}{These results reflect the fact that purely covariance or nearest-neighbour based spatial models are only useful for prediction locations that are `close' to data points. Spatial models that include covariates are typically much more effective at predicting over regions of data paucity.}

% Note that all run times increased under the MB scheme; however, \pkg{FRK}~v2 increased by a factor of less than 2, while increases in the run times for other packages were between a factor of 3 and 10. 
% Given that the training sample size is significantly larger in the MB scheme than under the MR scheme (32,850 pixels compared to 6,000 pixels), this suggests that \pkg{FRK}~v2 is well suited to fitting and predicting with large sample sizes.  
 \Copy{ElaborateAnalyses}{Overall, these results suggest that \pkg{FRK}~v2 is comparable to, or favourable to, other packages in this application where point-referenced spatial data are featured. 
 However, the main advantages of \pkg{FRK}~v2 lie in the ease with which it does more elaborate analyses with spatial or spatio-temporal non-Gaussian data of differing support, as shown in the next sections.}

\subsection{Block prediction: Contaminated soil}\label{sec:block_prediction}

%(The Paul and Cressie plots only go to a block size of about 190, while mine extend to a block size of 250. I manually computed the square root of the area of their largest block (being conservative with the boundaries), and found it to be 245. So, I believe that either (i) the Paul and Cressie plots are cut off at a certain point, (ii) a mistake has been made in computing the block sizes, or (iii) the block sizes are not computed simply by taking the square root of the area. I investigated the third point, and I think I can rule out the thought that they used the diagonal length of the block rather than the area, because I computed the diagonal length (just using Pythagoras' theorem) of the largest block and found it to be around 350.    )

\begin{figure}
    \centering
    \includegraphics[width = \linewidth]{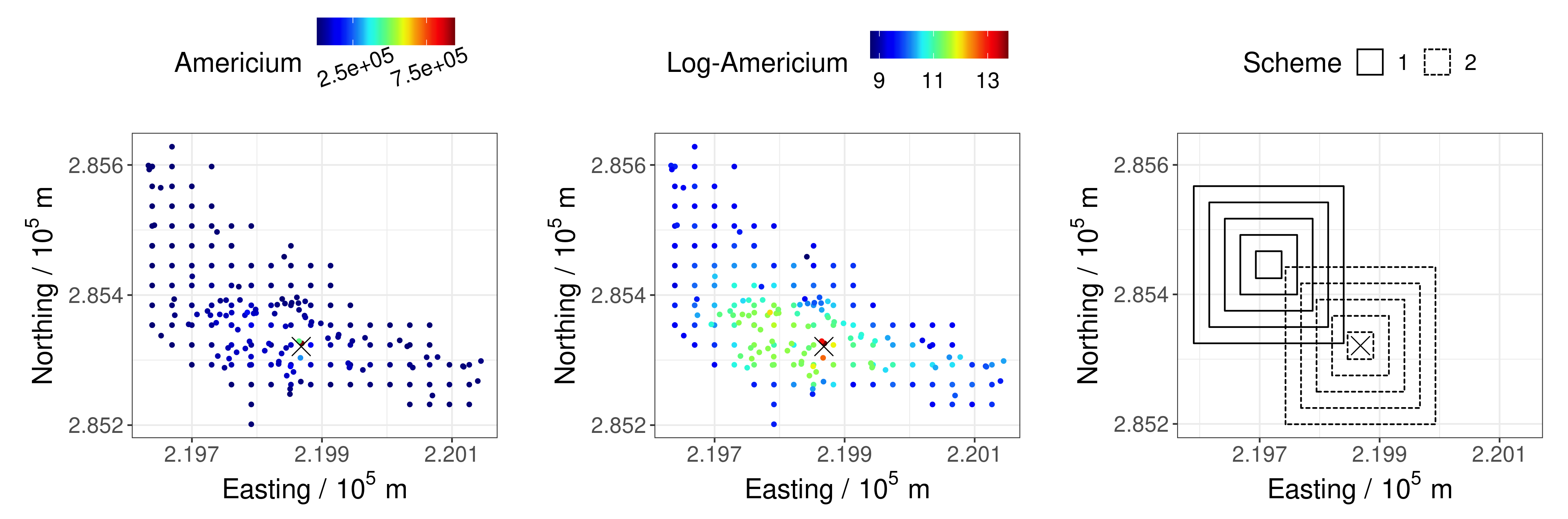}
        \vspace{-20pt}
    \caption{Americium (Am) soil data. The `$\times$' denotes Ground Zero (GZ), where the devices were detonated. (Left) Am concentrations on the original scale. (Centre) Am concentrations on the log scale. (Right) Blocking schemes: Scheme 1 (solid lines), centred away from GZ, and Scheme 2 (dashed lines), centred on GZ.}   
  \label{fig:Am_data}
\end{figure}

Between the years 1954 and 1963, nuclear devices were detonated at Area 13 of the Nevada Test Site in the United States, contaminating the surrounding soil with the radioactive element americium (Am). 
%In 1971, the Nevada Applied Ecology Group measured Am concentrations in a region surrounding Ground Zero (GZ), the location where the devices were detonated \citep{Paul_Cressie_2011_lognormal_kriging_block_prediction}. 
The data set we use in this example comprises Am concentrations (in $10^3$ counts per minute) in a spatial domain immediately surrounding Ground Zero (GZ, the location where the devices were detonated); it was previously analysed by \cite{Huang_2009_multivar_intrinsic_rand_functions_cokriging} and \cite{Paul_Cressie_2011_lognormal_kriging_block_prediction}. 
%Measurements were recorded at 196 unique locations, but the presence of several colocated measurements bring the total number of observations to 212. 
The total number of measurements (including some that are co-located) is 212.
The left and centre panels of Figure~\ref{fig:Am_data} show the data on the original scale and on the log scale, respectively.
\cite{Paul_Cressie_2011_lognormal_kriging_block_prediction} note that the Am concentrations are clearly lognormally distributed, and that soil remediation is often done by averaging the contaminant over pre-specified spatial regions of $D$ called blocks.
Hence, %soil remediation for 
 this application requires lognormal prediction over blocks, a task well suited to \pkg{FRK}~v2. 
% Note that \pkg{FRK}~v2 does not explicitly cater for a lognormal response, but an effective approach is to use \code{response = "Gaussian"} and \code{link = "log"}.  
 The right panel of Figure~\ref{fig:Am_data} shows two blocking schemes that we predict over: Both schemes contain five blocks, but one scheme is centred away from GZ, and the other is centred on GZ. 
 These blocking schemes represent the user-specified prediction regions, \mbox{$\{\tilde{R}_l : l = 1, \dots, 10\}$}.

As in \cite{Paul_Cressie_2011_lognormal_kriging_block_prediction}, we use a piece-wise linear trend in the `distance from GZ' as a fixed effect. Specifically, the log of the observations within a distance of 30.48m (100 ft) from GZ are assumed to follow a different trend to those observations beyond 30.48m from GZ, making up two regimes depending on distance from GZ. 
 In \FRKgeneric, covariates are provided with the BAU object and, hence, in this example, we must first construct the BAUs; here, we do this using the helper function \fct{auto\_BAUs}: 
% Note that the simulated examples in Section~\ref{SEC:IllustrativeExample} did not contain covariates, and \fct{auto\_BAUs} was called internally within \fct{FRK}. 

\begin{Code}
R> BAUs <- auto_BAUs(manifold = plane(), type = "grid", data = Am_data)
\end{Code} 
 
 The following code constructs the covariates that are needed to fit this piece-wise linear trend: \code{BAUs\$x1} and \code{BAUs\$x3} are indicator variables used to model the intercepts in each regime, and \code{BAUs\$x2} and \code{BAUs\$x4} are used to model the slopes of the trend in each regime.

\begin{Code}
R> d_BAU   <- distR(coordinates(BAUs), Ground_Zero)
R> BAUs$x1 <- d_BAU < 30.48
R> BAUs$x2 <- d_BAU * BAUs$x1
R> BAUs$x3 <- d_BAU >= 30.48
R> BAUs$x4 <- d_BAU * BAUs$x3
\end{Code}

Spatial statistical modelling for this problem is done by setting \code{response = "Gaussian"} and \code{link = "log"} when calling \fct{FRK}. 
\Copy{lognormalkriging}{In order to mimic lognormal block kriging, which models the response as a lognormal process, here we fix the measurement-error variance to a value that is small relative to the total variance of the data (specifically, we set the measurement-error variance equal to 1, while $\sd{\vec{Z}} = 77300$) prior to model fitting.} 
 In summary, the data model we use is Gaussian, $Z_j \mid \vec{\mu} \sim \text{Gau}(\mu_{Z_j}, 1)$ for $j = 1, \dots, m$, and we use a log link function, $g(\cdot) = \log(\cdot)$.
 Note that in the code below, we suppress the global intercept since we model the region-specific intercepts separately through \code{x1} and \code{x3}. 
\begin{Code}
R> Am_data$std <- 1 
R> S <- FRK(f = Am ~ -1 + x1 + x2 + x3 + x4, 
+    data = Am_data, BAUs = BAUs,
+    response = "gaussian", link = "log", est_error = FALSE) 
\end{Code}
%$

\begin{figure}
    \centering
    \includegraphics[width = \linewidth]{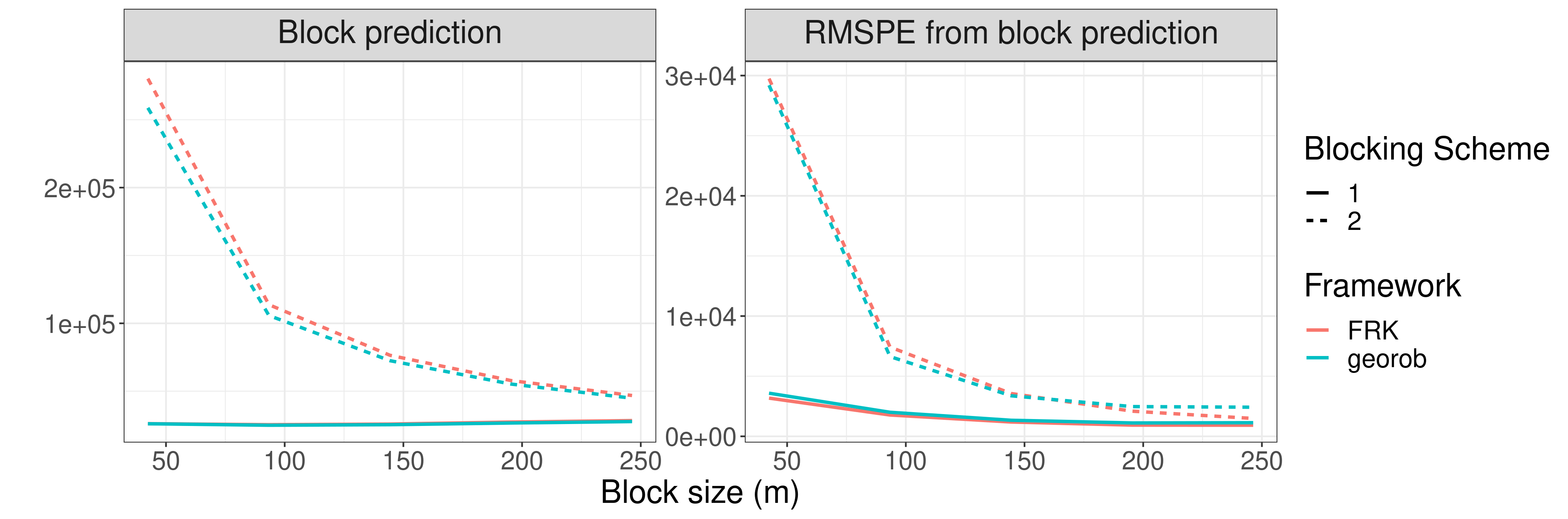}
        \vspace{-20pt} 
    \caption{Plots of block-predictions (left) and RMSPE (right) of Am concentrations against block size, $|B|^{1/2}$. Both quantities are in units of $10^3$ counts per minute. In both panels, the solid line corresponds to Scheme 1, and the dashed line corresponds to Scheme 2; the red line corresponds to \pkg{FRK}~v2, and the blue line corresponds to \pkg{georob}.}   
  \label{fig:Americium_block_preds_by_size}
\end{figure}

%By predicting over the BAUs, predictions over the entire spatial domain can be generated. %as depicted in Figure~\ref{fig:Am_BAU_predictions}.
%Alternatively, by passing a \class{SpatialPolygonsDataFrame} object into the \code{newdata} argument of \fct{predict}, one may straightforwardly generate block-level predictions.
% \citet*[ch.~3]{Cressie_1993_stats_for_spatial_data}
% \cite[][App.~C]{Cressie_2006_block_kriging_lognormal_spatial_processes}

In order to generate block-level predictions, we pass a \class{SpatialPolygonsDataFrame} object into the \code{newdata} argument of \fct{predict}. 
 In the following code, \code{blocks} is a \class{SpatialPolygonsDataFrame} object containing the polygons corresponding to the two blocking schemes shown in Figure~\ref{fig:Am_data}:

\begin{Code}
R> pred <- predict(S, newdata = blocks)
\end{Code}

%Another \proglang{R} package thats performs lognormal block predictions is \pkg{georob} \citep{georob}, which, using methods proposed by \cite{Cressie_2006_block_kriging_lognormal_spatial_processes}, implements an approximately unbiased back-transformation of kriging predictions of log-transformed data.
To validate the \pkg{FRK}~v2 predictions, we used the \proglang{R} package \pkg{georob} version 0.3.14 \citep{georob}, which implements an unbiased back-transformation of kriging predictions of log-transformed data \citep{Cressie_2006_block_kriging_lognormal_spatial_processes}.  
 Computation times for kriging do not scale well with sample size, however %as
 the size of this data set is sufficiently small for straightforward kriging to be possible. %, \pkg{georob} can be used to validate the predictions and associated uncertainty quantification obtained using \pkg{FRK}~v2. 
 The package \pkg{georob} provides users with two approaches to lognormal block kriging; 
%the first assumes permanence-of-lognormality, that is, that both point and block values follow log-normal laws, which cannot strictly hold, whilst the second, referred to as the optimal predictor, involves averaging back-transformed point predictions over the blocks. 
% The assumption of permanence-of-lognormality does not have a significant impact on the back-transformation when the blocks are small \citep{Cressie_2006_block_kriging_lognormal_spatial_processes}, however, for larger blocks, it is recommended to use the so-called optimal predictor. 
 we used the `optimal predictor', as recommended by the \pkg{georob} manual when predicting over large blocks.
%as the bias introduced by the assumption of permanence-of-lognormality typically has a greater impact on the back-transformation when the blocks are large \citep{Cressie_2006_block_kriging_lognormal_spatial_processes}. 
%Hence, as the blocking schemes in this application contain relatively large blocks, we will use the optimal predictor.
Figure~\ref{fig:Americium_block_preds_by_size} shows the block predictions and associated RMSPE obtained using \pkg{FRK}~v2 and \pkg{georob} for the two blocking schemes shown in Figure~\ref{fig:Am_data}.
It can be clearly seen that the results are very similar despite the use of dimension reduction in \pkg{FRK}~v2. 

%The similarity in results lends confidence that the predictions and associated prediction standard errors obtained using \pkg{FRK}~v2 are reasonable.
%, and suggests that, despite the fact that \pkg{FRK}~v2 uses low-rank approximations, the results are comparable to those obtained using `exact' methods.
%Figure~\ref{fig:Americium_block_preds_by_size} shows that, as one may expect, the block predictions of Scheme 1 (centred away from GZ) are smaller than the block predictions of Scheme 2 (centred on GZ). 
%In Scheme 1, the block predictions gradually increase with block size, as the blocks become closer to GZ; in Scheme 2, we witness the opposite behaviour, with the block predictions dropping rapidly with block size, as the area immediately surrounding GZ becomes an increasingly small percentage of the blocks. 
%In both schemes, the RMSPE decreases with block size, but it is consistently larger in Scheme 2 than in Scheme 1. 
%% The results obtained using \pkg{FRK}~v2 and \pkg{georob} do not corroborate those of \cite{Paul_Cressie_2011_lognormal_kriging_block_prediction}.
%Whilst the predictions obtained using \pkg{FRK}~v2 and \pkg{georob} are (somewhat) similar to those provided by \cite{Paul_Cressie_2011_lognormal_kriging_block_prediction}, there is a significant discrepancy between the reported RMSPE. In particular, \cite{Paul_Cressie_2011_lognormal_kriging_block_prediction} report a larger RMSPE (by a factor of between 1.5 and 10) for both blocking schemes, and, in Scheme 2, their RMSPE decreases at a slower rate, with the RMSPE decreasing linearly with the block-size.

\subsection{Spatial change-of-support: Poverty in Sydney}\label{sec:spatialCOS}

 The Australian Statistical Geography Standard (ASGS) defines a series of nested geographical areas in Australia known as Statistical Area Levels.  
%Statistical Area Level 1 (SA1) regions have an average population of 400 people each, Statistical Area Level 2 (SA2) regions have an average population of about 10000 people each, and the Statistical Area Level 3 (SA3) regions have a population range of between 30,000 and 130,000 people. 
 Statistical Area Level 3 (SA3) regions are aggregations of Statistical Area Level 2 (SA2) regions, and SA2 regions are aggregations of  Statistical Area Level 1 (SA1) regions. 
%For more information on Statistical Areas, see the ASGS webpage (\url{https://www.abs.gov.au/websitedbs/d3310114.nsf/home/australian+statistical+geography+standard+(asgs)}), and for the shapefiles of each Statistical Area Level, see the Australian Bureau of Statistics webpage
%(\url{https://www.abs.gov.au/AUSSTATS/abs@.nsf/DetailsPage/1270.0.55.001July\%202011}).
 In this example, we consider a region of the state of New South Wales in Australia, which contains 7,909 SA1 regions, 180 SA2 regions, and 31 SA3 regions, and we aim to infer `poverty' levels at the SA1 and SA3 regions from a data set containing mostly SA2 data and a small amount of SA1 data. 
 The data were collected in the Australian Census of 2011, and they consist of the number of families of various types within a range of weekly income brackets; in Appendix~\ref{Appendix:Sydney_data_description}, we provide further details on the way in which we define the poverty line for each family type. 
% Note that data at the SA1 and SA3 regions are available, and we use these to validate our down-scaled and up-scaled predictions.
 Note that data at the SA1 regions are available, and we use these to validate our down-scaled predictions.

\begin{figure}[t!]
    \centering
\includegraphics[width = \linewidth]{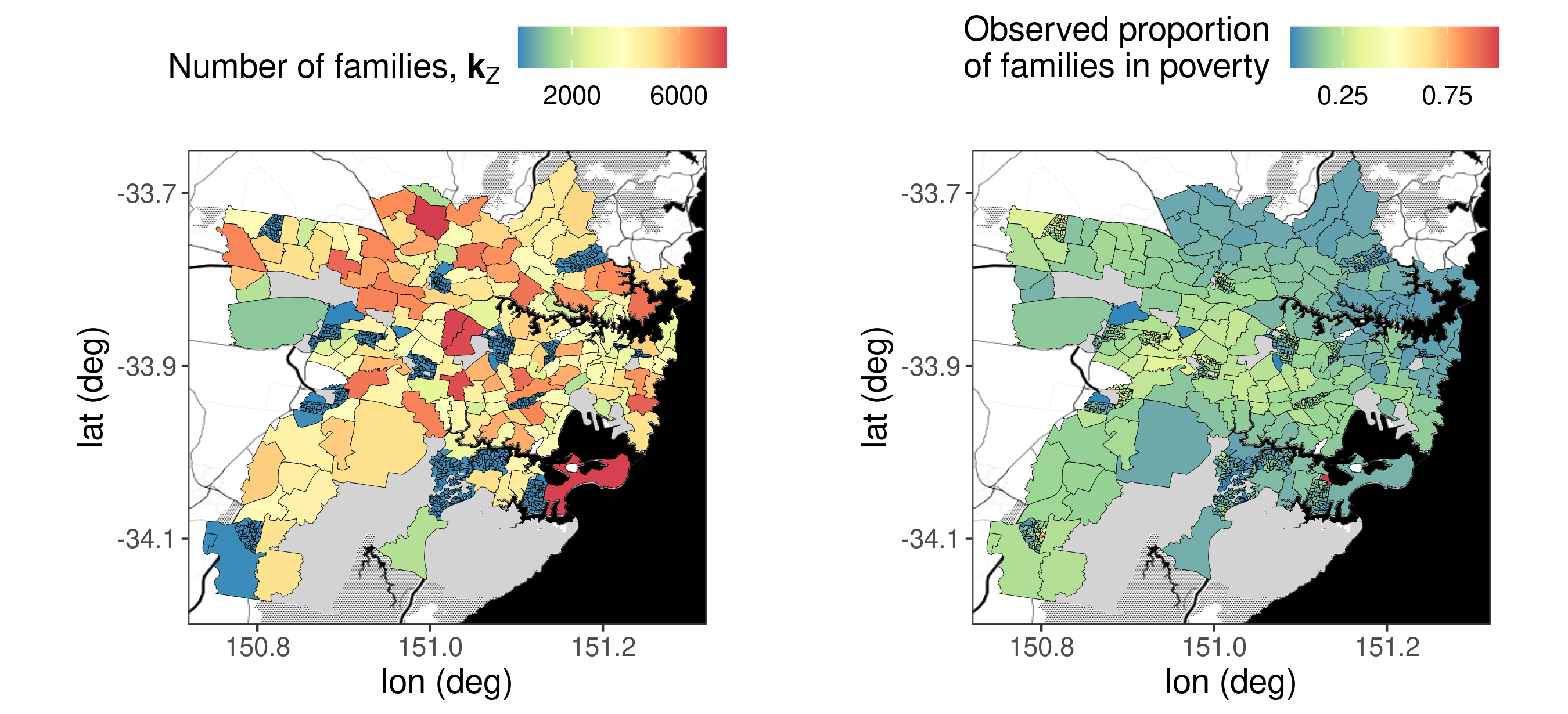}
    \vspace{-20pt}    
    \caption{
    Training data on SA1/SA2 regions used for modelling the proportion of families `in poverty' (see Appendix~\ref{Appendix:Sydney_data_description} for how we define `in poverty'). 
    (Left) The number of families, $\vec{k}_Z$. 
    (Right) The observed proportion of families in poverty computed by dividing the number of families in poverty by the total number of families, that is, $Z_j / k_{Z_j}$, for $j = 1, \dots, m$. 
    Solid-grey regions correspond to SA1/SA2 regions in which the total number of families is zero. The data are overlayed on a Stamen base map, where the textured grey areas correspond to bushland, large black areas correspond to ocean or large bodies of water, and solid-black lines correspond to major arterial roads. Map tiles by Stamen Design, under CC BY 3.0. Data by OpenStreetMap, under ODbL.}   
  \label{fig:SA2_level_data}
\end{figure}

 Sampling once from a large area is often relatively inexpensive compared to acquiring multiple samples from small areas. 
 Our training data, shown in Figure~\ref{fig:SA2_level_data}, is reflective of such a scenario. It includes mostly SA2 regions, but some SA1 regions have also been included.

 We use SA2-region (and some SA1-region) data for model fitting, and we use the SA1 regions as the BAUs. 
%SA2 regions are aggregations of SA1 regions, so many of the observation supports encompass multiple BAUs. 
% Hence, the BAU-level size parameters $\vec{k}$, here the number of families in each SA1 region, are needed and must be provided with the BAU object (see Section~\ref{sec:Distributions with size parameters}).
Since many of the observation supports encompass multiple BAUs (SA2 regions are aggregations of SA1 regions), the BAU-level size parameters $\vec{k}$, here the number of families in each SA1 region, are needed and must be provided with the BAU object (see Section~\ref{sec:Distributions with size parameters}).

 \begin{Code}
R> SA1s$k_BAU <- SA1s$number_of_families
 \end{Code}

 Our data model is $Z_j \mid \vec{\mu}, \vec{k}_Z \sim \text{Bin}(k_{Z_j}, \pi_{Z_j})$, for $j = 1, \dots, m$, and we use the logit link function, $f(\pi_i) = \log(\frac{\pi_i}{1 - \pi_i})$ in (\ref{eqn:pi_i_equals_f_Y_i}). 
 The model is established and fit in the code below. 
\begin{Code}
R> S <- FRK(f = number_of_families_in_poverty ~ 1, 
+    data = SA2s_and_some_SA1s, BAUs = SA1s, 
+    response = "binomial", link = "logit")
\end{Code}

Spatial prediction over all of the SA1 regions is obtained as follows:
\begin{Code}
R> SA1_predictions <- predict(S)
\end{Code}
 
\begin{figure}[t!]
    \centering
    \includegraphics[width = \linewidth]{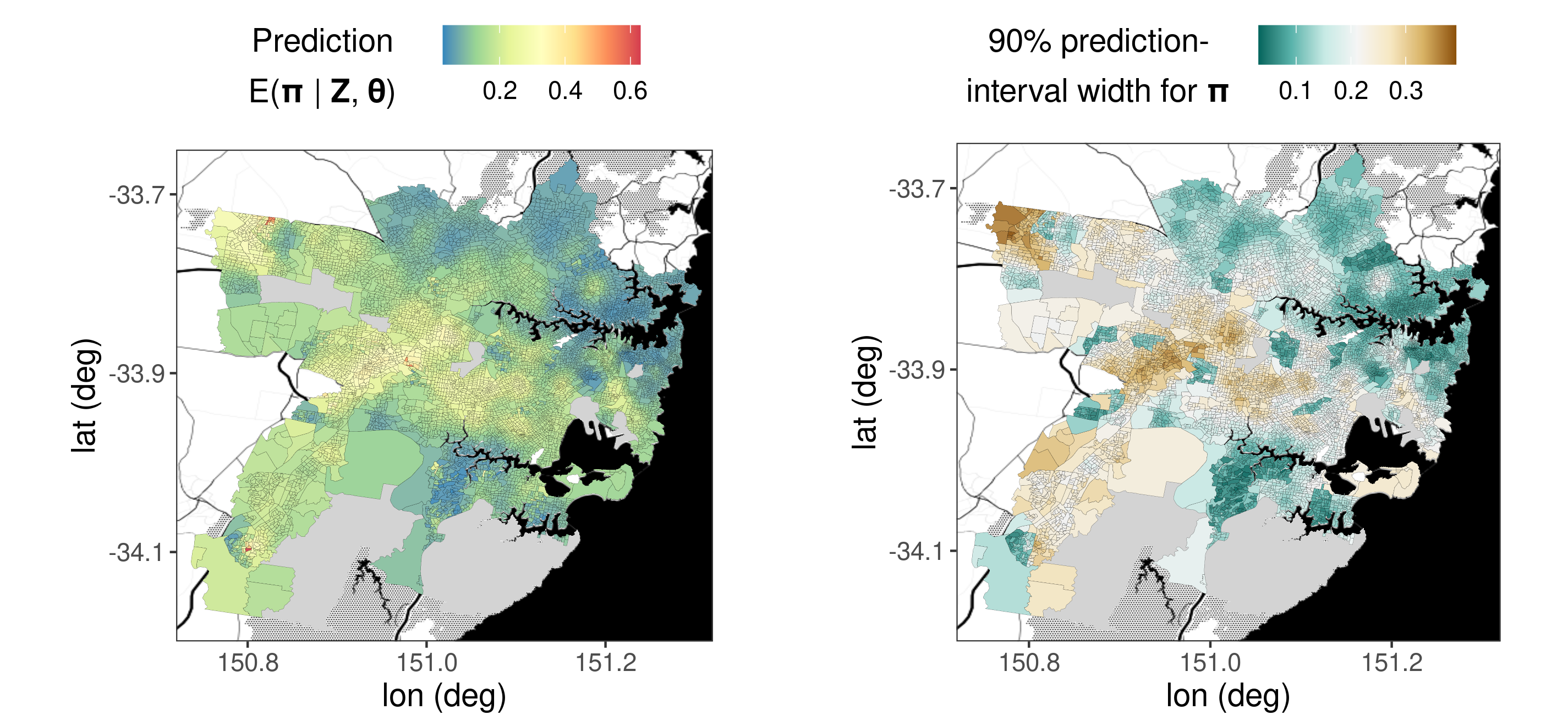}
        \vspace{-15pt}
    \caption{
    SA1-level predictions. 
    (Left) Prediction of the probability process, $\vec{\pi}$, 
%    over the SA1 regions, 
    representing the proportion of families in poverty in each SA1 region. 
    (Right) The 90\% prediction-interval width for each element of $\vec{\pi}$.
    Solid-grey regions correspond to SA2 regions in which the total number of families is zero, and hence they are omitted from the study. 
    For details on the underlying Stamen base map, refer to the caption of Figure~\ref{fig:SA2_level_data}. 
}   
  \label{fig:SA1_predictions}
\end{figure}

 Recall that, by default, \fct{predict} returns predictions of the mean process, $\vec{\mu}$, and, if applicable, the probability process, $\vec{\pi}$; here, we focus on $\vec{\pi}$, which can be interpreted as the proportion of families living below the poverty line. 
 The predictions and associated uncertainty over the SA1 regions are shown in Figure~\ref{fig:SA1_predictions}, which was generated using \fct{plot}. 
 Predicting over different spatial supports is straightforward with \pkg{FRK}~v2. 
 We predict over the SA3 regions by passing them as a \class{SpatialPolygonsDataFrame} object to the argument \code{newdata}.

\begin{Code}
R> SA3_predictions <- predict(S, newdata = SA3s)
\end{Code}

\begin{figure}[t!]
    \centering
    \includegraphics[width = \linewidth]{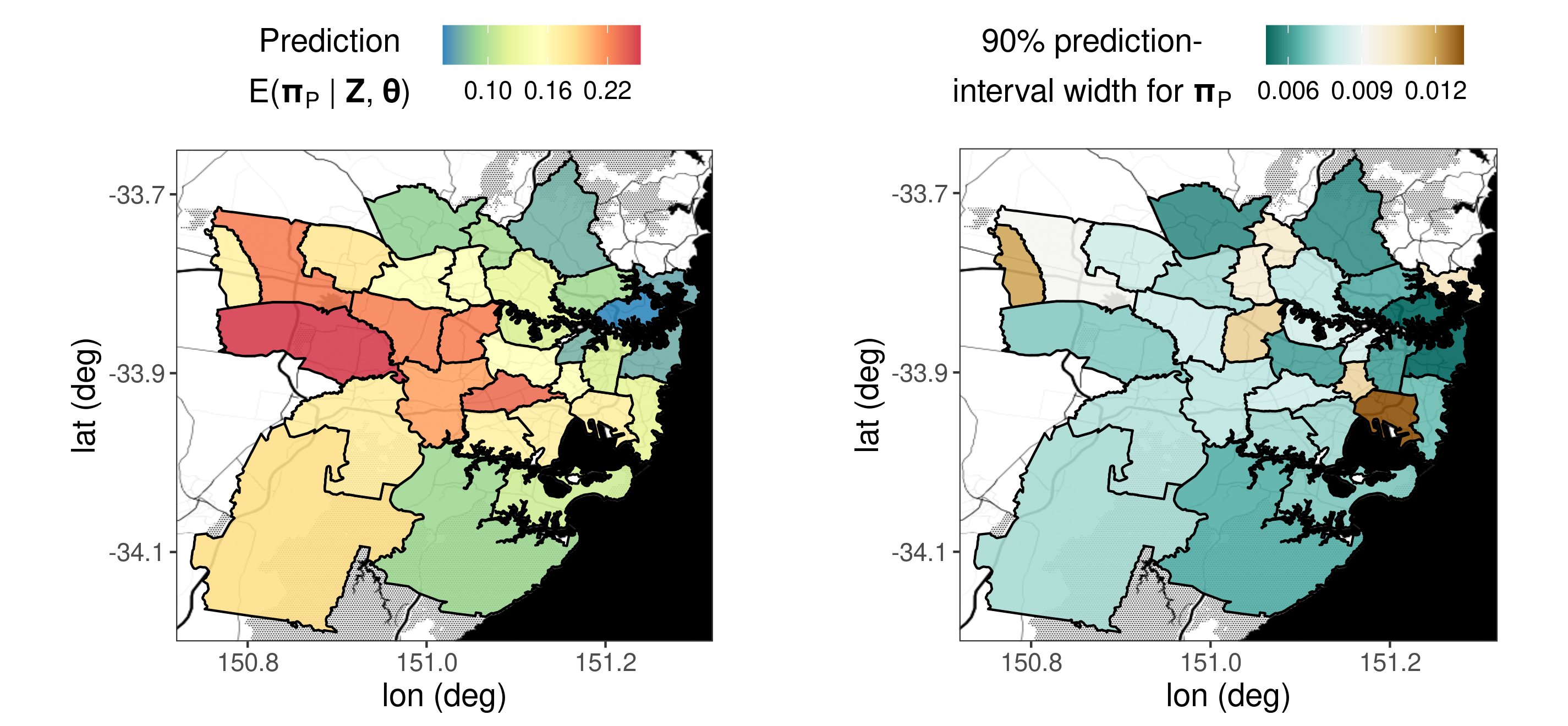}
        \vspace{-15pt}
    \caption{
        SA3-level predictions. 
    (Left) Prediction of the probability process, $\vec{\pi}_P$, 
%    over the SA3 regions, 
	representing the proportion of families in poverty in each SA3 region. 
    (Right) The 90\% prediction-interval width for each element of $\vec{\pi}_P$. 
    For details on the underlying Stamen base map, refer to the caption of Figure~\ref{fig:SA2_level_data}.
}   
  \label{fig:SA3_predictions}
\end{figure}

 Figure~\ref{fig:SA3_predictions} shows the SA3-region predictions and associated prediction-interval widths for the probability process, $\vec{\pi}_P$. 
 Again, this graphic was generated using \fct{plot}. 

 We assessed the model's ability to quantify uncertainty over the SA1 regions by computing the empirical coverage from nominal 90\% prediction intervals obtained via simulated predictive data at the SA1 level. 
 We found the empirical coverage to be 90.8\%, which is very close to the nominal value of 90\%. 
 Note that the inclusion of some fine-scale data (SA1 region data) greatly aids in the estimation of the fine-scale variance parameter, $\sigma^2_\xi$, and hence in providing valid prediction intervals when downscaling. 
 If only coarse-resolution data were available (i.e., all data supports were associated with multiple BAUs), to avoid identifiability issues, \pkg{FRK}~v2 automatically fixes $\sigma^2_\xi$ before fitting the model with \pkg{TMB}, to a rough, and possibly unreliable, estimate. 
 If one does know $\sigma^2_\xi$ or can obtain a reliable estimate of it (e.g., using past census data), one may specify it using the argument, \code{known\_sigma2fs}. 
 
% We conclude this example by noting that, although the prediction polygons, data supports, and BAUs in this study have a nested relationship, in general these elements can be entirely unrelated: Prediction in \pkg{FRK} can be implemented over any arbitrary, user-specified polygons; see Section~\ref{sec:03-03:negative-binomial} for an illustration. 
 
%% A naive predictor of the SA1 level proportion of families in poverty can be obtained by interpolating the SA2 level proportions over its constituent SA1s. 
%In the absence of covariate information, predictive performance in spatial change-of-support problems is not necessarily improved by using \pkg{FRK}~v2 over a naive interpolation method: For instance, the resulting RMSPE when predicting the SA1 level data with \pkg{FRK}~v2 was 7.07, whilst the naive predictor scored 7.03.
%However, \pkg{FRK}~v2 can provide reliable uncertainty quantification and it allows covariate information to be considered, the inclusion of which would likely improve prediction over a naive predictor.

\subsection{Spatio-temporal data: Crime in Chicago}\label{sec:ST_example}

 The city of Chicago is divided into 77 so-called community areas (CAs). 
An attractive property of CAs is their relative consistency, with 
boundaries that have changed little since their inception in the 1920s \citep{Chicago_library_census_data_info}. 
 %Shapefiles for the CAs are available for download from the open data source website, Plenario (\url{http://plenar.io/explore/discover}), or directly from the city of Chicago website (\url{https://data.cityofchicago.org/}). 
In this study, we model the number of crimes in each CA between the years 2001 and 2019 inclusive.
A full list of crimes committed in Chicago during this period is provided by the Chicago Police Department; the data %were originally downloaded from the open-data-source website Plenario \citep{Plenario_open_source_website} and 
 are available at \mbox{\url{https://hpc.niasra.uow.edu.au/ckan/en_AU/dataset/chicago_crime_dataset}}. 
%The full data set has over 7 million crimes; however, here we consider only violent, non-sexual crimes, which we define as crimes labelled as assault or battery. 
%We considered only violent, non-sexual crimes, which we defined as crimes labelled as assault or battery 
%%(we excluded homicide, which is comparatively rare, with 10,324 recorded instances in our data set).
%(we excluded homicide, which is comparatively rare, accounting for less than 1\% of the violent crimes in our data).
%The data considered in this analysis consisted of 1,748,360 crimes. 
 We considered only crimes labelled as `assault' or `battery'; there were roughly 1.75 million such crimes in total between 2001 and 2019. 
 We are interested in modelling the total number of crimes in each CA and in each year; once binned into CA-year bins (done automatically by \FRKgeneric), the number of aggregated-level data is $76 \times 19 = 1444$. 
 The CA containing O'Hare airport is non-populous and is almost disjoint from the other CAs; for simplicity, we excluded it from this analysis, leaving 76 CAs.

In this example, we use the CAs as our spatial BAUs, which are read in from a shapefile and stored as a \class{SpatialPolygonsDataFrame} object. 
Spatio-temporal BAUs may then be constructed by passing the CAs and data into \fct{auto\_BAUs}. 
 In this case, the spatio-temporal BAUs are space-time volumes constructed by taking all combinations of the spatial BAU footprints with the yearly intervals that make up the 19-year period of interest. 
\begin{Code}
R> ST_BAUs <- auto_BAUs(manifold = STplane(), data = chicago_crimes_fit,
+    spatial_BAUs = community_areas, tunit = "years") 
\end{Code}

When modelling crime, it is natural to include population, or population density, as a covariate. 
As the CAs are of unequal area, we use population rather than population density. This covariate was obtained from the % Combined Community Data Snapshots provided by the 
\cite{Chicago_community_data_snapshots}. It is difficult to obtain population data for every year so, for simplicity, we assume that population was constant over the time-span of the data.
% We observed a distinct negative trend when plotting the total number of crimes in each year; hence, we also include time as a covariate.
%Figure~\ref{fig:chicago_temporal_trend} shows the total number of crimes committed across Chicago in each year, and suggests that a piecewise, linear temporal model split by the year 2014 is appropriate. 

Next, we generate spatio-temporal basis functions automatically using \fct{auto\_basis}.  
% Note that each observation in our data set corresponds to an occurrence of a crime, and these crimes are indexed by the coordinates (longitude, latitude) and date at which the crime occurred; that is, we have unstructured spatio-temporal data in which observations may be recorded at any point in time and location in space. 
%Therefore, we store the full data set, which we label as \mbox{\code{chicago\_crimes}}, as an object of class \class{STIDF}. 
%In a pre-processing step, we aggregated the number of crimes occurring at every unique coordinate and date combination, and stored this number as a field called \mbox{\code{number\_of\_crimes}}; the majority of entries in this field are 1, but it is possible that multiple crimes are recorded at the same location and time (for example, if there are multiple offenders, or if an offender is charged with multiple crimes in the same incident).
%A subset of the \mbox{\code{chicago\_crimes}} data, specifically, all of the data excluding the years 2010 and 2019, was used for training the model and automatically constructing the basis functions. %; this training subset is labelled \mbox{\code{chicago\_crimes\_fit}}.
%\begin{Code}
%R> basis <- auto_basis(manifold = STplane(), data = chicago_crimes_fit, 
%+    tunit = "years")
%\end{Code}
 Then, we initialise and fit the \class{SRE} object using \fct{FRK}, setting \code{response = "poisson"} and \code{link = "log"} in order to use a Poisson data model, $Z_j \mid \vec{\mu} \sim \text{Poi}(\mu_{Z_j})$, for $j = 1, \dots, m$, and a log link function, $g(\cdot) = \log(\cdot)$. 
 Setting \code{sum\_variables = "number\_of\_crimes"} indicates that we wish to sum (rather than average) the individual crimes into aggregated-level data representing the total count in each CA and each year.
 As the number of spatial BAUs (the CAs) is relatively low, and we have observed each spatial BAU multiple times, we may attribute to each spatial BAU its own fine-scale variance parameter by setting \mbox{\code{fs\_by\_spatial\_BAU = TRUE}} (see Section~\ref{sec:spatio-temporal}). 
 We excluded the years 2010 and 2019 from the training data and used them to evaluate crime predictions and forecasts, respectively.  
%R> S <- FRK(f = number_of_crimes ~ log(population) + x1 + x2 + x3,   
\begin{Code}
R> S <- FRK(f = number_of_crimes ~ log(population),   
+    data = chicago_crimes_fit, basis = basis, BAUs = ST_BAUs,         
+    response = "poisson", link = "log", 
+    sum_variables = "number_of_crimes", fs_by_spatial_BAU = TRUE) 
\end{Code}

\begin{figure}[t!]
    \centering
    \includegraphics[width = \linewidth]{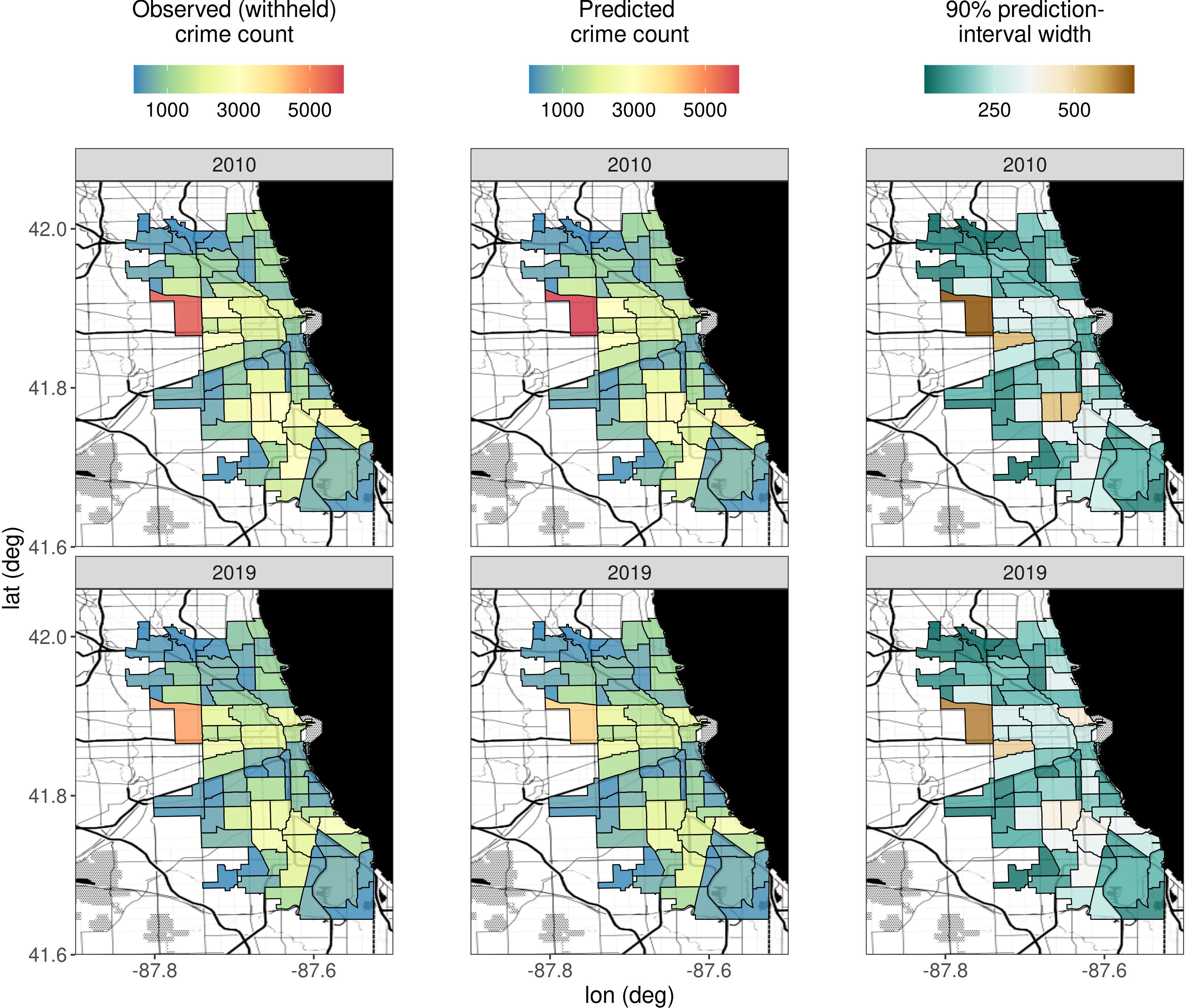}
        \vspace{-15pt}
    \caption{Observed (withheld) number of crimes, predictions, and prediction-interval width over Chicago in the prediction (2010) and forecast (2019) years. The first row corresponds to the year 2010, and the second row corresponds to the year 2019. The first column shows the observed (withheld) number of crimes; the second column shows the predicted number of crimes; and the third column shows the width of a prediction interval with a nominal coverage of 90\%. For details on the underlying Stamen base map, refer to the caption of Figure~\ref{fig:SA2_level_data} (the large black region in this figure correspond to Lake Michigan).
}   
  \label{fig:chicago_validation_predictions}
\end{figure}

%\Copy{ChicagoInference}{%
%  We may now conduct inference with the fitted model. For instance, the estimated effect of log-population, extracted with \fct{coef}, is $0.99$, so that a unit increase in log-population increases the expected number of crimes by a factor of $\exp(0.99) \approx 2.7$. 
%}
%we generate predictive data over the spatio-temporal BAUs using \fct{predict}, and plot the results using \fct{plot}. 
Finally, we generate predictive data over the spatio-temporal BAUs using \fct{predict}, and plot the results using \fct{plot}. 

\begin{figure}[t!]
    \centering
    \includegraphics[width = \linewidth]{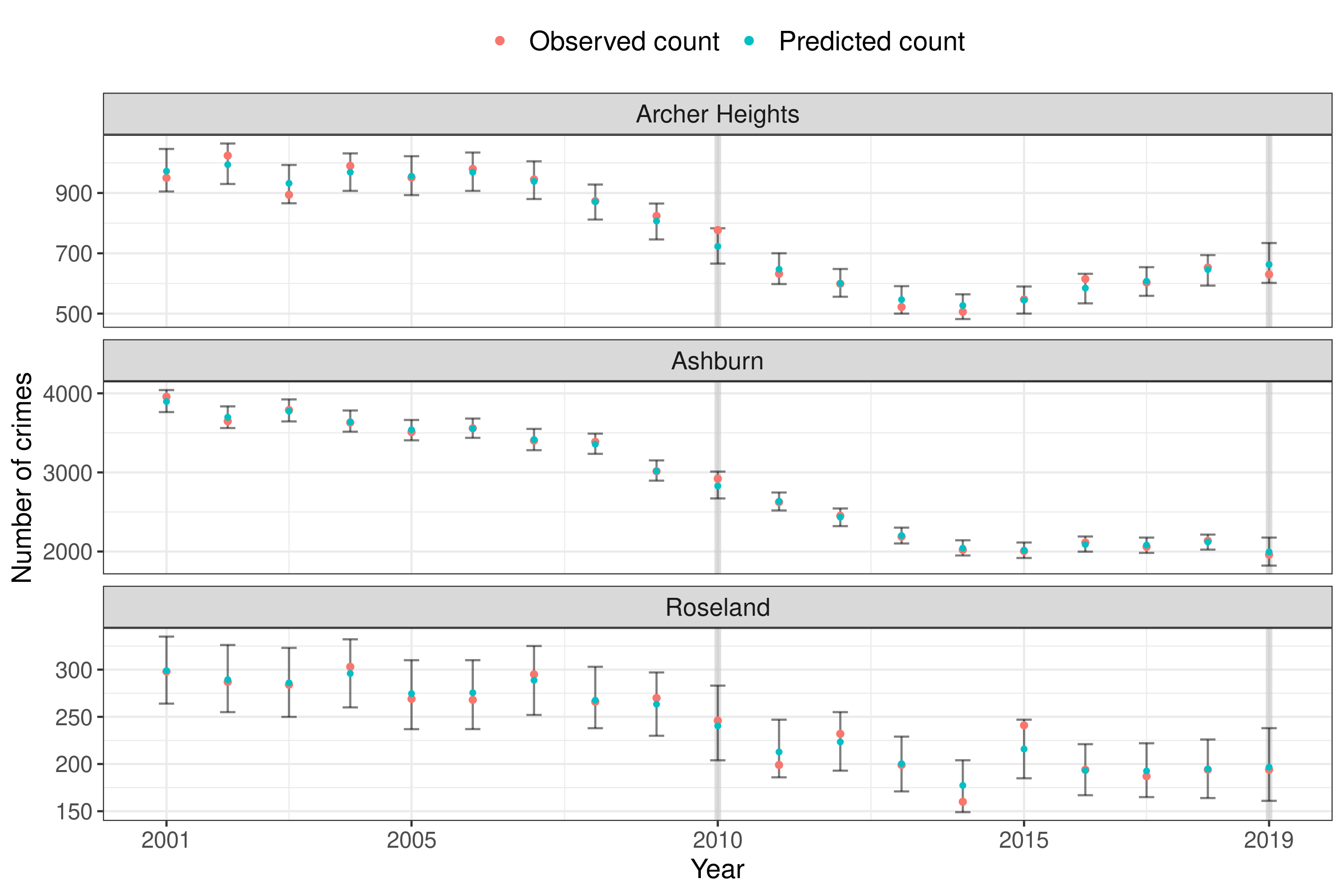}
        \vspace{-15pt}
    \caption{Time-series plots of predicted and observed number of crimes for three randomly-selected CAs. The prediction (2010) and forecast (2019) years are highlighted in light-grey. The observed number of crimes in each year is indicated by a red dot, whilst the predicted number of crimes is indicated by a blue dot. The error bars represent a 90\% prediction interval. We note that the prediction intervals are slightly wider in validation years (2010 and 2019) than in observed years, and that the observed number of crimes is contained within the prediction interval in all years for these CAs. 
}   
  \label{fig:chicago_time_series}
\end{figure}

%\begin{figure}[t!]
%    \centering
%    \includegraphics[width = \linewidth]{results/4_4_Chicago_focused_CAs_predictive_distributions.png}
%    \vspace{-15pt}    
%    \caption{Predictive distributions in the prediction (2010) and forecast (2019) years for three randomly-selected CAs (Archer Heights, Ashburn, and Roseland). 
% In each panel, the red line corresponds to the observed (withheld) number of crimes and the blue line corresponds to the predicted number of crimes. 
%    The first row corresponds to the year 2010 and the second row corresponds to the year 2019. 
%    The first, second, and third columns correspond to Archer Heights, Ashburn, and Roseland, respectively.}   
%  \label{fig:chicago_predictive_distributions}
%\end{figure}

The observed (withheld) number of crimes, predicted number of crimes, and prediction uncertainty for the prediction (2010) and forecast (2019) years are shown in Figure~\ref{fig:chicago_validation_predictions}. 
For both years, Figure~\ref{fig:chicago_validation_predictions} shows agreement between the predicted and observed number of crimes. Furthermore, the prediction uncertainty is roughly proportional to the predicted value, as expected when counts are modelled.
For the prediction year and forecast year, we also computed the empirical coverage when using 90\%, 80\%, 70\%, and 60\% prediction intervals, and the mean absolute percentage error (MAPE; see Appendix~\ref{app:ScoringRules}). 
We consistently observed that the $x$\% empirical coverage in the year 2010 was slightly higher (4\% on average) than the $x$\% nominal coverage, whilst it was lower (13\% on average) in the forecast year.
We observed MAPE scores of 4\% and 9\% in the years 2010 and 2019, respectively.
The slightly worse results in the year 2019 are expected, as forecasting into the future is a harder task than predicting within the time span of the data.
Nonetheless, these predictions/forecasts are cause for optimism given the complexity of modelling crime in a spatio-temporal setting. 

Next, we next focus on three randomly-selected CAs: Archer heights, Ashburn, and Roseland. The time series of the observed data, predictions, and 90\% prediction intervals for these CAs, are shown in Figure~\ref{fig:chicago_time_series}. 
%The prediction intervals are slightly wider in validation years (2010 and 2019) than in observed years. 
% The observed (withheld) number of crimes is contained within the prediction interval for all time points for these CAs.
The prediction intervals are slightly wider in validation years (2010 and 2019) than in observed years, and the observed (withheld) number of crimes is contained within the prediction interval for all time points for these CAs.
%The predictive distributions in the validation years for the three CAs of interest are shown in Figure~\ref{fig:chicago_predictive_distributions}.
%The forecasts for Ashburn and Roseland in the year 2019 are particularly accurate, with the forecasted number of crimes very close to the observed (withheld) number of crimes. 

\section{Conclusion}\label{SEC:Conclusion}

 In this paper, we have described  \pkg{FRK}~v2, which is a major upgrade to \pkg{FRK}~v1. Substantial enhancements allow for the spatial and spatio-temporal modelling of, and large-scale prediction from, big, non-Gaussian data sets. 
Using a GLMM model and the software \pkg{TMB}, \pkg{FRK}~v2 can now cater for many distributions within the exponential family, as well as many link functions. % (Table~\ref{table:response_and_links}).  
 \Copy{MoreBasisFunctions}{Furthermore, \pkg{FRK}~v2 allows for the use of many more basis functions when modelling the spatial process, and hence it can often achieve more accurate predictions than \pkg{FRK}~v1 (Table~\ref{tab:Heaton_comparison:full}).} 
 The existing functionality of \pkg{FRK}~v1 is retained with this extension; in particular, the package makes use of automatic basis-function construction, it is capable of handling both point-referenced and areal data, and it resolves the spatial change-of-support problem through the use of BAUs.
 The current version now provides a highly accessible and user-friendly approach to spatial and spatio-temporal modelling of big data in both a Gaussian and non-Gaussian setting.

 One requirement of the framework is that covariates need to be known for every BAU, which may not be the case if covariates are recorded only at the data-support level. 
 Spatial interpolation of the covariates can be used to address this problem. 
 Another requirement is the practical necessity to fix the fine-scale variance parameter in spatial change-of-support applications; note that this is not an issue if one is able to obtain a reliable estimate through other means (e.g., via previously sampled data or via quality-control experiments). 
 We are currently exploring avenues to relax this requirement via the provision of a robust offline estimate.
 Despite the added flexibility of \pkg{FRK}~v2, several models of interest, such as the zero-inflated Poisson, are still not catered for. 
 Future work will see the introduction of such models.
 \Copy{FutureWork}{Further future work includes adding support for multivariate responses, for which basis-function models are well suited \citep[for an overview, see][Sec.~4.1]{Cressie_2021_review_spatial-basis-function_models}, and for spatially varying regression coefficients, which are sometimes needed in spatial analyses.}

\section*{Acknowledgments}

Matthew Sainsbury-Dale's research was supported by an Australian Government Research Training Program Scholarship.
Andrew Zammit-Mangion's and Noel Cressie's research was supported by an Australian Research Council (ARC) Discovery Project, DP190100180. Andrew Zammit-Mangion's research was also supported by an ARC Discovery Early Career Research Award, DE180100203.  
The authors would like to thank Rajib Paul for providing the Americium data analysed in Section~\ref{sec:block_prediction}, Michael Bertolacci for his discussion surrounding the MODIS comparison study, and Yi Cao for his technical assistance during the project. 
 We are also grateful to two anonymous reviewers whose suggestions improved our article.

%% -- Bibliography -------------------------------------------------------------
%% - References need to be provided in a .bib BibTeX database.
%% - All references should be made with \cite, \citet, \citep, \citealp etc.
%%   (and never hard-coded). See the FAQ for details.
%% - JSS-specific markup (\proglang, \pkg, \code) should be used in the .bib.
%% - Titles in the .bib should be in title case.
%% - DOIs should be included where available.

\bibliography{FRKv2}

\begin{thebibliography}{54}
\newcommand{\enquote}[1]{``#1''}
\providecommand{\natexlab}[1]{#1}
\providecommand{\url}[1]{\texttt{#1}}
\providecommand{\urlprefix}{URL }
\expandafter\ifx\csname urlstyle\endcsname\relax
  \providecommand{\doi}[1]{doi:\discretionary{}{}{}#1}\else
  \providecommand{\doi}{doi:\discretionary{}{}{}\begingroup
  \urlstyle{rm}\Url}\fi
\providecommand{\eprint}[2][]{\url{#2}}

\bibitem[{Bachl \emph{et~al.}(2019)Bachl, Lindgren, Borchers, and
  Illian}]{Bachl_2019_inlabru}
Bachl FE, Lindgren F, Borchers DL, Illian JB (2019).
\newblock \enquote{\pkg{inlabru}: An \proglang{R} package for {Bayesian}
  Spatial Modelling from Ecological Survey Data.}
\newblock \emph{{Methods in Ecology and Evolution}}, \textbf{10}, 760--766.

\bibitem[{Bates \emph{et~al.}(2019)Bates, Maechler, and Davis}]{Matrix_Package}
Bates D, Maechler M, Davis TA (2019).
\newblock \emph{\pkg{Matrix}: Sparse and Dense Matrix Classes and Methods}.
\newblock \proglang{R} package version 1.2-17,
  \urlprefix\url{http://Matrix.R-forge.R-project.org/}.

\bibitem[{Bell(2005)}]{CppAD_Package}
Bell BM (2005).
\newblock \enquote{\pkg{CppAD}: A Package for \proglang{C++} Algorithmic
  Differentiation.}
\newblock \url{http://www.coin-or.org/CppAD}.
\newblock Accessed: 2021-08-10.

\bibitem[{Box(1980)}]{Box_1980_simulation-based_model_checking}
Box GEP (1980).
\newblock \enquote{Sampling and {B}ayes' Inference in Scientific Modelling and
  Robustness.}
\newblock \emph{Journal of the Royal Statistical Society A}, \textbf{143}, 383
  -- 430.

\bibitem[{Bradley \emph{et~al.}(2018)Bradley, Holan, and
  Wikle}]{Bradley_2018_computationally_efficient_multivariate_ST_models_for_high-dimensional_count-valued_data}
Bradley JR, Holan SH, Wikle CK (2018).
\newblock \enquote{Computationally Efficient Multivariate Spatio-Temporal
  Models for High-Dimensional Count-Valued Data (With Discussion).}
\newblock \emph{Bayesian Analysis}, \textbf{13}, 253--310.

\bibitem[{Bradley \emph{et~al.}(2016)Bradley, Wikle, and
  Holan}]{Bradley_2016_Bayesian_spatial_COS_lattice_data}
Bradley JR, Wikle CK, Holan SH (2016).
\newblock \enquote{Bayesian Spatial Change of Support for Count-Valued Survey
  Data with Application to the {A}merican {C}ommunity {S}urvey.}
\newblock \emph{Journal of the American Statistical Association}, \textbf{111},
  472--487.

\bibitem[{Bradley \emph{et~al.}(2019)Bradley, Wikle, and
  Holan}]{Bradley_2019_ST_models_for_big_multinomial_data_using_conditional_multivariate_logit_beta_distribution}
Bradley JR, Wikle CK, Holan SH (2019).
\newblock \enquote{Spatio-Temporal Models for Big Multinomial Data Using the
  Conditional Multivariate Logit Beta Distribution.}
\newblock \emph{Journal of Time Series Analysis}, \textbf{50}, 363--382.

\bibitem[{{Chicago Metropolitan Agency for
  Planning}(2017)}]{Chicago_community_data_snapshots}
{Chicago Metropolitan Agency for Planning} (2017).
\newblock \enquote{Chicago Community Data Snapshots.}
\newblock \emph{Technical report}.
\newblock URL
  \url{https://www.cmap.illinois.gov/documents/10180/126764/_Combined_AllCCAs.pdf/}.
  Accessed: 2021-08-10.

\bibitem[{Cox and Snell(1968)}]{Cox_Snell_1968_residuals}
Cox DR, Snell EJ (1968).
\newblock \enquote{A General Definition of Residuals (With Discussion).}
\newblock \emph{Journal of the Royal Statistical Society B}, \textbf{30},
  248--275.

\bibitem[{Cressie(1993)}]{Cressie_1993_stats_for_spatial_data}
Cressie N (1993).
\newblock \emph{{Statistics for Spatial Data}}.
\newblock Revised edition. John Wiley \& Sons, Hoboken, NJ.
\newblock ISBN 0387310738.

\bibitem[{Cressie(2006)}]{Cressie_2006_block_kriging_lognormal_spatial_processes}
Cressie N (2006).
\newblock \enquote{Block Kriging for Lognormal Spatial Processes.}
\newblock \emph{{Mathematical Geology}}, \textbf{38}, 413--443.

\bibitem[{Cressie and Johannesson(2008)}]{Cressie_Johannesson_2008_FRK}
Cressie N, Johannesson G (2008).
\newblock \enquote{Fixed Rank Kriging for Very Large Spatial Data Sets.}
\newblock \emph{{Journal of the Royal Statistical Society B}}, \textbf{70},
  209--226.

\bibitem[{Cressie \emph{et~al.}(2021)Cressie, Sainsbury-Dale, and
  Zammit-Mangion}]{Cressie_2021_review_spatial-basis-function_models}
Cressie N, Sainsbury-Dale M, Zammit-Mangion A (2021).
\newblock \enquote{Basis-Function Models in Spatial Statistics.}
\newblock \emph{Annual Review of Statistics and its Applications}, \textbf{9},
  373--400.

\bibitem[{Datta \emph{et~al.}(2016)Datta, Banerjee, Finley, and
  Gelfand}]{Datta_2016_NNGP_spatial}
Datta A, Banerjee S, Finley AO, Gelfand AE (2016).
\newblock \enquote{Hierarchical Nearest-Neighbour Gaussian Process Models for
  Large Geostatistical Datasets.}
\newblock \emph{Journal of the American Statistical Association}, \textbf{111},
  800--812.

\bibitem[{Diggle \emph{et~al.}(1998)Diggle, Tawn, and
  Moyeed}]{Diggle_1998_spatial_GLMM}
Diggle PJ, Tawn JA, Moyeed RA (1998).
\newblock \enquote{Model-Based Geostatistics.}
\newblock \emph{Journal of the Royal Statistical Society C}, \textbf{47},
  299--350.

\bibitem[{Dunn and Smyth(1996)}]{Dunn_Smyth_1996_randomised_qunatile_residuals}
Dunn KP, Smyth GK (1996).
\newblock \enquote{Randomized Quantile Residuals.}
\newblock \emph{Journal of Computational and Graphical Statistics}, \textbf{5},
  1--10.

\bibitem[{Finley \emph{et~al.}(2015)Finley, Banerjee, and
  Gelfand}]{Finley_2015_spBayes}
Finley AO, Banerjee S, Gelfand AE (2015).
\newblock \enquote{{\pkg{spBayes}} for Large Univariate and Multivariate
  Point-Referenced Spatio-Temporal Data Models.}
\newblock \emph{Journal of Statistical Software}, \textbf{63}(13), 1--28.

\bibitem[{Finley \emph{et~al.}(2020)Finley, Datta, and
  Banerjee}]{Finley_2020_spNNGP}
Finley AO, Datta A, Banerjee S (2020).
\newblock \enquote{{\pkg{spNNGP} \proglang{R} Package for Nearest Neighbour
  Gaussian Process Models}.}
\newblock \textit{arXiv:2001.09111}.

\bibitem[{Furrer \emph{et~al.}(2006)Furrer, Nychka, and
  Genton}]{Furrer_2006_CovarianceTapering}
Furrer R, Nychka D, Genton MG (2006).
\newblock \enquote{Covariance Tapering for Interpolation of Large Spatial
  Datasets.}
\newblock \emph{Journal of Computational and Graphical Statistics},
  \textbf{15}, 502--523.

\bibitem[{Gelman and Hill(2007)}]{Gelman_Hill_2007}
Gelman A, Hill J (2007).
\newblock \emph{Data Analysis using Regression and Multilevel/Hierarchical
  Models}.
\newblock Cambridge University Press, Cambridge, England.

\bibitem[{Gelman \emph{et~al.}(1996)Gelman, Meng, and
  Stern}]{Gelman_1996_posterior_predictive_assessment}
Gelman A, Meng XL, Stern H (1996).
\newblock \enquote{Posterior Predictive Assessment of Model Fitness via
  Realized Discrepancies.}
\newblock \emph{Statistica Sinica}, \textbf{6}, 733 -- 807.

\bibitem[{Gneiting \emph{et~al.}(2007)Gneiting, Balabdaoui, and
  Raftery}]{Gneiting_2007_scoring_rules}
Gneiting T, Balabdaoui F, Raftery AE (2007).
\newblock \enquote{Probabilistic Forecasts, Calibration and Sharpness.}
\newblock \emph{Journal of the Royal Statistical Society B}, \textbf{69},
  243--268.

\bibitem[{Guennebaud \emph{et~al.}(2010)Guennebaud, Jacob
  \emph{et~al.}}]{Eigen}
Guennebaud G, Jacob B, \emph{et~al.} (2010).
\newblock \enquote{{\pkg{Eigen} v3}.}
\newblock \url{http://eigen.tuxfamily.org}.
\newblock Accessed: 2021-08-10.

\bibitem[{Hartig(2022)}]{DHARMa}
Hartig F (2022).
\newblock \emph{\pkg{DHARMa}: Residual Diagnostics for Hierarchical
  (Multi-Level/Mixed) Regression Models}.
\newblock \proglang{R} package version 0.4.5,
  \urlprefix\url{https://CRAN.R-project.org/package=DHARMa}.

\bibitem[{Hastie \emph{et~al.}(2009)Hastie, Tibshirani, and
  Friedman}]{Hastie_2009_Elements_Statistical_Learning}
Hastie T, Tibshirani R, Friedman J (2009).
\newblock \emph{The Elements of Statistical Learning}.
\newblock Second edition. Springer-Verlag, New York.

\bibitem[{Heaton \emph{et~al.}(2019)Heaton, Datta, Finley, Furrer, Guinness,
  Guhaniyogi, Gerber, Gramacy, Hammerling, Katzfuss, Lindgren, Nychka, Sun, and
  Zammit-Mangion}]{Heaton_2019_comparative_study}
Heaton MJ, Datta A, Finley AO, Furrer R, Guinness J, Guhaniyogi R, Gerber F,
  Gramacy RB, Hammerling D, Katzfuss M, Lindgren F, Nychka DW, Sun F,
  Zammit-Mangion A (2019).
\newblock \enquote{A Case Study Competition Among Methods for Analyzing Large
  Spatial Data.}
\newblock \emph{Journal of Agricultural, Biological and Environmental
  Statistics}, \textbf{24}, 398--425.

\bibitem[{Hersbach(2000)}]{Hersbach_2000_CRPS}
Hersbach H (2000).
\newblock \enquote{Decomposition of the Continuous Ranked Probability Score for
  Ensemble Prediction Systems.}
\newblock \emph{American Meteorological Society}, \textbf{15}, 559--570.

\bibitem[{Huang \emph{et~al.}(2009)Huang, Yao, Cressie, and
  Hsing}]{Huang_2009_multivar_intrinsic_rand_functions_cokriging}
Huang C, Yao Y, Cressie N, Hsing T (2009).
\newblock \enquote{Multivariate Intrinsic Random Functions for Cokriging.}
\newblock \emph{{Mathematical Geosciences}}, \textbf{41}, 887--904.

\bibitem[{Hughes(2014)}]{ngspatial_2014}
Hughes J (2014).
\newblock \enquote{{\pkg{ngspatial}: A Package for Fitting the Centered
  Autologistic and Sparse Spatial Generalized Linear Mixed Models for Areal
  Data}.}
\newblock \emph{{The \proglang{R} Journal}}, \textbf{6}, 81--95.

\bibitem[{Kristensen \emph{et~al.}(2016)Kristensen, Nielsen, Berg, Skaug, and
  Bell}]{Kristensen_2016_TMB}
Kristensen K, Nielsen A, Berg CW, Skaug H, Bell BM (2016).
\newblock \enquote{{\pkg{TMB}: Automatic differentiation and Laplace
  approximation}.}
\newblock \emph{Journal of Statistical Software}, \textbf{70}(5), 1--21.

\bibitem[{Lee and
  Park(2020)}]{Lee_2020_partitioned_domain_basis_function_non_Gaussian}
Lee BS, Park J (2020).
\newblock \enquote{A Scalable Partitioned Approach to Model Massive
  Nonstationary Non-gaussian Spatial Datasets.}
\newblock \textit{arXiv:2001.09111}.

\bibitem[{Leroux \emph{et~al.}(2000)Leroux, Lei, and
  Breslow}]{Leroux_2000_two_parameter_autoregressive_model}
Leroux B, Lei X, Breslow N (2000).
\newblock \enquote{Estimation of Disease Rates in Small Areas: A New Mixed
  Model for Spatial Dependence.}
\newblock In M~Halloran, D~Berry (eds.), \emph{Statistical Models in
  Epidemiology, the Environment and Clinical Trials}, pp. 179--191.
  Springer-Verlag, New York.

\bibitem[{Lindgren and Rue(2015)}]{Lindgren_2015_R-INLA}
Lindgren F, Rue H (2015).
\newblock \enquote{Bayesian Spatial Modelling with \proglang{R}-\pkg{INLA}.}
\newblock \emph{Journal of Statistical Software}, \textbf{63}(19), 1--25.

\bibitem[{Lindgren \emph{et~al.}(2011)Lindgren, Rue, and
  Lindstr{\"o}m}]{Lindgren_Rue_2011_GF_GMRF_SPDE}
Lindgren F, Rue H, Lindstr{\"o}m J (2011).
\newblock \enquote{An Explicit Link Between {Gaussian fields and Gaussian
  Markov} Random Fields: The Stochastic Partial Differential Equation
  Approach.}
\newblock \emph{Journal of the Royal Statistical Society B}, \textbf{73},
  423--498.

\bibitem[{Lopes \emph{et~al.}(2011)Lopes, Gamerman, and
  Salazar}]{Lopes_2011_spatial_GLMM_reduced_rank_factor_analytic_model}
Lopes HF, Gamerman D, Salazar E (2011).
\newblock \enquote{Generalized Spatial Dynamic Factor Models.}
\newblock \emph{Computational Statistics and Data Analysis}, \textbf{55},
  1319--1330.

\bibitem[{McCullagh and Nelder(1989)}]{McCullagh_Nelder_1989_GLM}
McCullagh P, Nelder JA (1989).
\newblock \emph{{Generalized Linear Models}}.
\newblock Chapman \& Hall, London, UK.

\bibitem[{{Melbourne Institute of Applied Economic and Social
  Research}(2011)}]{MIAESR_poverty_guidelines_2011}
{Melbourne Institute of Applied Economic and Social Research} (2011).
\newblock \enquote{Poverty Lines: {Australia}, {March} Quarter 2011.}
\newblock \emph{Technical report}.
\newblock URL
  \url{https://melbourneinstitute.unimelb.edu.au/assets/documents/poverty-lines/2017/Poverty-Lines-Australia-March-Quarter-2011.pdf}.
  Accessed 2021-08-10.

\bibitem[{{MODIS Characterization Support Team}(2015)}]{MODIS_satelitte}
{MODIS Characterization Support Team} (2015).
\newblock \enquote{{MODIS} 500m Calibrated Radiance Product. {NASA MODIS}
  Adaptive Processing System, {Goddard Space Flight Center, USA}.}
\newblock \url{https://mcst.gsfc.nasa.gov/}.

\bibitem[{Nychka \emph{et~al.}(2016)Nychka, Hammerling, Sain, and
  Lenssen}]{Nychka_2016_LatticeKrig}
Nychka D, Hammerling D, Sain S, Lenssen N (2016).
\newblock \emph{LatticeKrig: Multiresolution Kriging Based on Markov Random
  Fields}.
\newblock \proglang{R} package version 6.2,
  \urlprefix\url{www.image.ucar.edu/LatticeKrig}.

\bibitem[{Papritz(2020)}]{georob}
Papritz A (2020).
\newblock \emph{\pkg{georob}: Robust Geostatistical Analysis of Spatial Data}.
\newblock \proglang{R} package version 0.3-13,
  \urlprefix\url{https://cran.r-project.org/web/packages/georob/index.html}.

\bibitem[{Paul and
  Cressie(2011)}]{Paul_Cressie_2011_lognormal_kriging_block_prediction}
Paul R, Cressie N (2011).
\newblock \enquote{Lognormal Block Kriging for Contaminated Soil.}
\newblock \emph{{European Journal of Soil Science}}, \textbf{62}, 337--345.

\bibitem[{Pebesma and Bivand(2005)}]{Pebesma_2005_sp_package}
Pebesma EJ, Bivand RS (2005).
\newblock \enquote{Classes and Methods for Spatial Data in \proglang{R}.}
\newblock \emph{\proglang{R} News}, \textbf{5}, 9--13.

\bibitem[{{\proglang{R} Core Team}(2021)}]{Rcoreteam_2021}
{\proglang{R} Core Team} (2021).
\newblock \emph{\proglang{R}: A Language and Environment for Statistical
  Computing}.
\newblock \proglang{R} Foundation for Statistical Computing, Vienna, Austria.

\bibitem[{Rubin(1984)}]{Rubin_1984_model_selection}
Rubin DR (1984).
\newblock \enquote{{B}ayesianly Justifiable and Relevant Frequency Calculations
  for the Applied Statistician.}
\newblock \emph{Annals of Applied Statistics}, \textbf{12}, 1151 -- 1172.

\bibitem[{Rue and Martino(2007)}]{Rue_Martino_2007}
Rue H, Martino S (2007).
\newblock \enquote{Approximate {Bayesian} Inference for Hierarchical {Gaussian
  Markov} random field models.}
\newblock \emph{Journal of Statistical Planning and Inference}, \textbf{137},
  3177--3192.

\bibitem[{Rue \emph{et~al.}(2009)Rue, Martino, and Chopin}]{Rue_2009_INLA}
Rue H, Martino S, Chopin N (2009).
\newblock \enquote{Approximate {Bayesian} Inference for Latent {Gaussian}
  Models by Using Integrated Nested {Laplace} Approximations.}
\newblock \emph{Journal of the Royal Statistical Society B}, \textbf{71},
  319--392.

\bibitem[{Sengupta and Cressie(2013)}]{Sengupta_Cressie_2013_spatial_GLMM_FRK}
Sengupta A, Cressie N (2013).
\newblock \enquote{Hierarchical Statistical Modelling of Big Spatial Datasets
  Using the {Exponential Family} of Distributions.}
\newblock \emph{Spatial Statistics}, \textbf{4}, 14--44.

\bibitem[{{The University of Chicago
  Library}(2020)}]{Chicago_library_census_data_info}
{The University of Chicago Library} (2020).
\newblock \enquote{Spatially Referenced Census Data for the City of {Chicago}:
  Sources Available at or Through the {University of Chicago} Library.}
\newblock URL
  \url{https://www.lib.uchicago.edu/e/collections/maps/censusinfo.html}.
\newblock Accessed: 2021-08-10.

\bibitem[{Tierney and Kadane(1986)}]{Tierney_1986_Laplace_approx}
Tierney L, Kadane JB (1986).
\newblock \enquote{Accurate Approximations for Posterior Moments and Marginal
  Densities.}
\newblock \emph{Journal of the American Statistical Association}, \textbf{81},
  82--86.

\bibitem[{Wang and Furrer(2021)}]{spatialfusion}
Wang C, Furrer R (2021).
\newblock \enquote{Combining Heterogeneous Spatial Datasets With Process-Based
  Spatial Fusion Models: A Unifying Framework.}
\newblock \emph{Computational Statistics \& Data Analysis}, \textbf{161},
  107240.
\newblock \doi{10.1016/j.csda.2021.107240}.

\bibitem[{Wickham(2016)}]{Wickham_2016_ggplot2}
Wickham H (2016).
\newblock \emph{\pkg{ggplot2}: Elegant Graphics for Data Analysis}.
\newblock Springer-Verlag, New York, NY.
\newblock \urlprefix\url{https://ggplot2.tidyverse.org}.

\bibitem[{Wood(2017)}]{Wood_2017_GAM:R}
Wood S (2017).
\newblock \emph{Generalized Additive Models: An Introduction with
  \proglang{R}}.
\newblock Second edition. Chapman and Hall/CRC, Boca Raton, FL.

\bibitem[{Zammit-Mangion and Cressie(2021)}]{FRK_paper}
Zammit-Mangion A, Cressie N (2021).
\newblock \enquote{\pkg{FRK}: An \proglang{R} Package for Spatial and
  Spatio-Temporal Prediction with Large Datasets.}
\newblock \emph{Journal of Statistical Software}, \textbf{98}(4), 1--48.

\bibitem[{Zhang and Cressie(2020)}]{Zhang_2020_spatio-temporal_Arctic_sea_ice}
Zhang B, Cressie N (2020).
\newblock \enquote{Bayesian Inference of Spatio-Temporal Changes of {Arctic}
  Sea Ice.}
\newblock \emph{Bayesian Analysis}, \textbf{15}, 605--631.

\end{thebibliography}

\newpage

\addtocontents{toc}{\protect\setcounter{tocdepth}{1}} % set only sections
\numberwithin{equation}{section}

\begin{appendix}

\section{Construction of basis functions and BAUs}\label{Appendix:Basis_fns_and_BAUs}

In this appendix, we review the construction of basis functions and BAUs with \pkg{FRK}. Further details for each function discussed below are available in the package manual. 

In \pkg{FRK}, several standard basis functions, including the compactly-supported bi-square basis functions (default), may be constructed automatically from the data using \fct{auto\_basis}, or manually using \fct{local\_basis}. These functions produce an object of class \class{Basis}, and arbitrary, user-defined basis functions may be constructed with the constructor \fct{Basis}. In a spatio-temporal setting, \pkg{FRK} accommodates spatio-temporal data by using spatio-temporal basis functions constructed via a tensor product of spatial and temporal basis functions; this is achieved with the function \fct{TensorP}, which c%omputes the tensor product of a \class{Basis} object over some spatial manifold and a \class{Basis} object over the real line (corresponding to the temporal dimension), and 
returns an object of class \class{TensorP\_Basis}. %Note that \fct{auto_basis} can also be used in a spatio-temporal setting. 
 Useful methods include \fct{eval\_basis}, which evaluates a set of basis functions over arbitrary points or polygons, and \fct{show\_basis}, which visualizes a set of basis functions.

 The primary function for constructing BAUs is \fct{auto\_BAUs}, which automatically constructs the BAUs as an object of class \class{SpatialPixelsDataFrame} or \class{STFDF}, depending on whether the data is spatial or spatio-temporal. 
 In a spatio-temporal setting, the user may provide spatial BAUs to \fct{auto\_BAUs} via the argument \code{spatial\_BAUs}, which can be useful when the spatial domain can be partitioned using some real-world boundaries (e.g., the example in Section~\ref{sec:ST_example}).  
 The function \fct{BAUs\_from\_points} constructs BAUs from point-level data, which can be useful for replicating traditional geostatistical analyses.

\section[Parameterisations of K and Q]{Parameterisations of $\vec{K}$ and $\vec{Q}$}\label{Appendix:CovarianceTapering}

Recall from Section~\ref{subsection:04-01:ProcessLayer} that \pkg{FRK}~v2 allows the covariance matrix of basis-function coefficients, $\vec{\eta}$, to be parameterised using either a covariance matrix, $\vec{K}$, or using a precision matrix, $\vec{Q}$.  
 In this appendix, we describe the parameterisation of these matrices.
 Both $\vec{K}$ and $\vec{Q}$ are block-diagonal matrices, wherein basis-function coefficients within a basis-function resolution are dependent, but independent between different resolutions. Hence, $\vec{K}$ and $\vec{Q}$ are fully defined via their intra-resolution dependencies. 
 
\subsection[Covariance matrix K]{Covariance matrix $\vec{K}$}\label{Appendix:CovarianceTapering:covariance_matrix_K}

 Let $K_k(\vec{s}, \vec{s}^*)$ denote the covariance function associated with the basis-function coefficients corresponding to the $k$th basis-function resolution. 
 In \FRKgeneric, we let $K_k(\vec{s}, \vec{s}^*)$ be the exponential covariance function, that is, 
\begin{equation}\label{eqn:02-01:K_covariance_function}
    K_k(\vec{s}, \vec{s}^*)  = \sigma^2_k \exp \left\{\frac{-d(\vec{s},\vec{s}^*)}{\tau_k}\right\}, 
\end{equation}
where $d(\vec{s},\vec{s}^*)$ is the distance between two basis-function centroids $\vec{s}, \vec{s}^* \in D$, $\sigma^2_k$ is a variance parameter, and $\tau_k$ is a length-scale parameter.  The $k$th sub-block of $\vec{K}$ is formed by evaluating (\ref{eqn:02-01:K_covariance_function}) for all pairs of basis-function centroids at the $k$th resolution.

Clearly (\ref{eqn:02-01:K_covariance_function}) is always non-zero for $\sigma^2_k > 0$, however 
 it is often reasonable to assume that coefficients associated with fine-resolution basis functions separated by medium-to-large distances are uncorrelated. 
To increase sparsity, \pkg{FRK}~v2 now allows covariance tapering \citep{Furrer_2006_CovarianceTapering} of the intra-resolution covariance function. 
 Noting that (\ref{eqn:02-01:K_covariance_function}) is a special case of the Mat\'ern covariance function with smoothness parameter $\nu = 0.5$, we follow the recommendation of \cite{Furrer_2006_CovarianceTapering} and use the spherical taper: 
\begin{equation}\label{eqn:02-01:taper_function}
%K_{\beta_k}(\vec{s}, \vec{s}^*) 
T_{\beta_k}(\vec{s}, \vec{s}^*) 
=
\left\{1-\frac{d(\vec{s},\vec{s}^*)}{\beta_k}\right\}^2_{+}   \left\{1+\frac{d(\vec{s},\vec{s}^*)}{2\beta_k}\right\},
\end{equation}
where $x_{+} \equiv \max(0, x)$, and $\beta_k$ is a resolution-dependent tapering parameter controlling the strength of the taper. 
 In \pkg{FRK}~v2, we let $\beta_k$ be proportional to the minimum distance between basis-function centroids; specifically, we set $\beta_k = \texttt{taper} \times \text{mindist}(k)$, where $\text{mindist}(k)$ is the minimum distance between the centroids of basis functions at the $k$th resolution and \code{taper} is a user-specified argument.
The tapered covariance function is obtained by taking the product of the original covariance function (\ref{eqn:02-01:K_covariance_function}) and the taper function (\ref{eqn:02-01:taper_function}).
% Element wise product of the covaiance matrix and taper matrix

\subsection[Precision matrix Q]{Precision matrix $\vec{Q}$}\label{Appendix:SparsePrecisionMatrix}

\pkg{FRK}~v2 offers two types of sparse precision matrices: One is for regularly spaced basis functions, and the other is for irregularly spaced basis functions. 
 This choice is determined by the slot \code{regular} in the \class{Basis} object.

%When the basis functions are regularly spaced (\code{regular = TRUE}), \pkg{FRK}~v2 uses a precision matrix that is related to that used in the \proglang{R} package \pkg{LatticeKrig} \citep{Nychka_2016_LatticeKrig}. 
% Let $\mathcal{N}_{i,k}$ denote the set of first-order horizontal and vertical neighbouring basis functions of the $i$th basis function of resolution $k$, 
%%The number of elements in $\mathcal{N}_{ik}$, $|\mathcal{N}_{i,k}|$, will be four for interior basis functions, three for basis functions on edges, and two for basis functions on corners when the basis functions are arranged on a regular grid in $\mathbb{R}^2$.
% and let $\vec{Q}_k$ denote the precision matrix of the basis-function coefficients at resolution $k$. 
 
 \looseness=-1
 When the basis functions are regularly spaced (\code{regular = TRUE}), \pkg{FRK}~v2 uses a precision matrix based on the Leroux model \citep{Leroux_2000_two_parameter_autoregressive_model}. 
Let $\mathcal{N}_{i,k}$ denote the set of first-order horizontal and vertical neighbouring basis functions of the $i$th basis function of resolution $k$, 
%The number of elements in $\mathcal{N}_{ik}$, $|\mathcal{N}_{i,k}|$, will be four for interior basis functions, three for basis functions on edges, and two for basis functions on corners when the basis functions are arranged on a regular grid in $\mathbb{R}^2$.
 and let $\vec{Q}_k$ denote the precision matrix of the basis-function coefficients at resolution $k$. We model the elements of $\vec{Q}_k$ as
\begin{equation}\label{eqn:Q_k}
    \{\vec{Q}_{k}\}_{i, j}
=
\begin{cases}
\kappa_k + \rho_k |\mathcal{N}_{i,k}|   & i = j\\
-\rho_k & j \in \mathcal{N}_{i,k} \\
0 & \text{otherwise}
\end{cases},
\end{equation}
where $\kappa_k$ and $\rho_k$ are parameters that are estimated. 
%The full precision matrix $\vec{Q}$ is then $\text{bdiag}(\{\vec{Q}_k: k = 1,...,l\})$, where $l$ is the number of basis-function resolutions, and $\text{bdiag}(\cdot)$ returns a block-diagonal matrix from its arguments. 
We note that $\vec{Q}_k$ is diagonally dominant, and hence it is positive-definite. 
 This formulation implies that the coefficient of a given basis function is conditionally independent of all other basis-function coefficients given the coefficients of its first-order vertical and horizontal neighbours. 
% Note that \cite{Leroux_2000_two_parameter_autoregressive_model} used a model that is closely related to (\ref{eqn:Q_k}), and that \pkg{LatticeKrig} uses $\vec{Q}_k^\tp\vec{Q}_k$ as the precision matrix blocks, whereas \pkg{FRK}~v2 uses $\vec{Q}_k$.  
Note that \pkg{LatticeKrig} \citep{Nychka_2016_LatticeKrig} uses $\vec{Q}_k^\tp\vec{Q}_k$ as the precision matrix blocks, whereas \pkg{FRK}~v2 uses $\vec{Q}_k$.

 To cater for irregularly-spaced basis functions, \pkg{FRK}~v2 also offers a sparse precision matrix based on the distance between basis-function centroids:  
\begin{equation}\label{eqn:K_type_precision-block-exponential}
\{\vec{Q}_{k}\}_{i, j}
=
\begin{cases}
\kappa_k -\sum_{l \neq i}\{\vec{Q}_{k}\}_{i, l}  & i = j\\
-\rho_k \exp \left\{\frac{-d(\vec{s}_{i, k}, \vec{s}_{j, k})}{\tau_k}\right\}
    T_{\beta_k}(\vec{s}_{i, k}, \vec{s}_{j, k})  & i \neq j\\
\end{cases},
\end{equation}
where $\kappa_k$, $\rho_k$, and $\tau_k$ are parameters that are estimated, and $T_{\beta_k}(\cdot, \cdot)$ is defined as in Appendix~\ref{Appendix:CovarianceTapering:covariance_matrix_K}. 
 Again, this matrix is diagonally dominant and hence is positive-definite. 
 This formulation implies that the partial correlation between basis-function coefficients decays exponentially with distance until a point (controlled by the tapering parameter $\beta_k$) at which the basis-function coefficients are conditionally independent.

\section[Incidence matrices: CZ and CP]{Incidence matrices: $\vec{C}_Z$ and $\vec{C}_P$}\label{Appendix:Incidence matrices: Cz and Cp}

Recall from Section~\ref{subsection:DataLayer} that $\vec{C}_Z$ aggregates the BAU-level mean process, $\vec{\mu}$, over the observation supports and, depending on the weights in (\ref{eqn:C_Z}), it can correspond to a weighted average or a weighted sum over the BAUs. 
 In \pkg{FRK}~v2, the weights $w_{ij}$ may be controlled through the argument \mbox{\code{normalise\_wts}} and the \code{wts} field of the \class{SpatialPixelsDataFrame}/\class{SpatialPolygonsDataFrame} object \citep{Pebesma_2005_sp_package} used to store the BAUs.  
 Specifically, the \code{wts} field allows one to attribute each BAU to a \textit{relative} weight $v_i$, $i = 1, \dots, N$, such that $w_{ij} \propto v_i$, where the constant of proportionality can vary with $j$.   
 For example, if the BAUs are of unequal area, then one may wish to set $v_i = |A_i|$. 
 By default (and implicit in \pkg{FRK}~v1), each $v_i$ is set to 1. 
  The argument \mbox{\code{normalise\_wts}} controls whether $\vec{C}_Z$ corresponds to a weighted sum or a weighted average. If set to \code{FALSE}, then $w_{ij} = v_i$ for all $j$ (weighted sum); if set to \mbox{\code{TRUE}} (default and implicit in \pkg{FRK}~v1), then the $\{w_{ij}\}$ are normalised so that each row of $\vec{C}_Z$ sums to 1 (weighted average) and $w_{ij} = v_i / \sum_{l \in c_j} v_l$.   
%  Note that if $v_i = |A_i|$ and \mbox{\code{normalise\_wts = TRUE}}, the $j$th row sum of $\vec{C}_Z$ is $\sum_{i \in c_j} |A_i| = |B_j|$, so that the normalised weights are $w_{ij} = |A_i|/|B_j|$.  
  Note that if $v_i = |A_i|$ and \mbox{\code{normalise\_wts = TRUE}}, the normalised weights are $w_{ij} = |A_i|/|B_j|$, since the BAUs are disjoint and $\sum_{i \in c_j} |A_i| = |B_j|$.

Recall from Section~\ref{subsection:Prediction} that $\vec{C}_P$ aggregates the BAU-level mean process, $\vec{\mu}$, over the prediction regions and, depending on the weights in (\ref{eqn:C_P}), it can correspond to a weighted average or a weighted sum over the BAUs. 
 Like $\vec{C}_Z$, the relative weights, $\{\tilde{v}_i: i = 1, \dots, N\}$, such that $\tilde{w}_{ik} \propto \tilde{v}_i$, are controlled by the \code{wts} field of the BAU object, and the argument \mbox{\code{normalise\_wts}} is used to control whether $\vec{C}_P$ represents a weighted sum or a weighted average. 
 For consistency between the model fitting and prediction stages, 
 \pkg{FRK}~v2 enforces the use of the same relative weights, $\tilde{v}_i = v_i$ for $i = 1, \dots, N$, and the same setting of \code{normalise\_wts}, in construction of both $\vec{C}_Z$ and $\vec{C}_P$.

 Recall from Section~\ref{sec:Distributions with size parameters} that in most applications that consider binomial or negative-binomial data models, 
 the conditional mean of an observation is treated as a simple aggregate of the underlying mean process. 
 Therefore, with these distributions, \pkg{FRK}~v2 enforces the matrix $\vec{C}_Z$ in (\ref{eqn:C_Z}) to be constructed with the relative weights \mbox{$\{v_i = 1$ : $i = 1, \dots, N\}$} and with \mbox{\code{normalise\_wts = FALSE}}, and hence $w_{ij} = 1$ in (\ref{eqn:C_Z}). 
 Then, the mapping from the BAU-level mean, $\vec{\mu}$, to the data-level mean, $\vec{\mu}_Z$, in (\ref{eqn:mu_Z}) is a simple, unweighted summation over the BAUs.  
 Since \pkg{FRK}~v2 enforces the use of the same relative weights, $\tilde{v}_i = v_i$ for $i = 1, \dots, N$, and the same setting of \code{normalise\_wts}, in construction of both $\vec{C}_Z$ and $\vec{C}_P$, the mapping from $\vec{\mu}$ to $\vec{\mu}_P$ in (\ref{eqn:mu_P}) is also a simple, unweighted summation over the BAUs.

\section[Distributions with size parameters: Linking pi to mu]{Distributions with size parameters: Linking $\vec{\pi}$ to $\vec{\mu}$}\label{Appendix:Distributions with size parameters}

%Recall from Section~\ref{sec:Distributions with size parameters} that two distributions considered in this framework, namely, the binomial distribution and the negative-binomial distribution, have an assumed-known `size' parameter and a `probability of success'
%parameter. 
% Further recall that the logit, probit, and complementary log-log `link' functions have a special interpretation in the framework used by \pkg{FRK}~v2; namely, they are used to link the latent spatial process, $Y(\cdot)$, to the probability process, $\pi(\cdot)$: 
% \begin{equation}
%    f(\pi(\vec{s})) = Y(\vec{s}), \quad \vec{s} \in D,
%\end{equation}
%where $f(\cdot)$ is one of the aforementioned functions whose inverse has a range of $(0, 1)$. 
% The BAU-level probability process is $\vec{\pi} \equiv (\pi_i: i = 1, \dots, N)^\tp$, where $\pi_i = f^{-1}(Y_i)$,  $i = 1, \dots, N$.
% Next, we link the BAU-level mean process to the BAU-level probability process, 
%  \begin{equation}
% h(\mu_i; k_i) = \pi_i, \quad i = 1, \dots, N, 
% \end{equation}
% where $h(\cdot\,; \cdot)$ is determined solely by the response distribution, and 
%% \mbox{$\vec{k} \equiv (k_1, \dots, k_N)^\tp$} 
%\mbox{$\vec{k} \equiv (k_i: 1, \dots, N)^\tp$} 
% is the vector of BAU-level size parameters. 

%Recall from Section~\ref{sec:Distributions with size parameters} that $h(\cdot\,; \cdot)$ is a function used in (\ref{eqn:h_mu_equals_pi}) to link the BAU-level probability process, $\vec{\pi}$, to the BAU-level mean process, $\vec{\mu}$. 
 Recall from Section~\ref{sec:Distributions with size parameters} that $h(\cdot\,; \cdot)$ is a function that links the probability process, $\pi(\cdot)$, to the mean process, $\mu(\cdot)$. 
 In this appendix, we give its derivation.  
  The expectation of a binomial random variable, $Z \mid \pi, k \sim \text{Bin}(k, \pi)$, is $\E{Z} = k\pi$, which motivates the use of
  \begin{equation}\label{eqn:h_binomial}
 h(\mu; k) = \frac{\mu}{k},
 \end{equation}
 when the response distribution is binomial. 
 The expectation of a negative-binomial random variable, $Z \mid \pi, k \sim\text{NB}(k, \pi)$, 
%  The expectation of a negative-binomial random variable $Z$, with a target number of successes $k$ and with probability of success parameter $\pi$,     
 is $\E{Z} = \frac{k(1 - \pi)}{\pi}$, which motivates the use of
  \begin{equation}\label{eqn:h_neg_binomial}
 h(\mu; k) = \frac{k}{\mu + k},
 \end{equation}
 when the response distribution is negative-binomial. 
 By construction, using $h(\cdot\,; \cdot)$ to link the probability process to the mean process in (\ref{eqn:h_mu_equals_pi}), ensures that the range of the mean process is appropriate for modelling the expectation of a binomial distributed random variable, 
 where the mean must lie in the range $(0, k)$, %%% PREVENT ORPHAN/WIDOW
 or a negative-binomial distributed random variable,  
  where the mean must lie in the range $(0, \infty)$. %%% PREVENT ORPHAN/WIDOW

\section{Scoring rules}\label{app:ScoringRules}

Suppose that we have a discrete validation domain $D^* \subset D$, which is used for model validation.
As prediction-performance measures for the examples in this paper, we considered the following diagnostics (for simplicity, we describe the diagnostics in terms of prediction of the continuous mean process):

\begin{itemize}
    \item (Empirical) root-mean-squared prediction error (RMSPE): Let $\hat{\mu}(\vec{s})$ denote a point-predictor of $\mu(\vec{s})$, where $\mu(\vec{s})$ is the true value of the mean process evaluated at $\vec{s}\in D^*$. Then the empirical RMSPE, used to assess point-wise predictive performance, is
    \begin{equation*}
        \textrm{RMSPE}
        \equiv
        \sqrt{\frac{1}{|D^*|}\sum_{\vec{s} \in D^*}(\hat{\mu}(\vec{s}) - \mu(\vec{s}))^2}.
    \end{equation*}
    \item (Empirical) mean-absolute error (MAE): Also used to assess point-wise predictive performance, the empirical MAE is
    \begin{equation*}
        \textrm{MAE}
        \equiv
        \frac{1}{|D^*|}\sum_{\vec{s} \in D^*}|\hat{\mu}(\vec{s}) - \mu(\vec{s})|.
    \end{equation*}
    \item (Empirical) mean-absolute percentage error (MAPE): This is similar to the empirical MAE, but considers relative error instead; the empirical MAPE is
    \begin{equation*}
        \textrm{MAPE}
        \equiv
        \frac{1}{|D^*|}\sum_{\vec{s} \in D^*}\left|\frac{\hat{\mu}(\vec{s}) - \mu(\vec{s})}{\mu(\vec{s})}\right|.
    \end{equation*}
    \item (Averaged) continuous ranked probability score \citep[CRPS;][sec 4.2.]{Gneiting_2007_scoring_rules}: 
    The averaged CRPS is used to evaluate the predictive cumulative distribution function (CDF) of the mean process, $F(\mu; \vec{s}, \vec{Z})$,  over all $\vec{s} \in D^*$, and is defined as
    \begin{equation*}
%    \textrm{CRPS}(F, \mu(\vec{s})) 
    \textrm{CRPS}
    \equiv \frac{1}{|D^*|}\sum_{\vec{s} \in D^*}
    \int_{-\infty}^\infty (F(x; \vec{s}, \vec{Z}) - \mathbbm{1}\{x \geq \mu(\vec{s})\})^2 \d x,
    \end{equation*}
    where $\mathbbm{1}\{ \cdot \}$ denotes an indicator function that takes the value 1 if its argument is true, and 0 otherwise. 
    For some predictive CDFs (in particular, the Gaussian and log-Gaussian), there exist closed-form expressions to compute the CRPS. However, in general, no closed-form expression exists, in which case we may use an \textit{empirical} predictive CDF from a sample (e.g., a Monte Carlo sample) to evaluate the CRPS in terms of the respective order statistics \citep{Hersbach_2000_CRPS}. 
    \item (Averaged) interval score \citep[IS;][sec.~6.2]{Gneiting_2007_scoring_rules}: Given a set of purported $(1-\alpha)\times 100$\% prediction intervals for $\mu(\vec{s})$, $\vec{s} \in D^*$, the averaged IS is defined as
    \begin{align*}
    \text{IS}_\alpha
    \equiv 
    \frac{1}{|D^*|}\sum_{\vec{s} \in D^*}
    \bigg(
    & U(\vec{s}) - L(\vec{s}) 
    + \\
    &\frac{2}{\alpha}(L(\vec{s})-\mu(\vec{s})) \mathbbm{1}\{\mu(\vec{s}) < L(\vec{s})\}
    + \frac{2}{\alpha}(\mu(\vec{s})- U(\vec{s})) \mathbbm{1}\{\mu(\vec{s}) > U(\vec{s})\} \bigg),
    \end{align*}
    where $L(\vec{s})$ and $U(\vec{s})$ are the lower and upper bounds of the prediction interval at location $\vec{s}$.
    The IS rewards narrow prediction intervals and penalises instances in which an observation misses the interval, with the size of the penalty depending on $\alpha$. 
    \item (Empirical) coverage: The empirical coverage of the prediction intervals is defined as 
    \begin{equation*}
    \text{Cvg} \equiv \frac{1}{|D^*|}\sum_{\vec{s} \in D^*} \mathbbm{1}\{L(\vec{s}) \leq \mu(\vec{s})  \leq U(\vec{s})\}
    \end{equation*}
    If the intervals are indeed $(1-\alpha)\times 100$\% prediction intervals throughout $D^*$, the empirical coverage should be approximately equal to $1-\alpha$.
    \item Brier score \citep[Sec. 3]{Gneiting_2007_scoring_rules}: The Brier score %, applicable in a binary setting, 
    is defined as 
    \begin{equation*}
    \text{Brier Score} \equiv
    \frac{1}{|D^*|}\sum_{\vec{s} \in D^*} (Z_{\vec{s}} - \hat{\pi}(\vec{s}))^2,
    \end{equation*}
    where $Z_{\vec{s}}$ denotes the validation datum at $\vec{s}$ (taking a value of 0 or 1), and $\hat{\pi}(\vec{s})$ denotes a point-prediction of the probability process at $\vec{s}$.
\end{itemize}

\section{Sydney poverty lines}\label{Appendix:Sydney_data_description}

Here we provide some details on how we define the poverty lines for the data in Section~\ref{sec:spatialCOS}. 
We base our definitions of poverty lines on a Melbourne Institute of Applied Economic and Social Research (MIAESR) report that was published in March 2011 \citep{MIAESR_poverty_guidelines_2011}.
However, the family units in the 2011 Australian Census do not align exactly with those used by the MIAESR and, since this example is shown for purely illustrative purposes, we make several assumptions.  
First, we assume `families with children' in the Census data consist of exactly two parents and two children.
Second, since `other families' in the Census is difficult to interpret and categorise appropriately in the context of the MIAESR guidelines, we exclude `other families' from the study (less than 2\% of all families). 
Third, the Census data do not provide exact income figures, but rather they provide income brackets of width \$200; we thus round the MIAESR guidelines to the nearest \$200. 
Fourth, the Census data do not make clear whether the head of the family is in the workforce; we therefore assume that the head of the family \textit{is} in the workforce, and hence we use the first half of Table 1 of the MIAESR report guidelines for defining poverty lines.
These assumptions lead us to define poverty lines (in Australian dollars) for each family unit considered in this study as weekly incomes of: \$600 for a couple with no children, \$800 for a couple with children, and \$600 for a one-parent family. The proportion of families we deem to be in poverty 
%in each region 
is based on their being below these thresholds.

\end{appendix}
\end{document}